\documentclass[12pt]{article}
\pdfoutput=1
\usepackage{putex}
\usepackage{comment}
\usepackage{graphicx}
\usepackage{caption}
\usepackage{amsmath}
\usepackage[usenames,dvipsnames,table]{xcolor}
\usepackage{array}
\usepackage{subcaption}
\usepackage{epstopdf}
\usepackage{enumerate}
\usepackage{cite}
\usepackage{tensor}
\usepackage{slashed}
\usepackage[export]{adjustbox}
\usepackage[aligntableaux=center]{ytableau}
\usepackage{dsfont}
\usepackage[utf8]{inputenc}
\usepackage[
      colorlinks=true,
      linkcolor=blue,
      urlcolor=blue,
      filecolor=black,
      citecolor=red,
      ]{hyperref}
\usepackage{footmisc}
\newcommand{\RNum}[1]{\uppercase\expandafter{\romannumeral #1\relax}}

\newcommand {\be} {\begin {equation}}
\newcommand {\ee} {\end {equation}}

\newcommand {\bes} {\begin {equation*}}
\newcommand {\ees} {\end {equation*}}

\newcommand{\beq}{\begin{equation}}
\newcommand{\eeq}{\end{equation}}

\def\ie{\begin{equation}\begin{aligned}}
\def\fe{\end{aligned}\end{equation}}

\numberwithin{equation}{section}

\def\<{\langle}
\def\>{\rangle}
\def \eps {\epsilon}
\def\ell{l}

\begin{document}



\institution{PU}{ $^{a}$ Department of Physics, Princeton University, Princeton, NJ 08544, USA }
\institution{}{ $^b$ Center for Cosmology and Particle Physics, New York University, New York, NY 10003, USA}

\title{RG Interfaces from Double-Trace Deformations} 

\authors{Simone Giombi$^a$, Elizabeth Helfenberger$^a$, and Himanshu Khanchandani$^b$}

\abstract{We study a class of interface conformal field theories obtained by taking a large $N$ CFT and turning on a relevant double-trace deformation over half space. At low energies, this leads to a conformal interface separating two CFTs which are related by RG flow. We set up the large $N$ expansion of these models by employing a Hubbard-Stratonovich transformation over half space, 
and use this approach to compute some of the defect CFT data. We also calculate the free energy of the theory in the case of spherical interface, which encodes a conformal anomaly coefficient for even dimensional interface, and the analog of the $g$-function for odd-dimensional interface. These models have a dual description in terms of a gravitational theory in AdS where a bulk scalar field satisfies different boundary conditions on each half of the AdS boundary. We review this construction and show that the results of the large $N$ expansion on the CFT side are in precise agreement with the holographic predictions.}

\date{}
\maketitle

\tableofcontents

\section{Introduction and Summary}
Conformal interfaces are codimension one defects separating two conformal field theories (CFTs). They constitute an interesting class of defect CFTs (for an introduction to defect CFT see \cite{Billo:2016cpy, Andrei:2018die}), and they have been extensively studied in a variety of different contexts, see for instance \cite{Bachas:2001vj, Bak:2003jk, Clark:2004sb, Sakai:2008tt, Brehm:2015lja, Wen:2017smb, Karch:2022vot, Karch:2023evr}. Of special interest among conformal interfaces are renormalization group (RG) interfaces, which separate two CFTs that are related by an RG flow \cite{Brunner:2007ur, Brunner:2007qu, Gaiotto:2012np}. They may be constructed by starting with a CFT and introducing a relevant deformation on half space, such that upon RG flow one gets an ``IR CFT" over one half space, and an ``UV CFT" (the original CFT) over the other half space. A simple example of such an object is a free field theory with a mass deformation turned on over half space. This results in a free scalar with Dirichlet boundary conditions, and the boundary can be thought of as an interface between the free scalar and the empty theory. RG interfaces have been extensively studied in two dimensions \cite{Brunner:2007ur, Brunner:2007qu, Gaiotto:2012np,Konechny:2014opa,Konechny:2016eek, Poghosyan:2014jia,Brunner:2015vva, Cogburn:2023xzw}. In this paper, we study RG interfaces in higher dimensions, which are relatively less well studied \cite{Dimofte:2013lba, Gliozzi:2015qsa, Melby-Thompson:2017aip}.

In the case of a $d$-dimensional CFT with a large $N$ expansion, a well-known example of RG flow is obtained by turning on a relevant double-trace deformation \cite{Klebanov:1999tb, Witten:2001ua, Gubser:2002vv, Berkooz:2002ug, Mueck:2002gm, Gubser:2002zh, Hartman:2006dy, Diaz:2007an, Giombi:2018vtc}
\begin{equation}
S = S_{CFT} + \lambda \int d^d x \ O^2 
\label{double-trace-def}
\end{equation}
where $O$ is a single-trace operator in the original CFT with scaling dimension $\Delta < d/2$. This relevant perturbation triggers an RG flow to an IR CFT in which the operator $O$ has scaling dimension $d - \Delta + O(1/N)$, while the scaling dimensions and OPE coefficients involving all single trace operators other than $O$ receive corrections to subleading order in the large $N$ expansion. The CFT data in the IR may be computed in the $1/N$ expansion using the standard Hubbard-Stratonovich transformation. 

To obtain a double-trace RG interface, one simply turns on the double-trace deformation only over half of the space\footnote{For a related application of double-trace deformations to defect CFTs, see \cite{Brax:2023goj}.}
\begin{equation}
S = S_{CFT} + \lambda \int _{z>0}d^dx\, O^2\,,
\label{double-trace-def-half}
\end{equation}
where $z$ denotes the coordinate transverse to the plane at $z=0$ separating the left and right halves. Upon RG flow, one then finds that the local CFT data (the scaling dimensions and the ``bulk" OPE coefficients) are that of the IR CFT for $z>0$, and that of the UV CFT for $z<0$. At $z=0$, one has a conformal interface separating the two CFTs. Then, in addition to the local ``bulk" data, one also has a spectrum of defect operators living on the interface, as well as new ``bulk-defect" OPE data. In Section \ref{Sec:LargeN}, we set up the large $N$ expansion of the system by using a Hubbard-Stratonovich transformation, and compute some of the defect CFT data, including certain one-point and two-point correlation functions. For the case where the interface is spherical, we also compute the free energy of the theory, to the first non-trivial order in the $1/N$ expansion. For an even-dimensional interface, this free energy can be used to extract one of the conformal anomaly coefficients associated with the interface, while for an odd-dimensional interface it encodes an analog of the $g$-function studied in 2d CFTs \cite{PhysRevLett.67.161, Friedan:2003yc}.  

Well-known examples of CFTs like the critical $O(N)$ vector model and the Gross-Neveu model can be thought of as arising from a double-trace deformation (\ref{double-trace-def}) of free scalar and free fermion models respectively (in this case, the ``single trace" operator $O$ corresponds to the bilinears $\phi^i \phi^i$ or $\bar\psi_i \psi^i$). When the deformation is turned on over half space, one then gets an RG interface separating ``free" and ``critical" scalar or fermionic vector models.  In Section \ref{sec:examples} we apply our general large $N$ results to these specific examples, and check them against the prediction of the epsilon expansion (the RG interface between the free scalar and the Wilson-Fisher fixed point was studied in $d=4-\epsilon$ in \cite{Gliozzi:2015qsa}).

The holographic description of a double-trace deformation is well-known \cite{Klebanov:1999tb, Witten:2001ua}: it corresponds to a change in boundary conditions for the bulk scalar field in AdS which is dual to the single trace operator $O$. 
The dimension $\Delta$ of a CFT operator corresponding to an AdS scalar field $\varphi$ with mass $m$ is given by the relation $m^2=\Delta(\Delta-d)$ (we set the AdS radius to one). The two solutions to this equation correspond to the possible boundary behaviors $\varphi \sim z^{\Delta_{\pm}}$ as $z\rightarrow 0$ (here $z$ denotes the holographic AdS coordinate in the usual Poincare metric $ds^2=(dz^2+dx^{\mu} dx^{\mu})/z^2$). In the range $d/2 - 1 < \Delta < d/2$ (or $-d^2/4<m^2<-d^2/4+1$), both behaviors lead to possible unitary boundary conditions for the scalar field: the choice $\varphi \sim z^{\Delta}$ corresponds to the UV CFT (where $O$ has dimension $\Delta<d/2$), while the choice $\varphi \sim z^{d - \Delta}$ corresponds to the IR CFT, where the dual operator has dimension $d-\Delta$. Note that both the UV and IR CFT are described by the same dual theory in AdS, the only difference is in the choice of boundary conditions for a bulk scalar field. 

The above dictionary has a natural generalization to the case of the double-trace interface defined by (\ref{double-trace-def-half}). As discussed in \cite{Melby-Thompson:2017aip}, one should divide the boundary of AdS$_{d+1}$ in two regions (corresponding to the UV and IR CFTs), and impose ``inhomogeneous" boundary conditions on the scalar field dual to $O$: namely, one imposes $\varphi \sim z^{\Delta}$ asymptotics as we approach one of the boundary regions, and $\varphi \sim z^{d-\Delta}$ as we approach the other region.\footnote{This is an example of a ``mixed boundary value problem", but to avoid confusion with the notion of mixed boundary conditions \cite{Witten:2001ua, Hartman:2006dy} (which are imposed over the full boundary and describe the double-trace deformation (\ref{double-trace-def}) along the flow), we will refer to the present case as inhomogeneous boundary conditions. Of course, one may in principle also study the half-space deformation (\ref{double-trace-def-half}) along the flow, by imposing the mixed boundary conditions of \cite{Witten:2001ua} over half of the boundary, and ordinary boundary conditions over the other half.} In practice, this can be conveniently done using Janus coordinates \cite{Bak:2003jk}, as reviewed in Section \ref{Sec:HolographicSetup} (see Figure \ref{holographic-sketch} for a sketch of the setup). Note that the geometry in the bulk remains that of pure AdS, and the pattern of conformal symmetry breaking $SO(d+1,1)\rightarrow SO(d,1)$ associated with the interface is realized in the bulk just via the choice of inhomogenous boundary conditions on the scalar field (there is no backreaction on the classical geometry). In \cite{Melby-Thompson:2017aip}, this holographic setup was used to compute the free energy and some of the defect CFT data of the interface CFT. We review these calculations below and show that they reproduce, as expected, the results of Section \ref{Sec:LargeN} obtained directly on the CFT side. We show that the agreement between CFT and AdS calculations essentially follows from certain identities satisfied by the scalar Green's function with the inhomogeneous boundary conditions (generalizing similar arguments in \cite{Hartman:2006dy, Giombi:2011ya, Giombi:2018vtc}), and discuss in detail various examples of one-, two-, and three-point functions. 

An explicit realization of this holographic setup is provided by the scalar and fermionic vector models mentioned above. In this case, the dual is expected to be the Vasiliev higher spin gravity in AdS (see \cite{Giombi:2012ms, Giombi:2016ejx} for reviews of the higher spin/vector model duality), with inhomogeneous boundary conditions imposed on the bulk scalar field dual to the $\phi^i\phi^i$ or $\bar\psi_i \psi^i$ operators. In addition to the conformal symmetry breaking, which leads to the protected displacement operator, in this case one also expects a tower of protected higher spin operators on the interface, as we discuss briefly in Section \ref{sec:examples}. It would be interesting to further study properties of the RG interface in the dual higher spin gravity theories.  

The rest of this paper is organized as follows: in Section \ref{Sec:LargeN}, we set up the large $N$ expansion of the double-trace RG interface introducting the Hubbard-Stratonovich auxiliary field, and compute some of the correlation functions and defect CFT data of the theory. In Section \ref{Sec:FreeEnergy}, we use the large $N$ approach to compute the free energy for a spherical interface, to the first subleading order at large $N$. In Section \ref{sec:examples}, we discuss various explicit examples involving free fields on one side of the interface, apply our general large $N$ results to these models, and make some 
cross-checks using the Wilson-Fisher epsilon expansion. In Section \ref{Sec:HolographicSetup}, we review the holographic description of the double-trace interfaces, and then in Section \ref{sec:equivalence} we discuss the agreement between the CFT and holographic calculations. Several technical details and calculations are collected in the Appendices.

\section{Large $N$ Setup} \label{Sec:LargeN}
We start by describing the setup for double trace RG interfaces in a general large $N$ CFT, and then calculate the two-point functions in the presence of the interface. These two-point functions were obtained in \cite{Melby-Thompson:2017aip} by doing a bulk AdS calculation, but here we derive them from a purely CFT calculation.

Consider a CFT in $d$ dimensions with a sensible large $N$ limit which contains a single trace operator $O$ with scaling dimension $\Delta<d/2$. Then we can define an RG interface in flat space $(\mathbf{x}, z)$ placed at $z=0$ by the following action 
\begin{equation}
S = S_0 + \lambda \int _{\mathbb{R}_+^d}\ O^2
\end{equation}
where ${\mathbb{R}_+^d}  = \{(\mathbf{x},z): z>0\}$ and $S_0$ is the action of the original CFT (this will be referred to as the UV CFT below). We will use the convention that the right half space is $z > 0$ and left half space is $z < 0$. The partition function in the presence of a source for the operator $O$ is 
\begin{equation}
\begin{split}
Z[J] &= \int [\mathcal{D} \phi] e^{-S_0 [\phi] -\lambda \int_{\mathbb{R}_+^d}   O^2  \  + \ \int_{\mathbb{R}^d} J O}\\
&= Z_{0} \langle e^{-\lambda \int_{\mathbb{R}_+^d}    O^2 \ + \ \int_{\mathbb{R}^d}  J O} \rangle_{0}
\end{split}
\end{equation}
where $\langle \dots \rangle_{0}$ denotes the expectation value in the UV theory with $J=0$ and $\lambda = 0$. We proceed as in \cite{Gubser:2002vv, Klebanov:2011gs} and perform a large $N$ expansion by introducing a Hubbard-Stratonovich auxiliary field $\sigma$ 
\begin{equation}
\begin{split}
\frac{Z[J]}{Z_{0}} &= \int [\mathcal{D} \sigma]  \langle e^{ - \int_{\mathbb{R}_+^d} d^d x \ \left(\sigma O -\frac{\sigma^2}{4\lambda} \right) +\int_{\mathbb{R}^d}  d^d x J O } \rangle_{0} \\
\label{HS-trick}
\end{split}
\end{equation}
Since $O$ is a single trace operator and we have a large $N$ CFT, the higher point correlation functions of $O$ factorize at large $N$ into products of two-point functions:
\begin{equation}
\langle e^{-\int (\sigma \theta(z) - J) O } \rangle_{0} = e^{\frac{1}{2} \int d^d x_1 d^d x_2 (\theta(z_1) \sigma-J)(x_1)(\theta(z_2)\sigma -J)(x_2) \langle O(x_1) O(x_2) \rangle_{0}} + O(1/N).
\end{equation}
where $\theta(z)$ is the Heaviside step function. The two-point function of $O$ in the undeformed theory has the following normalization
\begin{equation}
\begin{split}
G_{12} \equiv \langle O(x_1) O(x_2) \rangle_{0} &= \frac{C_{O}}{x_{12}^{2\Delta}}. 
\label{G0-norm}
\end{split}
\end{equation}
At this point it is more convenient to work in the folded picture, taking $z \to -z$ for the unperturbed (left) side, which turns the interface into a boundary of the product theory $\overline{\text{CFT}}_\text{UV} \times \text{CFT}_\text{IR}$. We may now express the action as an integral over $\mathbb{R}_+^d$ by reflecting points on the ``left'' side of the interface using $\mathcal{R}(\mathbf{x}, z) = (\mathbf{x}, -z)$, and we denote $J_{\mathcal{R}i} \equiv J(\mathcal{R} x_i)$. The partition function is Gaussian upon shifting $\sigma$, which gives the following result after integrating out $\sigma$
\begin{equation}\label{eq:partition}
\begin{split}
&\frac{Z[J]}{Z_{0}} = e^{-\frac{1}{2}\log\det(G_{12}+\frac{\delta(x_{12})}{2\lambda})}\exp\biggr[ \frac{1}{2}\int_{\mathbb{R}_+^d} d^d x_1 d^d x_2\biggr( J_1 G_{12} J_2 + 2J_1 G_{1,\mathcal{R} 2} J_{\mathcal{R}2} + J_{\mathcal{R}1}G_{\mathcal{R}1\mathcal{R}2}J_{\mathcal{R}2} \\
&\qquad -\int_{\mathbb{R}_+^d}  d^d x_3 d^d x_4 \left(J_1 G_{13} + J_{\mathcal{R}1} G_{\mathcal{R}1,3}\right) \left(G_{34}+\frac{\delta(x_{34})}{2\lambda}\right)_{+}^{-1}\left( G_{42} J_2 + G_{4,\mathcal{R}2}J_{\mathcal{R}2}\right)\biggr)\biggr]
\end{split}
\end{equation}
where $\left(G(x-x_3)+\frac{\delta(x-x_3)}{2\lambda}\right)_{+}^{-1}$ is the inverse of $G(x-x_3)+\frac{\delta(x-x_3)}{2\lambda}$ over positive half space:
\begin{equation}
\int_{\mathbb{R}_+^d} d^d x   \left(G(x_1-x)+\frac{\delta(x_1-x)}{2\lambda}\right)_{+}^{-1} \left(G(x-x_2)+\frac{\delta(x-x_2)}{2\lambda}\right) = \delta(x_1-x_2) \hspace{0.5cm} \text{ if }z_1,z_2>0 .
\end{equation}
We may then take the functional derivative of \eqref{eq:partition} with respect to $J$ and take IR limit $\lambda \to \infty$. The functional derivative with respect to $J_{\mathcal{R}}$ and $J$ bring down factors of $O$ and $\sigma/ 2 \lambda$ respectively (this is because $O$ on the right side is identified with $\sigma/(2\lambda)$, as follows from how the auxiliary field was introduced in (\ref{HS-trick})). This gives us the following two-point functions for $O$ and $\sigma$ (using the ``L'' subscript to denote points on the left half of space in the unfolded theory):
\begin{equation}
\begin{split}
\langle \sigma(x_1)\sigma(x_2) \rangle  &= -G_{+}^{-1}(x_1-x_2) \\
\langle \sigma(x_1)O(x_{L2}) \rangle &= \int_{\mathbb{R}_+^d} d^d x_3 G_{+}^{-1}(x_1-x_3)G(x_3- x_{L2}) \\ 
&= -\int_{\mathbb{R}_+^d} d^d x_3 \langle \sigma(x_1)\sigma(x_2)\rangle \,\langle O(x_{L2})O(x_3)\rangle_0 \\
\langle O( x_{L1}) O(x_{L2}) \rangle  &= \langle O(x_{L1}) O( x_{L2}) \rangle_{0} \\
&\qquad +  \int_{\mathbb{R}_+^d} d^d x_3 d^d x_4 \langle O( x_{L1}) O(x_3)\rangle_{0}   \langle \sigma(x_3) \sigma(x_4) \rangle \langle O(x_4) O(x_{L2}) \rangle_{0}   \\
\label{basic-2pt}
\end{split}
\end{equation}
where here and below a correlation function with no subscript $\langle \dots \rangle$ is understood to be a correlation function in the interface theory in the IR limit. Note that when the operator $O$ appears in correlation functions, it is always inserted on the right side of the interface (the UV theory side), while $\sigma$ plays the role of $O$ on the right side (the IR theory side).  
Let us note that the above results may also be obtained in a quicker way by taking the IR limit directly in the action by dropping the quadratic term $\sigma^2/(2\lambda)$, and using the action 
\begin{equation}
S = S_{0} + \int_{\mathbb{R}_+^d}  d^d x \ \sigma O 
\label{IR-action}
\end{equation}
to develop the $1/N$ expansion. This approach is well-known in the study of double-trace deformations (see e.g. \cite{Giombi:2018vtc}), and it works essentially the same way in our case where the perturbation is only on half-space. The action yields an ``induced" propagator for $\sigma$, given by the first line of (\ref{basic-2pt}), and correlation functions may be computed in the $1/N$ expansion by treating the second term in (\ref{IR-action}) as an interaction. One can readily obtain the second and third line of (\ref{basic-2pt}) proceeding this way. 

Using (\ref{basic-2pt}), we can explicitly calculate the two-point functions involving $O$ and $\sigma$ in the presence of an interface. The two-point function of $\sigma$ can be found by inverting $-G(x_1-x_2)$ on positive half space: 
\begin{equation}\label{eq:sigmadef}
\int_{\mathbb{R}_+^d} d^d x \langle O(x_1) O(x) \rangle_{0} \langle \sigma (x) \sigma(x_2) \rangle = -\delta(x_1-x_2), \qquad x_1, x_2 \in \mathbb{R}_+^d .
\end{equation}
One can perform the inversion using the methods in \cite{McAvity:1995zd}. The details of the calculation are given in Appendix \eqref{App:SigmaPropagator}, and here we quote the result:
\begin{equation}\label{eq:SigmaPropagator}
\begin{split}
\langle \sigma(x_1) \sigma(x_2) \rangle 
\equiv \frac{h(\xi_{12})}{(4 z_1 z_2)^{d-\Delta}} = \frac{\tilde{C}_\sigma}{(4 z_1 z_2)^{d-\Delta}} \xi_{12}^{-\frac{d}{2}} {}_2 F_1 \left( \frac{d}{2},  - \frac{d}{2} + \Delta, 1 - \frac{d}{2} + \Delta, -\frac{1}{\xi_{12}} \right) 
\end{split}
\end{equation}
where we introduced 
\begin{equation}
\xi_{12} = \frac{(x_1-x_2)^2}{4 z_1 z_2}, \hspace{1cm} \tilde{C}_\sigma = \frac{\Gamma (\Delta) \sin \pi\left( \Delta - \frac{d}{2} \right) \Gamma \left( \frac{d}{2} \right)  }{C_O \pi^{d + 1} \Gamma \left( \Delta - \frac{d}{2} \right)} . 
\label{cross-r}
\end{equation}
This result can then be used to find the two-point function of $O$ and $\sigma$: 
\begin{equation}\label{eq:Osigma}
\begin{split}
\langle O(x_{L1}) \sigma(x_2) \rangle &= -\int_{\mathbb{R}_+^d} d^d x \langle O(x_{L1})O(x) \rangle_{0} \langle \sigma(x)\sigma(x_2) \rangle \\
&= \frac{1}{(-2z_{L1})^{\Delta} (2z_2)^{d-\Delta}} \frac{\Gamma(\tfrac{d}{2}) \sin\pi(\tfrac{d}{2}-\Delta)}{\pi^{d/2+1}(-\xi_{12})^{d/2}}
\end{split}
\end{equation}
The details of the integral are given in \eqref{App:OSigmaPropagator}. Finally, the two-point function of $O$ is then
\begin{equation}\label{eq:OO}
\begin{split}
\langle O(x_{L1}) O(x_{L2}) \rangle &= \langle O(x_{L1}) O(x_{L2}) \rangle_{0} - \int_{\mathbb{R}_+^d} d^d x  \langle O(x_{L1})O(x) \rangle_{0}  \langle \sigma(x) O(x_{L2})\rangle   \\
&= \frac{1}{(4 z_1 z_2)^{\Delta}} \frac{C_O  \Gamma \left( \frac{d}{2} \right)  }{\Gamma(\Delta) \Gamma \left( \frac{d}{2} + 1 - \Delta \right) \xi^{ \frac{d}{2}}} {}_2 F_1 \left( \frac{d}{2},  \frac{d}{2} - \Delta, 1 + \frac{d}{2} - \Delta, -\frac{1}{\xi_{12}} \right).
\end{split}
\end{equation}
The integral was performed using the techniques described in \eqref{App:OPropagator}. Note that this is the same as the $\sigma$ two-point function upon sending $\Delta \to d-\Delta$ and $C_O \to C_\sigma$, where $C_\sigma$ is the normalization of $\langle \sigma \sigma \rangle$ in the theory without the interface (i.e. if the double trace deformation were defined on whole space, see for instance \cite{Giombi:2018vtc}):
\begin{equation}\label{eq:Csig-norm}
C_\sigma = -\frac{\pi ^{-d} \Gamma (\Delta ) \Gamma (d-\Delta )}{C_O \Gamma \left(\frac{d}{2}-\Delta \right) \Gamma \left(\Delta -\frac{d}{2}\right)}.
\end{equation}
All of the three two-point functions in (\ref{basic-2pt}) are in agreement with what was obtained from AdS calculation in \cite{Melby-Thompson:2017aip}, as reviewed in Section \ref{Sec:HolographicSetup} below. 
\subsection{Interface CFT data}\label{Sec:InterfaceData}
Having obtained the basic two-point functions, we now extract various pieces of interface CFT data. In the presence of a defect, there is a larger set of CFT data due to local excitations $\hat{\Phi}$ on the defect which couple to bulk operators $\Phi$ (see \cite{Billo:2016cpy} for a review). For scalar operators, the bulk-to-defect two-point function is constrained by conformal symmetry to take the form 
\begin{equation}
\langle \Phi_\Delta (\mathbf{x}_1, z_1) \hat{\Phi}_{\hat{\Delta}}(\mathbf{x}_2 ) \rangle = \frac{\mathcal{N}_\Phi \mu_{\Phi\hat{\Phi}}}{|2 z_1|^{\Delta - \hat{\Delta}} |x_{12}|^{2\hat{\Delta}}}.
\end{equation}
where $\mathcal{N}_\Phi $ is the normalization of $\Phi$ in the UV theory on whole space. Taking $ \hat{\Phi}$ to be the identity operator on the defect, we see that bulk operators can have a one-point function
\begin{equation}
\begin{array}{lll}
\langle \Phi_\Delta (\mathbf{x}_1, z_1)  \rangle &= \dfrac{\sqrt{\mathcal{N}_\Phi } a_{\Phi,L}}{(-2z_1)^{\Delta}}, \quad & z_1 < 0 \\
& & \\
  &=  \dfrac{\sqrt{\mathcal{N}_\Phi }  a_{\Phi,R}}{(2z_1)^{\Delta}}, \quad & z_1 > 0 .
\end{array}
\end{equation}
We will work for notational convenience in the folded theory $\overline{\text{CFT}}_\text{UV} \times \text{CFT}_\text{IR}$ so that we may treat the interface as a boundary, and we use the notation
\begin{equation}
\tilde{\xi} = \begin{cases} 
\xi_{12} |_{z_1 \to - z_1} = -(1+ \xi_{12}) &\qquad \xi_{12} < 0 \\
 \xi_{12} & \qquad \xi_{12} > 0 
\end{cases}.
\end{equation}
Two-point functions in the bulk may be decomposed into bulk and boundary channel conformal blocks as \cite{McAvity:1995zd,Liendo:2012hy}
\begin{equation}
\begin{split}
\label{eq:Two-point-Decom}
\langle \Phi_{1} (x) \Phi_{2} (x') \rangle &= \frac{\tilde{\xi}^{-(\Delta_1 + \Delta_2)/2}}{(2z)^{\Delta_1}(2 z ')^{\Delta_2}} \sqrt{\mathcal{N}_{\Phi_1} \mathcal{N}_{\Phi_2} }\left[ \delta_{12} + \sum_k \lambda_{12 k} a_k f_\text{bulk} (\Delta_{12}, \Delta_k; \tilde{\xi}) \right]  \\
&= \frac{1}{(2z )^{\Delta_1}(2 z ')^{\Delta_2}} \sqrt{\mathcal{N}_{\Phi_1} \mathcal{N}_{\Phi_2} } \left[ a_{1} a_{2} +\sum_l \mu_{1l} \mu_{2l} f_\text{bdy} (\hat{\Delta}_l ; \tilde{\xi})  \right] 
\end{split}
\end{equation}
where the bulk and boundary conformal blocks are
\begin{equation}
\begin{split}
f_\text{bulk} (\Delta_{12} , \Delta_k ; \tilde{\xi}) &= \tilde{\xi}^{\Delta_k/2} {}_2 F_1 \left(\frac{\Delta_k + \Delta_{12}}{2}, \frac{\Delta_k - \Delta_{12}}{2}; \Delta_k + 1 - \frac{d}{2}; -\tilde{\xi} \right) \\
f_\text{bdy} (\hat{\Delta}_l, \tilde{\xi}) &= \tilde{\xi}^{-\hat{\Delta}_l} {}_2 F_1 \left(\hat{\Delta}_l, \hat{\Delta}_l + 1- \frac{d}{2}; 2 \hat{\Delta}_l + 2 -d ; -\frac{1}{\tilde{\xi}} \right).  
\end{split}
\end{equation}
One can obtain the classical spectrum of single-trace operators on the interface by expanding either $\langle \sigma \sigma \rangle$ or $\langle O O \rangle$ in large $\tilde{\xi}$, which gives
\begin{equation}
\hat{\Delta}_l = \frac{d}{2} + l, \quad l = 0, 1, 2, \dots .
\end{equation}
The OPE coefficients for bulk-to-defect correlation functions between interface operators and $O$ and $\sigma$ were found to be at leading order in $1/N$ \cite{Melby-Thompson:2017aip}
\begin{equation}
\begin{split}
\mu^2_{\sigma \hat{\Delta}_{d/2+l}} &= \frac{\Gamma (l+1) \Gamma \left(\frac{d}{2}+l\right) \Gamma \left(\frac{d}{2}+l-\Delta+1\right)}{(2 l)! \Gamma \left(\frac{d}{2}-\Delta +1\right) \Gamma (d-\Delta ) \Gamma \left(-\frac{d}{2}+l+\Delta +1\right)} \\
\mu^2_{O \hat{\Delta}_{d/2+l}} &= \mu^2_{\sigma \hat{\Delta}_{d/2+l}} \biggr |_{\Delta \to d- \Delta} \\
&= \frac{\Gamma (l+1) \Gamma \left(\frac{d}{2}+l\right) \Gamma \left(-\frac{d}{2}+l+\Delta+1\right)}{(2 l)! \Gamma (\Delta ) \Gamma \left(-\frac{d}{2}+\Delta +1\right) \Gamma\left(\frac{d}{2}+l-\Delta +1\right)}.
\end{split}
\end{equation}
The large $\tilde{\xi}$ expansion of $\langle O \sigma \rangle$ also gives an expression for $\mu_{O \hat{\Delta}_l}\mu_{\sigma \hat{\Delta}_l}$. We must be careful about the expression for $\langle O \sigma \rangle$ since we are in the folded theory:
\begin{equation}\label{eq:OsigFoldedOPE}
\begin{split}
\langle \frac{O}{\sqrt{C_O}} (x_1) \frac{\sigma}{\sqrt{C_\sigma}} (x_2) \rangle_\text{folded} &= \frac{\sqrt{\frac{2}{\pi }} \Gamma \left(\frac{d}{2}\right)}{\sqrt{(d-2 \Delta ) \Gamma (\Delta ) \csc \left(\frac{1}{2} \pi  (d-2 \Delta )\right) \Gamma (d-\Delta )}} \frac{(1+ \tilde{\xi}_{12})^{-d/2} }{(2z_1)^\Delta (2z_2)^{d-\Delta} }\\
&= \frac{1}{(2z_1)^\Delta (2z_2)^{d-\Delta} } \left[ a_O a_\sigma + \sum_{l} \mu_{O l} \mu_{\sigma l} f_\text{bdy} (\hat{\Delta}_l, \tilde{\xi} )\right]
\end{split}
\end{equation}
Expanding \eqref{eq:OsigFoldedOPE} in large $\tilde{\xi}$ again gives us a spectrum of defect operators with dimension $\frac{d}{2} + l$, and the OPE coefficients at leading order in $1/N$ are \cite{Melby-Thompson:2017aip}
\begin{equation}
\begin{split}
\mu_{O \hat{\Delta}_{d/2+l}}\mu_{\sigma \hat{\Delta}_{d/2+l}} &= (-1)^l \frac{\sqrt{\sin \left(\frac{1}{2} \pi  (d-2 \Delta )\right)}}{\sqrt{(d-2 \Delta ) \Gamma (\Delta )\Gamma (d-\Delta )}} \frac{2^{\frac{1}{2}-2 l} \Gamma \left(\frac{d}{2}+l\right)}{\Gamma \left(l+\frac{1}{2}\right)} \\
&= (-1)^l \sqrt{\mu_{O \hat{\Delta}_{d/2+l}}^2 \mu_{\sigma \hat{\Delta}_{d/2+l}}^2 }.
\end{split}
\end{equation}
Taking the coincident point limit of the $O$ two-point function, we obtain the one-point function of $O^2$ on the unperturbed side:
\begin{equation}\label{eq:OOCoincident}
\langle O^2 (x_{L}) \rangle = -\frac{C_O \Gamma \left(\frac{d}{2}\right)}{\Gamma (\Delta +1) \Gamma
   \left(\frac{d}{2}-\Delta \right) (2 z_L)^{2\Delta}}.
\end{equation}
Similarly, the coincident point limit of the $\sigma$ two-point function gives the one-point function of $\sigma^2$ on the perturbed side:
\begin{equation}\label{eq:sigmaCoincident}
\langle \sigma^2 (x_{R}) \rangle = \frac{ (d-2 \Delta ) \Gamma \left(\frac{d}{2}\right) \Gamma (\Delta )
   \sin \left(\frac{1}{2} \pi  (d-2 \Delta )\right) }{2 \pi^{d+1} C_O (\Delta -d) \Gamma \left(\Delta -\frac{d}{2}\right) (2 z_R)^{2(d-\Delta)}} .
\end{equation}
Below we will compute further examples of one-point functions in the interface CFT, working directly with the action (\ref{IR-action}) in the IR limit, and to leading order in $1/N$. 
\subsubsection{$\langle \Phi \rangle$}
We start by considering the one-point function coefficient for a single-trace operator $\Phi$ which is not $O$ or $\sigma$. When $\Phi$ is placed on the unperturbed (``left'') side of the interface (at a point $x_L = (\mathbf{x}_L, -z_L)$, $z_L > 0$), we have
\begin{equation}\label{eq:PhiOnePointLeft}
\begin{split}
&\langle \Phi (\mathbf{x}_L, -z_L) \rangle =\frac{\sqrt{\mathcal{N}_\Phi} a_{\Phi, L}}{(2 z_L)^{\Delta_{\Phi}}}  \\
&=  \frac{1}{2} \int_{\mathbb{R}_+^d} d^d x_1 d^d x_2 \langle \Phi (x_L) O (x_1) O(x_2) \rangle_{0} \langle \sigma(x_1) \sigma(x_2) \rangle \\
&= \frac{C_{\Phi OO} \tilde{C}_\sigma}{2} \int_0 ^\infty dz_1 dz_2 \int d^{d-1} \mathbf{x}_1 d^{d-1} \mathbf{x}_2 \frac{1}{(4 z_1 z_2)^{d-\Delta} }  \\
&\times \frac{ \xi_{12}^{-\frac{d}{2}} {}_2 F_1 \left( \frac{d}{2},  - \frac{d}{2} + \Delta, 1 - \frac{d}{2} + \Delta, -\frac{1}{\xi_{12}} \right) }{((\mathbf{x}_1 - \mathbf{x}_L)^2+(z_1+z_L)^2)^{\tfrac{\Delta_\Phi }{2}}((\mathbf{x}_2-\mathbf{x}_L)^2+(z_2+z_L)^2)^{\tfrac{\Delta_\Phi }{2}}((\mathbf{x}_2-\mathbf{x}_1)^2+(z_2-z_1)^2)^{\Delta-\tfrac{\Delta_\Phi }{2}}} .
\end{split}
\end{equation}
One may extract the coefficient $a_{\Phi, L}$ more easily by multiplying both sides by $z_L^{\Delta_{\Phi}-d}$, integrating out $(\mathbf{x}_L, z_L)$, and dividing by the regularized volume of hyperbolic space at the end. We discuss the details of the calculation in Appendix \eqref{App:PhiOnePoint} and here just quote the result:
\begin{equation}\label{eq:OnePointResultLeft}
\begin{split}
\sqrt{\mathcal{N}_\Phi} a_{\Phi, L}  &=-\frac{C_\sigma C_{\Phi OO}  \pi ^d  \Delta _{\Phi } \Gamma
   \left(\frac{d}{2}-\Delta \right) \Gamma \left(-\frac{d}{2}+\Delta +1\right)^2 \Gamma \left(\Delta
   _{\Phi }-d\right) \Gamma \left(\frac{1}{2} \left(d-2 \Delta +\Delta _{\Phi
   }\right)\right)}{ 2\Gamma \left(\frac{d}{2}-\Delta +1\right) \Gamma (d-\Delta ) \Gamma \left(\Delta
   -\frac{d}{2}\right) \Gamma \left(\frac{\Delta _{\Phi }}{2}+1\right){}^2 \Gamma \left(-d+\Delta
   +\frac{\Delta _{\Phi }}{2}+1\right)}.
\end{split}
\end{equation}
We also consider the one-point function coefficient of $\Phi$ when it is placed on the perturbed (``right'') side of the interface (at a point $x_R = (\mathbf{x}_R, z_R)$, $z_R>0$). Starting from the action (\ref{IR-action}) and working perturbatively in $1/N$, we have
\begin{equation}\label{eq:PhiOnePoint}
\langle \Phi (x_R) \rangle =  \frac{1}{2} \int_{\mathbb{R}_+^d} d^d x_1 d^d x_2 \langle \Phi (x_R) O (x_1) O(x_2) \rangle_{0} \langle \sigma(x_1) \sigma(x_2) \rangle.
\end{equation}
To obtain $a_{\Phi, R}$, we can exchange the roles of $\sigma$ and $O$ by taking $\Delta \to d-\Delta$, $C_\sigma \to C_O$, and $C_{\Phi OO}\to C_{\Phi \sigma \sigma}$. The result is 
\begin{equation}\label{eq:OnePointResultRight}
\sqrt{\mathcal{N}_\Phi} a_{\Phi,R} = \frac{ C_O C_{\Phi \sigma \sigma} \pi ^d  \Delta _{\Phi } \Gamma
   \left(\frac{d}{2}-\Delta +1\right) \Gamma \left(-\frac{d}{2}+\Delta +\frac{\Delta _{\Phi
   }}{2}\right) \Gamma \left(\Delta _{\Phi }-d\right)}{2 \Gamma (\Delta ) \Gamma \left(\frac{\Delta
   _{\Phi }}{2}+1\right){}^2 \Gamma \left(-\Delta +\frac{\Delta _{\Phi }}{2}+1\right)}.
\end{equation}
We can also express $a_{\Phi, R}$ in terms of $C_{\Phi O O}$ rather than $C_{\Phi \sigma \sigma}$, using the following expression given in \cite{Giombi:2017mxl}
\begin{equation}
C_{\Phi \sigma \sigma}^{IR} = \frac{C_{\Phi OO}^{UV}}{\pi^d C_O^2} \frac{\Gamma (\Delta )^2 \Gamma \left(d-\Delta -\frac{\Delta _{\Phi }}{2}\right) \Gamma \left(\frac{d}{2}-\Delta +\frac{\Delta _{\Phi
   }}{2}\right)}{\Gamma \left(\frac{d}{2}-\Delta \right)^2 \Gamma \left(\Delta -\frac{\Delta _{\Phi }}{2}\right) \Gamma
   \left(-\frac{d}{2}+\Delta +\frac{\Delta _{\Phi }}{2}\right)}.
\end{equation}
Plugging this into \eqref{eq:OnePointResultRight} gives
\begin{equation} \label{eq:OnePointResultRight2} 
\sqrt{\mathcal{N}_\Phi} a_{\Phi, R} = \frac{C_{\Phi OO}  \Delta _{\Phi } \Gamma (\Delta ) \Gamma
   \left(\frac{d}{2}-\Delta +1\right) \Gamma \left(d-\Delta -\frac{\Delta _{\Phi }}{2}\right) \Gamma
   \left(\Delta _{\Phi }-d\right) \Gamma \left(\frac{1}{2} \left(d-2 \Delta +\Delta _{\Phi
   }\right)\right)}{ 2 C_O \Gamma \left(\frac{d}{2}-\Delta \right)^2 \Gamma \left(\Delta
   -\frac{\Delta _{\Phi }}{2}\right) \Gamma \left(\frac{\Delta _{\Phi }}{2}+1\right){}^2 \Gamma
   \left(-\Delta +\frac{\Delta _{\Phi }}{2}+1\right)}.
\end{equation}
\subsubsection{$\langle O \rangle$ and $\langle \sigma \rangle$ }
We now consider the one-point function coefficient of $O$ when it is inserted on the unperturbed side of the interface. We have
\begin{equation}
\begin{split}
&\langle O(x_L) \rangle =  \frac{1}{2} \int d^d x_1 d^d x_2 \langle O(x_L) O(x_1) O(x_2) \rangle_{0} \langle \sigma(x_1) \sigma(x_2) \rangle \\
& +   \frac{1}{8} \int d^d x_1 d^d x_2 d^d x_3 d^d x_4 \langle O(x_1) O(x_2) O(x_3) O(x_4) O(x_L) \rangle_{0} \langle \sigma(x_1) \sigma(x_2) \rangle \langle \sigma(x_3) \sigma(x_4) \rangle + \dots \\
&=  \frac{1}{2} \int d^d x_1 d^d x_2 \langle O(x_L) O(x_1) O(x_2) \rangle_{0}  \langle \sigma(x_1) \sigma(x_2) \rangle \\
& -   \frac{1}{2} \int d^d x_1 d^d x_2 d^d x_3 \langle O(x_1) O(x_2) O(x_3) \rangle_{0} \langle O(x_L) \sigma(x_1) \rangle  \langle \sigma(x_2) \sigma(x_3) \rangle + \dots.
\label{O1pt-CFT}
\end{split}
\end{equation}
The first term has the same structure as $\langle \Phi (x_L) \rangle$, while the integrals over $x_2$ and $x_3$ in the second term have the same structure as $\langle \Phi (x_R) \rangle$. We can therefore use our results in \eqref{eq:OnePointResultLeft} and \eqref{eq:OnePointResultRight2} to write
\begin{equation}
\begin{split}
&\frac{\langle O(\mathbf{x}_L, -z_L) \rangle}{\sqrt{C_O}} = \frac{\sqrt{\mathcal{N}_\Phi}  a_{\Phi, L}}{(2 z_L)^\Delta} \Bigr |_{\substack{\Delta_\Phi \to \Delta \\ C_{\Phi OO} \to C_{OOO} }} \\
& -\int_0^\infty dz_1 \int d^{d-1} \mathbf{x}_1 \frac{\sqrt{\mathcal{N}_\Phi}  a_{\Phi, R}}{(2 z_1)^\Delta} \Bigr |_{\substack{\Delta_\Phi \to \Delta \\ C_{\Phi OO} \to C_{OOO}}}  \left(\frac{  \pi ^{-\frac{d}{2}-1} \Gamma \left(\frac{d}{2}\right) \sin \left(\frac{1}{2} \pi  (d-2 \Delta )\right) z_1^{\Delta -\frac{d}{2}} z_L^{\frac{1}{2} (d-2 \Delta )}}{ \left(\left(z_L+z_1\right)^2+(\mathbf{x}_1- \mathbf{x}_L)^2\right)^{d/2}}\right)
\end{split}
\end{equation}
where $z_L$ is a positive quantity. The integral over $x_1$ is easily evaluated, and we get 
\begin{equation}\label{eq:OOnePointIR}
\begin{split}
\langle O(\mathbf{x}_L, -z_L) \rangle &=\frac{C_{OOO}C_\sigma}{ (2 z_L)^{\Delta }} \frac{ \pi ^{d+1} \Delta  \csc \left(\frac{\pi 
   d}{2}\right) \Gamma \left(d-\frac{3 \Delta }{2}\right) \Gamma
   \left(-\frac{d}{2}+\Delta +1\right)}{\Gamma \left(\frac{\Delta }{2}+1\right)^2 \Gamma (d-\Delta
   +1)^2 \Gamma \left(\frac{\Delta -d}{2}\right)}.
\end{split}
\end{equation}
We can obtain the $\sigma$ one-point function by exchanging the roles of $O$ and $\sigma$, sending $\Delta \to d-\Delta$, $C_\sigma \to C_O$, and $C_{O OO}\to C_{\sigma \sigma \sigma}$:
\begin{equation}
\begin{split}
\langle \sigma (\mathbf{x}_R, z_R) \rangle &= \frac{C_{\sigma \sigma \sigma} C_O}{(2 z_R)^{d-\Delta}} \frac{ \pi ^{d+1}  (d-\Delta ) \csc \left(\frac{\pi 
   d}{2}\right) \Gamma \left(\frac{d}{2}-\Delta +1\right) \Gamma \left(\frac{3 \Delta
   }{2}-\frac{d}{2}\right) }{\Gamma \left(-\frac{\Delta }{2}\right) \Gamma (\Delta +1)^2 \Gamma \left(\frac{1}{2} (d-\Delta +2)\right)^2} .
\end{split}
\end{equation}
Using the following result from \cite{Giombi:2018vtc}, we can rephrase the sigma one-point function in terms of UV data:
\begin{equation}
C_{\sigma \sigma \sigma} = - \frac{C_{OOO}}{C_O^3} \frac{ \pi ^{-\frac{3 d}{2}-1} \Gamma (\Delta )^3 \sin \left(\frac{1}{2} \pi  (d-3 \Delta )\right) \Gamma \left(\frac{1}{2} (d-3
   \Delta +2)\right) \Gamma \left(d-\frac{3 \Delta }{2}\right) \Gamma \left(\frac{d-\Delta }{2}\right)^3}{\Gamma \left(\frac{\Delta
   }{2}\right)^3 \Gamma \left(\frac{d}{2}-\Delta \right)^3}.
\end{equation}
We have
\begin{equation}\label{eq:SigmaOnePointIR}
\langle \sigma (\mathbf{x}_R, z_R) \rangle = -\frac{C_{OOO}}{C_O^2 (2 z_R)^{d-\Delta}} \frac{ 2^{ \Delta -2} (d-2 \Delta ) \csc \left(\frac{\pi  d}{2}\right) \sin \left(\frac{\pi 
   \Delta }{2}\right) \Gamma \left(\frac{\Delta +1}{2}\right) \Gamma \left(d-\frac{3 \Delta }{2}\right) \Gamma \left(\frac{d-\Delta }{2}\right)}{ (d-\Delta ) \pi^{(d+1)/2} \Gamma \left(\frac{\Delta }{2}+1\right) \Gamma \left(\frac{d}{2}-\Delta \right)^2}.
\end{equation}
We could also determine $\langle \sigma \rangle$ directly from a diagrammatic computation. Namely,
\begin{equation}
\begin{split}
&\langle \sigma(x_R) \rangle =  -\frac{1}{2} \int d^d x_1 d^d x_2 d^d x_3 \langle O(x_1) O(x_2) O(x_3) \rangle_{0} \langle \sigma(x_R) \sigma(x_1) \rangle \langle \sigma(x_2) \sigma(x_3) \rangle + \dots \\
&= -  \int_0^\infty dz_1 \int d^{d-1} \mathbf{x}_1 \frac{\sqrt{\mathcal{N}_\Phi}  a_{\Phi, R}}{(2 z_1)^\Delta} \Bigr |_{\substack{\Delta_\Phi \to \Delta \\ C_{\Phi OO} \to C_{OOO}}}  \frac{\tilde{C}_\sigma}{(4 z_1 z_R)^{d-\Delta}} \xi_{1}^{-\frac{d}{2}} {}_2 F_1 \left( \frac{d}{2},  - \frac{d}{2} + \Delta, 1 - \frac{d}{2} + \Delta, -\frac{1}{\xi_{1}} \right) 
\label{sigma1pt-CFT}
\end{split}
\end{equation}
where we used the fact that the integral over $x_1$ and $x_2$ has the same structure as \eqref{eq:PhiOnePoint} and defined $\xi_1 = (x_1-x_R)^2/(4 z_1 z_R)$. Because the $z_1$-dependence of the integrand is $z_1^{-d}$, this is equivalent to an integral over $H^d$ in which the integrand is only a function of hyperbolic distance $\xi_1$. We can thus use hyperbolic ball coordinates and set one point at the center of the ball ($\mathbf{x} = 0, z = 1$). We also write the hypergeometric function in its Mellin-Barnes integral representation, which gives
\begin{equation}
\begin{split}
\langle \sigma(x_R) \rangle &= \frac{\tilde{C}_\sigma}{(2 z_R)^{d-\Delta}}\sqrt{\mathcal{N}_\Phi} a_{\Phi, R} \Bigr |_{\substack{\Delta_\Phi \to \Delta \\ C_{\Phi OO} \to C_{OOO}}} \frac{\pi ^{d/2} (d-2 \Delta )}{\Gamma \left(\frac{d}{2}\right)^2}  \\
&\times \int_{-\infty}^\infty \frac{ds}{2\pi i} \frac{\Gamma (-s) \Gamma \left(\frac{d}{2}+s\right) \Gamma \left(-\frac{d}{2}+s+\Delta \right)}{\Gamma \left(-\frac{d}{2}+s+\Delta +1\right)} \int_0^1 du \ u^{-2 s-1} \left(1-u^2\right)^{s-\frac{d}{2}} \\
&= \frac{\tilde{C}_\sigma}{(2 z_R)^{d-\Delta}} \sqrt{\mathcal{N}_\Phi} a_{\Phi, R} \Bigr |_{\substack{\Delta_\Phi \to \Delta \\ C_{\Phi OO} \to C_{OOO}}} \frac{\pi ^{\frac{d}{2}-1}  (d-2 \Delta ) \sin \left(\frac{\pi  d}{2}\right)}{2\Gamma \left(\frac{d}{2}\right)} \\
&\times \int_{-\infty}^\infty \frac{ds}{2\pi i} \frac{\Gamma (-s)^2 \Gamma \left(-\frac{d}{2}+s+1\right) \Gamma \left(\frac{d}{2}+s\right) \Gamma \left(-\frac{d}{2}+s+\Delta \right)}{\Gamma \left(-\frac{d}{2}+s+\Delta +1\right)} .
\end{split}
\end{equation}
Using Barnes' second lemma to perform the $s$ integral and plugging in the expression for $\sqrt{\mathcal{N}_\Phi} a_{\Phi,R}$, we obtain
\begin{equation}
\langle \sigma(x_R) \rangle =-\frac{C_{OOO}}{C_O^2 (2 z_R)^{d-\Delta}} \frac{ \pi ^{-\frac{d}{2}-\frac{1}{2}} 2^{ \Delta -2} (d-2 \Delta ) \csc \left(\frac{\pi  d}{2}\right) \sin \left(\frac{\pi 
   \Delta }{2}\right) \Gamma \left(\frac{\Delta +1}{2}\right) \Gamma \left(d-\frac{3 \Delta }{2}\right) \Gamma \left(\frac{d-\Delta }{2}\right)}{ (d-\Delta ) \Gamma \left(\frac{\Delta }{2}+1\right) \Gamma \left(\frac{d}{2}-\Delta \right)^2}
\end{equation}
which matches \eqref{eq:SigmaOnePointIR}. 

\section{Free energy} \label{Sec:FreeEnergy}
In this section, we calculate the free energy associated with our interface. In two dimensions, this is related to the ``boundary entropy" or the $g-$function \cite{PhysRevLett.67.161} (the exact relation between the free energy we define and the boundary $g-$function is spelled out below). The $g-$ function in two dimensions was originally proposed in \cite{PhysRevLett.67.161}, and later proved in \cite{Friedan:2003yc, Casini:2016fgb}, to decrease under RG flows localized on the defect. There are several similar proposals for RG monotones for codimension one defects in higher dimensions \cite{Yamaguchi:2002pa, Fujita:2011fp, Nozaki:2012qd, Gaiotto:2014gha, Estes:2014hka, Jensen:2015swa, Herzog:2017kkj, Kobayashi:2018lil, Casini:2018nym, Giombi:2020rmc}. In particular, in \cite{Jensen:2015swa}, it was proved that in $d = 3$ CFT in the presence of a boundary, the coefficients of the Euler density term in the boundary trace anomaly (called $b-$anomaly coefficient) decreases under a boundary RG flow (by the folding trick, the same result applies to the interface case). Here, we will first calculate the free energy of the CFT in the presence of the interface, and then show that this free energy can be used to extract the analog of the $g$-function in even $d$ (odd-dimensional interface), or the trace anomaly coefficient in odd $d$ (even dimensional interface). 
In the case of $d=3$, we will then check that the value of the $b$-anomaly coefficient we obtain is consistent with a simple RG flow in the example of vector $O(N)$ model.

The free energy of the large $N$ CFT in the presence of the interface can be written as 
\begin{equation}
F = F_{0, \text{UV}} + F_{\sigma} + O(\frac{1}{N})
\end{equation}
where $F_{0, \text{UV}}$ is the free energy of the UV CFT, and $F_{\sigma}$ denotes the contribution of the one-loop determinant arising from the path-integral over the auxiliary field $\sigma$. Explicitly, we have\footnote{Along the flow, one has $F_{\sigma} = \frac{1}{2} \text{tr} \log \left( G_{12}+\frac{\delta(x_{12})}{4\lambda}\right)$, but here we are interested in the IR limit, so we directly drop the $\lambda$-dependent term.} 
\begin{equation}
\begin{split}
F_\sigma &\equiv F - F_{0, \text{UV}} = -\log \left(\frac{Z}{Z_{0, \text{UV}}}\right) = 
\frac{1}{2} \text{tr} \log \left( G_{12}\right)\, \\
\end{split}
\end{equation}
where $G_{12}$ is the two-point function of the $O$ operator in the UV CFT, which defines the non-local ``kinetic term" for $\sigma$. Since the auxiliary field $\sigma$ is supported over half space ($\mathbb{R}_+^d$), the determinant of the non-local operator $G_{12}$ is to be computed over half space. For a CFT defined on the round sphere $S^d$, with the interface located on the $S^{d-1}$ equator, this means that the determinant is computed on the hemisphere. In practice, instead of working on the hemisphere, we find it more convenient to conformally map the problem to $H^d$, as explained in the next section. 
 
\subsection{Large $N$ calculation on $H_d$} \label{Sec:LargeNFreeEnergy}
As mentioned above, it turns out to be easier to compute the $\sigma$ determinant on the hyperbolic space rather than on the hemishpere. Therefore, we proceed by first doing a conformal transformation to the hyperbolic space (with the interface located on its spherical boundary), and compute the determinant on this space. 

For our original theory on a flat space, one can clearly map half of the space to $H^d$ by a Weyl transformation, as 
\begin{equation}\label{eq:WeylRescale}
ds^2 = dz^2 + d \textbf{x}^2 = z^2 (\frac{dz^2 + d \textbf{x}^2}{z^2}) = z^2 ds^2_{H^d}.
\end{equation}
where $z > 0$ denotes the half space where the $\sigma$ field is supported. The two-point function of $\sigma$ on this $H^d$ can be obtained from the flat space result \eqref{eq:SigmaPropagator} by applying the Weyl transformation: \begin{equation}
\begin{split}
&\langle \sigma(x_1)\sigma(x_2)\rangle = 4^{\Delta- d} h(\xi) \\
& \to h(\xi) \equiv \frac{4^{\Delta- d} \Gamma (\Delta) \sin \pi\left( \Delta - \frac{d}{2} \right) \Gamma \left( \frac{d}{2} \right)  }{C_O \pi^{d + 1} \Gamma \left( \Delta - \frac{d}{2} \right) \xi^{\frac{d}{2}}} {}_2 F_1 \left( \frac{d}{2},  - \frac{d}{2} + \Delta, 1 - \frac{d}{2} + \Delta, -\frac{1}{\xi} \right),
\label{sigsig-Hd}
\end{split}
\end{equation} 
where $\xi$ is the cross ratio defined in (\ref{cross-r}) and we redefine $h$ to absorb the factor of $4^{\Delta- d}$. 

For the calculation of the free energy, we should consider the hyperbolic ball coordinates of $H^d$, with spherical $S^{d-1}$ boundary:
\begin{equation}\label{eq:HyperbolicCoords}
ds^2 = d \eta^2 + \sinh^2 \eta ds^2_{S^{d-1}}\,\qquad 0 \le \eta < \infty \,.
\end{equation}
Of course, this metric is Weyl equivalent to the hemisphere, since 
\begin{equation}
 d \eta^2 + \sinh^2 \eta ds^2_{S^{d-1}} = \frac{1}{\cos^2 \theta}\left(d\theta^2+\sin^2\theta ds^2_{S^{d-1}} \right)\,,\qquad \tanh\eta = \sin\theta 
\end{equation}
In these coordinates of $H^d$, the two-point function of $\sigma$ takes the same form (\ref{sigsig-Hd}) in terms of the cross ratio $\xi$. If we place one of the $\sigma$ operators at the center of hyperbolic space $\eta = 0$, then the cross-ratio can just be expressed in terms of the $\eta$ coordinate of the second insertion 
\begin{equation}\label{eq:CrossRatioEta}
\xi = \sinh^2 \frac{\eta}{2}.
\end{equation}

To compute the determinant, note that the two-point function of $\sigma$ is simply the inverse of the operator $G_{12}$ (with the operator inversion defined on half space only). Therefore, we can find the relevant eigenvalues by decomposing the $\sigma$ two point function into eigenfunctions of the Laplacian on the hyperbolic space. 
We can expand the two-point function into eigenfunctions as follows
\begin{equation}
h(\xi) =  \sum_l \int d \nu h(\nu) D(\nu,l) \Phi_{\nu, l} (x_1) \Phi^*_{\nu, l} (x_2). 
\end{equation}
If we choose the coordinates $x_1, x_2$ as above, then only the $l = 0$ term of the above sum survives and we get \footnote{This is because only the $l = 0$ eigenfunction is non-zero at the center of the hyperbolic space.}
\begin{equation}
h(\xi = \sinh^2 \frac{\eta}{2}) =  \int d \nu h(\nu) D(\nu,0) \Phi_{\nu, 0} (\eta). 
\end{equation}
The required eigenfunction and degeneracies $D(\nu, 0)$ are given by \cite{1994JMP....35.4217C, Bytsenko:1994bc}
\begin{equation}
\begin{split}
\Phi_{\nu, 0} (\eta) &= {}_2 F_1 \left( i \nu + \frac{d - 1}{2}, -i \nu + \frac{d - 1}{2}, \frac{d}{2}, - \sinh^2 \frac{\eta}{2} \right)\\
 D(\nu,0) &= \frac{2  \left|\Gamma \left( i \nu + \frac{d - 1}{2} \right) \right|^2}{ (4 \pi)^{d/2} \Gamma (d/2)   \left| \Gamma \left( i \nu \right) \right|^2}. 
\end{split}
\end{equation}
These eigenfunctions satisfy the following orthogonality relation
\begin{equation}
\int d \eta (\sinh \eta)^{d-1} \Phi_{\nu, 0} (\eta) \Phi^*_{\nu', 0} (\eta)  = \frac{\Gamma \left(\frac{d}{2} \right)}{2 \pi^{d/2}}\frac{\delta(\nu - \nu')}{D(\nu,0)}.
\end{equation}
Using that, we can get the mode decomposition coefficients as 
\begin{equation}
h(\nu) = \frac{(4 \pi)^{\frac{d}{2}}}{\Gamma \left( \frac{d}{2} \right) } \int_0^{\infty} d \xi \left( \xi (1 + \xi) \right)^{\frac{d}{2} - 1} h (\xi) {}_2 F_1 \left( i \nu + \frac{d - 1}{2}, -i \nu + \frac{d - 1}{2}, \frac{d}{2}, - \xi \right)
\end{equation}
To compute the integral, we express both hypergeometric functions in terms of a contour integral, and then perform the integral over $\xi$
\begin{equation}
\begin{split}
&h(\nu) = -\frac{2^{2\Delta- d} \Gamma (\Delta) }{ C_O \pi^{\frac{d}{2}}   \Gamma \left( \Delta - \frac{d}{2} \right)^2 \Gamma \left( \frac{d}{2} - \Delta \right) \left|\Gamma \left( i \nu + \frac{d - 1}{2} \right) \right|^2} \int \frac{d s_1 d s_2}{(2 \pi i)^2} \times \\
&  \frac{ \Gamma(s_1) \Gamma(s_2) \left|\Gamma \left( i \nu + \frac{d - 1}{2} - s_1 \right) \right|^2}{\Gamma \left( \frac{d}{2} - s_1 \right)} \frac{\Gamma \left( \frac{d}{2} - s_2 \right) \Gamma \left(  \Delta - \frac{d}{2} - s_2 \right)}{\Gamma \left( 1 + \Delta -  \frac{d}{2}  - s_2\right)} \int_0^{\infty} d \xi \left( \xi (1 + \xi) \right)^{\frac{d}{2} - 1} \xi^{s_2 - \frac{d}{2} - s_1} \\
&=  -\frac{2^{2\Delta- d} \Gamma (\Delta) }{ C_O \pi^{\frac{d}{2}}   \Gamma \left( \Delta - \frac{d}{2} \right)^2 \Gamma \left( \frac{d}{2} - \Delta \right) \left|\Gamma \left( i \nu + \frac{d - 1}{2} \right) \right|^2  \Gamma \left( 1 - \frac{d}{2} \right) } \int \frac{d s_1 d s_2}{(2 \pi i)^2}  \\
& \times \frac{ \Gamma(s_1) \Gamma(s_2) \left|\Gamma \left( i \nu + \frac{d - 1}{2} - s_1 \right) \right|^2 \Gamma \left( \frac{d}{2} - s_2 \right) \Gamma \left( \frac{2\Delta - d}{2} - s_2 \right) \Gamma \left(  \frac{2-d}{2} + s_1  - s_2\right) \Gamma \left(  s_2 - s_1 \right) }{\Gamma \left( \frac{d}{2} - s_1 \right) \Gamma \left( 1 + \Delta -  \frac{d}{2}  - s_2\right) }.
\end{split}
\end{equation}
We can now repeatedly use Barnes second lemma to do the integral over  $s_1$ and $s_2$
\begin{equation}
h(\nu) = -\frac{2^{2\Delta- d}  \Gamma (\Delta) \left|\Gamma \left( i \nu + \frac{1}{2} \right) \right|^2  }{ C_O \pi^{\frac{d}{2} }  \Gamma \left( \frac{d}{2} - \Delta \right) \left|\Gamma \left( i \nu + \frac{1 - d}{2} + \Delta \right) \right|^2   } .
\end{equation}
Using this mode expansion, the contribution of the $\sigma$ operator to the free energy of the system is then given by 
\begin{equation}
\begin{split}
F_{\sigma}  &= - \frac{\textrm{vol} (H^d)}{2} \int d \nu D(\nu, 0) \log (h(\nu)) \\
&= \frac{\textrm{vol} (H^d)}{(4 \pi)^{d/2} \Gamma (d/2)} \int_0^{\infty} d \nu \frac{  \left|\Gamma \left( i \nu + \frac{d - 1}{2} \right) \right|^2}{    \left| \Gamma \left( i \nu \right) \right|^2} \log \left( \frac{\left|\Gamma \left( i \nu + \frac{1 - d}{2} + \Delta \right) \right|^2 }{\left|\Gamma \left( i \nu + \frac{1}{2} \right) \right|^2} \right) 
\end{split}
\end{equation}
It is easier to compute the derivative of this with respect to $\Delta$
\begin{equation}
\frac{\partial F_{\sigma} }{\partial \Delta} = \frac{\textrm{vol} (H^d)}{ 2 (4 \pi)^{d/2} \Gamma (d/2)} \int_{- \infty}^{\infty} d \nu \frac{  \left|\Gamma \left( i \nu + \frac{d - 1}{2} \right) \right|^2}{    \left| \Gamma \left( i \nu \right) \right|^2}  \left(\psi ^{(0)}\left(\frac{1 - d}{2}+\Delta +i \nu \right) +  c.c. \right) 
\end{equation}
Closing the contour in the lower-half plane and performing the integral by summing up the residues, we get 
\begin{equation}
\begin{split}
\frac{\partial F_{\sigma} }{\partial \Delta} &= - \frac{\textrm{vol} (H^d) (d - 2 \Delta) \Gamma (\Delta) \Gamma(d - \Delta) \left( \sin \left(\frac{\pi d}{2} \right) \cos \pi \left(\frac{d}{2} - \Delta \right) + \cos \left(\frac{\pi d}{2} \right) \sin \pi \left(\frac{d}{2} - \Delta \right) \right)}{ 4 (4 \pi)^{d/2} \Gamma \left(\frac{d}{2} + 1 \right) \sin \left(\frac{\pi d}{2} \right) } \\
&= - \frac{ (d - 2 \Delta) \Gamma (\Delta) \Gamma(d - \Delta) \left( \sin \left(\frac{\pi d}{2} \right) \cos \pi \left(\frac{d}{2} - \Delta \right) + \cos \left(\frac{\pi d}{2} \right) \sin \pi \left(\frac{d}{2} - \Delta \right) \right)}{ 4  \Gamma \left(d + 1 \right) \sin \left(\frac{\pi d}{2} \right) \cos \left(\frac{\pi d}{2} \right) } \\
&= -\frac{(d-2 \Delta )  \Gamma (\Delta ) \sin (\pi  (d-\Delta )) \Gamma (d-\Delta )}{2 \Gamma (d+1) \sin (\pi  d)} .
\end{split}
\end{equation}
We know that for $\Delta = d/2$ we have $F_{\sigma} = 0$. So we may compute the free energy as 
\begin{equation}\label{eq:LargeNFreeEnergy}
F_{\sigma}(\Delta) = \frac{1}{\Gamma (d+1) \sin (\pi  d)} \int_0^{\Delta - \frac{d}{2}} du \ u \  \Gamma \left(\frac{d}{2} + u \right) \Gamma \left(\frac{d}{2} - u \right) \sin \pi \left(\frac{d}{2} - u \right) . 
\end{equation}
This is our final result for the contribution of the $\sigma$ determinant to the free energy of our large $N$ interface CFT. 

In order to write the free energy of the interface CFT in a more symmetric form, we make the following observation. On one hand, the free energy can be written as
\begin{equation}
F = F_{0,\text{UV}} + F_\sigma (\Delta) + O(1/N)\,.
\label{F-fromUV}
\end{equation}
On the other hand, sending $\Delta \to d- \Delta$ leaves the full interface theory invariant while swapping the roles of the UV and IR theory,\footnote{This is essentially because at large $N$, according to (\ref{IR-action}), the IR theory can be thought of as the Legendre transform of the UV theory, and vice-versa. This is also transparent in the holographic dual description of the double-trace deformation \cite{Klebanov:1999tb}.} giving
\begin{equation}
F = F_{0,\text{IR}} + F_\sigma (d-\Delta) + O(1/N) 
\label{F-fromIR}
\end{equation}
where $F_{0,\text{IR}}$ is the free energy of the IR limit of the double-trace deformed theory on the full sphere. Therefore, we can write the free energy as 
\begin{equation}
F = \frac{1}{2}\left(F_{0,\text{UV}}+F_{0,\text{IR}}\right) + \log(g)+ O(1/N)
\label{F-final} 
\end{equation}
where we defined
\begin{equation}
\log(g) = \frac{1}{2}\left(F_\sigma (\Delta)+  F_\sigma (d-\Delta)\right) = 
\frac{1}{2}  \int_0^{\Delta - \frac{d}{2}} du \ \frac{u \cos (\pi  u) \Gamma \left(\frac{d}{2}-u\right) \Gamma \left(\frac{d}{2}+u\right)}{\cos \left(\frac{\pi  d}{2}\right)  \Gamma (d+1)}\,.
\label{log-g}
\end{equation}
The result (\ref{F-final}), (\ref{log-g}) precisely agrees with that obtained in \cite{Melby-Thompson:2017aip} from the holographic AdS calculation. The setup for the AdS calculation of the free energy and correlation functions is reviewed in detail in Section \ref{Sec:HolographicSetup}.

In the case of even $d$, i.e. an odd-dimensional interface, the first term in (\ref{F-final}) is logarithmically divergent (containing a $1/\epsilon$ pole in dimensional regularization) and gives an average of the conformal anomalies of the UV and IR CFTs,\footnote{Explicitly, one has 
\begin{equation}
\frac{1}{2}\left(F_{0,\text{UV}}+F_{0,\text{IR}}\right)\Big{|}_{d=2n-\epsilon} = \frac{(-1)^{\frac{d}{2}}}{2\epsilon}\left(a_{UV}+a_{IR}\right)+\ldots\,,
\end{equation}
where $a_{UV}$ and $a_{IR}$ are the $a$-anomaly coefficients of the UV and IR CFTs.} while the $\log(g)$ term is finite. In $d=2$, the latter gives the $g$-function of the interface, and in higher even $d$ it gives its natural generalization. For example, from (\ref{log-g}) one gets
\begin{equation}
\log(g)|_{d=2} = -\frac{\pi}{4} \int_0^{1-\Delta}du\, u^2 \cot(\pi u) \,\qquad \log(g)|_{d=4}=\frac{\pi}{48} \int_0^{2-\Delta}du\, u^2(1-u^2) \cot(\pi u) \,\qquad \ldots 
\end{equation}
On the other hand, in odd $d$ (even-dimensional interface) the first term in (\ref{F-final}) is finite and gives the average of the $F$-coefficient \cite{Klebanov:2011gs} of the UV and IR CFTs, while the $\log(g)$ term is logarithmically UV divergent and encodes the interface anomaly coefficient associated with the Euler density (the $b$-anomaly coefficient in the case $d=3$). From (\ref{log-g}), we find for example
\begin{equation}
\begin{aligned}
&\log(g)|_{d=3-\epsilon} = \frac{(3-2 \Delta)^2 (4 (\Delta-3) \Delta+7)}{384 \epsilon}+O(\epsilon^0)\,,\\
&\log(g)|_{d=5-\epsilon} = \frac{(5-2 \Delta )^2 (4 (\Delta -5) \Delta  (4 (\Delta -5) \Delta +35)+277)}{46080 \epsilon }+O(\epsilon^0)\,.
\label{anomaly-examples}
\end{aligned}
\end{equation}
As usual, the $1/\epsilon$ pole in dimensional regularization corresponds to a logarithmic UV divergence, so that the coefficient of the pole is directly related to the interface anomaly coefficient. 
The relation to the interface trace anomaly is spelled out in more detail in the next section in the case of $d=3$. 

Let us finally comment on an observation relating our results to the change in free energy on the full sphere due to a double-trace deformation, which will serve as a non-trivial consistency check. From the equality of (\ref{F-fromUV}) and (\ref{F-fromIR}), one can deduce that 
\begin{equation}
F_{0,\text{IR}} - F_{0,\text{UV}} = F_\sigma (\Delta) - F_\sigma (d-\Delta). 
\end{equation}
The left-hand side of this equation is the change of the free energy under a double-trace deformation for the CFT defined on the full sphere. The result is well-known and has been computed in \cite{Gubser:2002vv, Diaz:2007an}. On the right-hand side, we have a difference of determinants computed on the half sphere, or hyberbolic space. 
Plugging in the result (\ref{eq:LargeNFreeEnergy}), we find
\begin{equation}
F_\sigma (\Delta) - F_\sigma (d-\Delta) = -\frac{1}{\sin \left(\frac{\pi  d}{2}\right) \Gamma (d+1)} \int_0^{\Delta-d/2} du \ u  \sin (\pi  u) \Gamma \left(\frac{d}{2}-u\right) \Gamma \left(\frac{d}{2}+u\right)\,.
\end{equation}
This indeed agrees with the well-known leading order result for $F_{0,\text{IR}} - F_{0,\text{UV}}$ under a double-trace flow \cite{Gubser:2002vv, Diaz:2007an}. 

\subsection{Extracting the $b$-anomaly coefficient in $d = 3$}
We now explain in more detail how the interface free energy can be used to extract the trace anomaly coefficients, focusing on the three dimensional case. For convenience, we will work in the folded picture, such that we have a boundary conformal field theory. We will follow a similar discussion in \cite{Giombi:2020rmc} where free energy on hyperbolic space was used to calculate trace anomaly coefficients. In the presence of a boundary, the trace anomaly of a 3d BCFT on a general curved space takes the following form \cite{Jensen:2015swa,Graham:1999pm,Herzog:2017kkj} 
\begin{equation}
\langle T^\mu_{ \ \mu} \rangle^{d=3} = \frac{\delta (z)}{4 \pi} \left( b \hat{\mathcal{R}} + d_1  \text{tr} \hat{K}^2 \right)
\end{equation}
where $\hat{\mathcal{R}}$ is the defect Ricci scalar and $\hat{K}_{ij}$ is the traceless part of the defect extrinsic curvature. More explicitly,
\begin{equation}
\hat{K}_{ij} = K_{ij} - \frac{1}{2} \gamma_{ij} K \implies \text{tr} \hat{K}^2 = \text{tr} K^2 - \frac{1}{2} K^2
\end{equation}
where $\gamma_{ij}$ is the metric induced on the interface. We consider the folded interface theory mapped to hyperbolic space, $H^3$. The boundary is a two-sphere, say of radius $R$. Then the Ricci scalar is $2/R^2$ while the extrinsic curvature is $K_{ij} = \frac{1}{2} \gamma_{ij} K$, meaning $\hat{K} = 0$. Thus we have
\begin{equation}
\langle T^\mu_{ \ \mu} \rangle^{d=3} = \frac{\delta (z)}{2 \pi R^2} b .
\end{equation}
We can determine $b$ by using the relationship between the stress tensor and the variation of the free energy under Weyl rescaling $g_{\mu \nu} \to e^{2 \alpha} g_{\mu \nu} $,  which is given by
\begin{equation}\label{eq:deltaW1}
\delta^W F = -\frac{1}{2} \int d^3 \mathbf{x} \sqrt{g} \delta g_{\mu \nu} \langle T^\mu_{ \ \nu} \rangle = - \alpha \int d^3 \mathbf{x} \sqrt{g} \langle T^\mu _{\ \mu} \rangle = -2 \alpha b.
\end{equation}
We can obtain an additional expression for $\delta^W F$ using \eqref{eq:LargeNFreeEnergy}. First we rewrite the free energy in terms of the volume of hyperbolic space, introduce a length scale $R$, and regulate using a radial cutoff, so vol$(H^3) = -2\pi \log(R/\epsilon)$:
\begin{equation}
\begin{split}
F_\sigma &=- \frac{ \text{vol}(H^3)}{12 \pi} \int_0^{\Delta - d/2} du \ u \frac{\Gamma(\frac{3}{2}+u)}{\Gamma(u-\frac{1}{2})}\\
&=\frac{(3-2 \Delta )^2 (4 (\Delta -3) \Delta +7)}{384 } \log(R/\epsilon).
\end{split}
\end{equation}
We then compute $\delta^W F$ by taking the variation of the above expression under Weyl rescaling, i.e. $R \to e^{\alpha} R$. Note that the part of the free energy from the UV CFT on whole space, $F_{0}$, should not contribute to this (there are no bulk conformal anomalies for a 3d theory). So we have
\begin{equation}\label{eq:deltaW2}
\delta^W F = \delta^W F_\alpha = \frac{1}{384}(3-2 \Delta )^2 (4 (\Delta -3) \Delta +7)\alpha.
\end{equation}
Comparing (\ref{eq:deltaW1}) and (\ref{eq:deltaW2}) we arrive at
\begin{equation}
b = -\frac{1}{768} (3-2 \Delta )^2 (4 (\Delta -3) \Delta +7). 
\end{equation}
As expected, this is the same (up to an overall factor of $-1/2$) as the coefficient of the $1/\epsilon$ pole in dimensional regularization, see eq. (\ref{anomaly-examples}). 

We may specify to the $O(N)$ or GN model examples by respectively taking $\Delta = d-2$ or $\Delta = d-1$. We then find $b_{O(N)} = b_{GN} = \frac{1}{768}$. We study these specific models in detail in  the next section.

Let us now look at a simple example of a RG flow triggered by a relevant operator on the interface. For instance, for the RG interface in the $O(N)$ model, we have the free vector model on one side and the critical $O(N)$ model on the other side. We may add a $\phi^2$ deformation on the interface under which we expect the system to flow to two decoupled theories: a theory of $N$ free scalar fields with Dirichlet boundary condition on one side and a critical $O(N)$ model with Dirichlet boundary condition (ordinary transition) on the other side. The $b-$ coefficients for these two decoupled theories are as follows \cite{Giombi:2020rmc}:
\begin{equation}
b_{N \textrm{ free-Dirichlet}} = - \frac{N}{96}, \hspace{1cm} b_{\textrm{critical } O(N)} = 0\, .
\end{equation} 
So we have $b_{\textrm{RG interface}} > b_{N \textrm{free- Dirichlet}}  + b_{\textrm{critical} O(N)}$ which is consistent with the $b-$theorem. 

\section{Specific models with free fields on one side of the interface}
\label{sec:examples}
In this section we discuss the case where the UV theory on one side of the interface is a free theory. We first make general arguments about the spectrum of operators on the interface, regardless of whether the interacting side is strongly or weakly coupled, which is mostly a review of the discussion in \cite{Gliozzi:2015qsa}. We then proceed to discuss interfaces for several specific models with a weakly coupled IR fixed point on half space. When we have a free theory on one side of the interface, the fields obey free equations of motion, which imposes constraints on the interface spectrum. In particular, we expect to have protected operators on the interface. This is similar to what happens in the case of free field theory with interactions localized on the boundary as explained in \cite{Giombi:2019enr}. Below we review and adapt the arguments from \cite{Giombi:2019enr} for why there should be protected operators on the interface. 

First let's look at a free scalar field $\phi$ on the free side of the interface. It has the following decomposition in terms of interface operators 
\begin{equation}
\phi (\mathbf{x}, z) = \sum_{\hat{\Phi}}  \frac{\mu_{\phi \hat{\Phi}}}{(2 z)^{\Delta - \hat{\Delta}}} D^{\hat{\Delta}} (z^2 \boldsymbol{\partial}^2) \hat{\Phi} (\mathbf{x}), \quad  D^{\hat{\Delta}} (z^2 \boldsymbol{\partial}^2) =  \sum_{m = 0}^{\infty} \frac{1}{m!} \frac{1}{(\hat{\Delta} +  \frac{3 - d}{2})_m} \left(- \frac{1}{4}z^2 \boldsymbol{\partial}^2 \right)^m
\end{equation} 
where the 
where $\mu_{\phi \hat{\Phi}}$ is the bulk-defect OPE coefficient and the differential operator $D^{\hat{\Delta}}$ can be fixed using conformal invariance as was done in \cite{McAvity:1995zd}. The symbol $(x)_m$ is the Pochhammer symbol defined by $(x)_m = \Gamma (x + m)/\Gamma(x)$. Applying the bulk equation of motion $\partial_{\mu} \partial^{\mu} \phi = 0$ to this OPE, one finds

\begin{equation}
\begin{split}
\partial_{\mu} \partial^{\mu} \phi &= \sum_{\hat{\Phi}} \frac{\mu_{\phi \hat{\Phi}}}{(2 z)^{\Delta - \hat{\Delta}}}  \sum_{m = 0}^{\infty} \frac{1}{m!} \frac{1}{(\hat{\Delta} +  \frac{ 3 - d}{2})_m} \bigg( (- \tfrac{1}{4}z^2)^m (\boldsymbol{\partial}^2)^{m + 1} \hat{\Phi} (\textbf{x})  \\
&+ (2 m - \Delta + \hat{\Delta})(2 m - 1 - \Delta + \hat{\Delta}) (- \tfrac{1}{4}\boldsymbol{\partial}^2)^m (z^2)^{m - 1}  \hat{\Phi} (\textbf{x}) \bigg) \\
&= \sum_{\hat{\Phi}} \frac{\mu_{\phi \hat{\Phi}}}{(2 z)^{\Delta - \hat{\Delta}}}  \sum_{m = 0}^{\infty} \frac{1}{m!} \frac{1}{(\hat{\Delta} + \frac{3 - d}{2})_m} \bigg(1 - \frac{(2 m + 2 - \Delta + \hat{\Delta})(2 m + 1 - \Delta + \hat{\Delta})}{4 (m + 1) (m  + \hat{\Delta} + \frac{3 - d}{2})} \bigg) \\
& \times \left(- \frac{1}{4}z^2 \boldsymbol{\partial}^2 \right)^m  \hat{\Phi} (\textbf{x}). 
\end{split}
\end{equation}
We must set this to $0$ for $z < 0$ because we have a free theory on the left side of the interface. Plugging in $\Delta = d/2-1$, this fixes $\hat{\Delta} = d/2- 1$ or $d/2$. So only primaries with scaling dimension equal to one of these two values can appear in the interface OPE of the bulk scalar. Let's call these operators $\hat{\phi}$ and $\widehat{\partial \phi}$ respectively. Moreover, these operators have protected dimension because this argument is independent of whether or not we have interactions on the right side of the interface.

The argument above also fixes the two point function of the scalar $\phi$ on the free side with any other operator on the free or interacting side, up to a constant. We now show how this works. We start by looking at the two-point function of the scalar $\phi$ with itself with both insertions on the free side. Using \eqref{eq:Two-point-Decom}, we may write this two-point function in the interface channel. It involves following two terms
\begin{equation}
\langle \phi (x_{L1}) \phi(x_{L2}) \rangle = \frac{\mathcal{N}_{\phi}}{(4 z_{L1} z_{L2})^{\frac{d}{2} - 1}} \left[ \frac{\mu_{\phi \hat{\phi}}^2}{2} \left( \frac{1}{\xi^{\frac{d}{2} - 1}} + \frac{1}{(1 + \xi)^{\frac{d}{2} - 1}} 
\right) + \frac{2 \mu_{\phi \widehat{\partial \phi}}^2}{d - 2} \left( \frac{1}{\xi^{\frac{d}{2} - 1}} - \frac{1}{(1 + \xi)^{\frac{d}{2} - 1}} 
\right)  \right].
\end{equation}
Above we used that the boundary channel conformal block simplifies for interface operators with dimensions $d/2 - 1$ and $d/2$. In the short distance limit, we expect the two point function to be $\mathcal{N}_{\phi}/ (x_{12}^2)^{d/2 - 1}$ which sets 
\begin{equation}
	\frac{\mu_{\phi \hat{\phi}}^2}{2} + \frac{2 \mu_{\phi \widehat{\partial \phi}}^2}{d - 2} = 1.
\end{equation}
Using this relation the above two-point function is 
\begin{equation} \label{eq:PhiPhiFreePhi}
\langle \phi (x_{L1}) \phi(x_{L2}) \rangle = \frac{\mathcal{N}_{\phi}}{(4 z_{L1} z_{L2})^{\frac{d}{2} - 1}} \left[ \frac{1}{\xi^{\frac{d}{2} - 1}} +  \frac{\mu_{\phi \hat{\phi}}^2 - 1}{(1 + \xi)^{\frac{d}{2} - 1}} \right].
\end{equation}
Next we look at the two-point function of $\phi$ with an arbitrary operator $\Phi$ other than $\phi$, but still inserted on the free side of interface. We can use the same reasoning as above, and the only thing that changes is that in the short distance limit, we expect the two point function to vanish. This sets 
\begin{equation}
	\frac{\mu_{\phi \hat{\phi}} \mu_{\Phi \hat{\phi}} }{2} + \frac{2 \mu_{\phi \widehat{\partial \phi}} \mu_{\Phi \widehat{\partial \phi}} }{d - 2} = 0.
\end{equation}
This fixes the two point function 
\begin{equation}
\langle \phi (x_{L1}) \Phi(x_{L2}) \rangle = \frac{\sqrt{\mathcal{N}_{\phi} \mathcal{N}_{\Phi} }}{(2 z_{L1})^{\frac{d}{2} - 1} (2 z_{L2})^{\Delta_\Phi}}  \frac{\mu_{\phi \hat{\phi}} \mu_{\Phi \hat{\phi}} }{(1 + \xi)^{\frac{d}{2} - 1}}.
\end{equation}
Finally, we consider the two-point function of $\phi$ with $\Phi$ when $\Phi$ is inserted on the interacting side. In this case, we have to look at the short distance limit of the correlator in the folded picture, i.e. when $z_1 \rightarrow -z_1$. In the folded picture we have a tensor product of the free and interacting CFT, and $\phi$ and $\Phi$ belong to different CFTs, so their two-point function should not have any short-distance singularity. This sets to zero the coefficient of $(1 + \xi)^{1 - \frac{d}{2}}$ 
\begin{equation}
\frac{\mu_{\phi \hat{\phi}} \mu_{\Phi \hat{\phi}} }{2} - \frac{2 \mu_{\phi \widehat{\partial \phi}} \mu_{\Phi \widehat{\partial \phi}} }{d - 2} = 0.
\end{equation}
This fixes the corresponding two-point function 
\begin{equation}
\label{eq:phiOFreephi}
\langle \phi (x_{L1}) \Phi(x_{R2}) \rangle = \frac{\sqrt{\mathcal{N}_{\phi} \mathcal{N}_{\Phi} }}{(2 z_{L1})^{\frac{d}{2} - 1} (2 z_{R2})^{\Delta_\Phi}}  \frac{\mu_{\phi \hat{\phi}} \mu_{\Phi \hat{\phi}} }{\xi^{\frac{d}{2} - 1}}.
\end{equation}
In this last case, when $\Phi$ is inserted on the interacting side, the final result also applies to the case when the operator $\Phi$ is the field $\phi$ inserted on the interacting side.

In addition to the protected interface operators appearing in the bulk-interface OPE of a scalar, there are also protected operators that appear in the interface OPE of bulk conserved currents. The simplest such operator is the displacement operator which is present in every defect CFT (see for instance \cite{Billo:2016cpy}). It may be defined by its appearance in the divergence of stress tensor  
\begin{equation}
\partial_{\mu}T^{\mu i} = 0, \quad \partial_{\mu} T^{\mu z} = D(\mathbf{x}) \delta(z)
\end{equation}
where $i$ is a direction parallel to the interface. This defines the displacement operator, and since the stress tensor is conserved and protected, the displacement operator must also have protected dimensions equal to $d$. Since we have a free theory on the left side, there is an infinite tower of higher spin currents $J^{\mu_1 \dots \mu_s}$ that are exactly conserved (see \cite{Giombi:2016ejx} for a review). These currents are bilinear in the scalar $\phi$ and contain $s$ derivatives, and hence have scaling dimension $d - 2 + s$. These give rise to higher-spin ``cousins'' of the displacement operator defined analogously as above 
\begin{equation}
	\partial_{\mu} J^{\mu_1 \dots \mu_{s - 2} z} = D^{\mu_1 \dots \mu_{s-2}} (\mathbf{x}) \delta(z).
\end{equation} 
Note that the current in the above equation must be located on the free side of the interface. Since the current is conserved and protected, this equation defines protected operator $D^{\mu_1 \dots \mu_{s-2}}$. Different $\mu's$ could be either equal to $i$ or $y$ and correspondingly we have protected operators on the interface for all transverse spin between $0$ and $s - 2$. 

In the rest of this section, we study specific examples where we have free scalars or fermions on the left side, and a weakly interacting CFT on the right. We study various models including the $O(N)$ model with quartic interaction at the Wilson-Fisher fixed point, and the Gross-Neveu/Gross-Neveu-Yukawa models. All these models are parametrized by the number of fields $N$, and at large $N$ the corresponding interface models must match our predictions from the large $N$ analysis of general double-trace interfaces. So we will use these models as a way to check our general results in the large $N$ analysis developed in the previous sections. 

\subsection{$O(N)$ model}\label{sec:ONModel}
We begin by considering the interface in the context of the $O(N)$ vector model, with $N$ free scalars $\phi^i$ perturbed on half space by a quartic interaction $(\phi^i \phi^i)^2$, which flows to a Wilson-Fisher fixed point. The study of this interface was initiated in \cite{Gliozzi:2015qsa} using conformal bootstrap techniques. The action is given by
\begin{equation}\label{eq:ActionON}
S = \int d^d x \left( \frac{1}{2}  (\partial_{\mu} \phi^i )^2 + \ \theta(z) \frac{\lambda}{4} (\phi^i \phi^i)^2 \right).
\end{equation}
At $d = 4-\epsilon$ there is a perturbative fixed point at coupling
\begin{equation}
\lambda_* = \frac{8 \pi^2}{N+8}\epsilon + O(\eps^2).
\end{equation}
We can perform a consistency check of our large $N$ results by checking if they agree with the results in \cite{Gliozzi:2015qsa} in overlapping regimes of validity. In particular, we do it for the one-point function coefficients of $\langle O^2 (x_{L}) \rangle$ and $\langle \sigma^2 (x_{R}) \rangle$ given in \eqref{eq:OOCoincident} and \eqref{eq:sigmaCoincident}. By taking $\Delta \to d-2$ for these results and expanding in $d = 4-\epsilon$, we have
\begin{equation}
\begin{split}
\frac{\langle O^2 (x_{L}) \rangle}{\sqrt{2} C_O} &\xrightarrow{d = 4-\epsilon}  \frac{1}{(2z_L)^{2 \Delta}}\left( - \frac{\epsilon}{4\sqrt{2}} \right)\\
\frac{\langle \sigma^2 (x_{R}) \rangle}{\sqrt{2} C_\sigma} &\xrightarrow{d = 4-\epsilon} \frac{1}{(2z_R)^{2(d-\Delta)}} \left( \frac{\epsilon}{4\sqrt{2}} \right).
\end{split}
\end{equation}
This matches the large $N$ limit of the perturbative results given in \cite{Gliozzi:2015qsa} for the critical $O(N)$ model. Namely, their $(\phi^i \phi^i)^2$ one-point function coefficients are given by\footnote{Comparing our notation to that of \cite{Gliozzi:2015qsa}, we have $a_{\phi^4, R}  = a_{\phi^4}^{IR}$ and $a_{\phi^4, L} = a_{\phi^4}^{UV}$.}
\begin{equation}
a_{\phi^4, R} = - a_{\phi^4, L} = \frac{\sqrt{2N(N+2)}}{8(N+8)}\epsilon  \xrightarrow{\text{Large } N}  \frac{\epsilon}{4\sqrt{2}} .
\end{equation}

In \cite{Gliozzi:2015qsa}, the one-point function of $\phi^2$ inserted on the free side of the interface was also computed. However, it appears to disagree with the large $N$ one-point function of $O$ in Eq. \eqref{eq:OOnePointIR}. Setting  $\Delta = d-2$ in the large $N$ expression for fixed $d$, we get a vanishing result, while the result in \cite{Gliozzi:2015qsa} has the behavior (for unit-normalized operator) $a_{\phi^2}\sim \epsilon^2/\sqrt{N}$, which does not vanish at the leading order at large $N$. This likely indicates that a subtle order of limits issue is at play, or possibly a problem with the regularization scheme. We leave further investigation of this disagreement for future work. 

\subsubsection{Two-point function from equations of motion}
Next, we look at the two-point function of the field $\phi^i$ in the $O(N)$ model. When one or both of the fields are inserted on the free side, we already discussed the result in \eqref{eq:phiOFreephi}  and \eqref{eq:PhiPhiFreePhi}. So here we look at the two-point function when both $\phi$ fields are inserted on the interacting side. Again by conformal invariance, it must take the following form
\begin{equation} \label{eq:TwoPointFlatForm}
\langle \phi^i(x_{R1}) \phi^j(x_{2R}) \rangle = \delta^{ij} \frac{G_{\phi} (\xi)}{(z_{R1} z_{R2})^{\Delta_{\phi}}}\,. 
\end{equation}   
Just like in \eqref{Sec:LargeNFreeEnergy}, it will be convenient to map 
the right half space (where the interaction is present) to the hyperbolic space. The two-point function on hyperbolic space is obtained by a Weyl transformation from \eqref{eq:TwoPointFlatForm} and is given by   $\delta^{ij} G_{\phi}$. We will determine the function $G_{\phi} (\xi)$ order by order in $\epsilon$
\begin{equation}
G_{\phi} (\xi) = G_{0} (\xi) + \epsilon G_{1} (\xi) + \epsilon^2 G_{2} (\xi) + O(\epsilon^3)  \,. 
\end{equation}
We will fix this function by using the equations of motion for the field $\phi$. This is reminiscent of what was done in \cite{Giombi:2020rmc} for the case of boundary conformal field theory, and the calculation here is very similar to what was done in that case. The field $\phi$ in hyperbolic space satisfies the following equation of motion
\begin{equation}
\left( \nabla^2 + \frac{d (d - 2)}{4} \right) \phi^i = \lambda_* \phi^i (\phi^k \phi^k) \,,
\end{equation}
where $\nabla^2$ is the scalar Laplacian on hyperbolic space and the other term comes from the conformal mass for the field $\phi$ on hyperbolic space. This will enforce a differential equation for the function $G_{\phi} (\xi)$ and will allow us to fix the function upto some constants. 

For the rest of this subsection, we work in  a normalization such that at short distances, i.e. in limit $\xi \rightarrow 0$, the function $G_{\phi} = \xi^{- \Delta_{\phi}}$. This is a particular choice of normalization for the field $\phi$. In this normalization, the fixed point value of the coupling is $\lambda_* = \epsilon/ (2 (N+ 8))$. The normalization choice immediately fixes 
\begin{equation}
G_{0} (\xi) = \frac{1}{\xi}.
\end{equation}
To get to the next order, we will apply the differential operator to the two-point function 
\begin{equation}
\label{eq:SecondOrderEOM}
\left( \nabla^2 + \frac{d (d - 2)}{4} \right) G_{\phi} (\xi)\left( \delta^{ij} G_{\phi} (\xi) \right) \equiv  D^{(2)} \left( \delta^{ij} G_{\phi} (\xi) \right) = \lambda_* \langle \phi^i (\phi^k \phi^k) (x_{R1}) \phi^j(x_{R2}) \rangle \, ,
\end{equation}
where we defined 
\begin{equation}
 D^{(2)} = \left( \xi (\xi + 1) \partial_{\xi}^2 + d  (\xi + \tfrac{1}{2} ) \partial_{\xi} +  \frac{d (d - 2)}{4} \right) \, .
\end{equation}
Since we have a factor of $\lambda_*$ on the term on the right hand side, we should calculate the correlator on the RHS in the free theory. But since the one-point functions in the free theory vanish, the RHS is just zero. Expanding the differential operator into powers of $\epsilon$ as $D^{(2)} = D^{(2)}_0 + \epsilon D^{(2)}_1$, we get the following equation for the first order correction to the two-point function 
\begin{equation}
D^{(2)}_0 G_{1} (\xi) = - D^{(2)}_1  G_{0} (\xi) \,.
\end{equation}
This equation can be solved to give 
\begin{equation}
G_{1} (\xi) = \frac{c_1}{\xi} + \frac{c_2}{1 + \xi} + \frac{\log \xi}{2 \xi} \,.
\end{equation}
Fixing the normalization as we fixed above sets $c_1 = 0$, so we have fixed the two-point function to order $\epsilon$ upto one undetermined constant, $c_2$. To go to next order in $\epsilon$, we have to apply the differential operator to both the fields $\phi^i$ and $\phi^j$ in the two-point function. Doing that gives the following fourth-order differential equation 
\begin{equation} 
\label{eq:FourthOrderEOM} 
\begin{split}
&\bigg[\xi (1 + \xi) \left( \xi (1 + \xi) \partial_{\xi}^4 + (d + 2)(1 + 2 \xi) \partial_{\xi}^3 \right) + \frac{\left( d(d + 2) + (8 + 6 d (d + 2)) \xi (1 + \xi) \right)}{4} \partial_{\xi}^2  \\
&+ \frac{d^3 (1 + 2 \xi)}{4} \partial_{\xi} + \frac{d^2 (d - 2)^2}{16} \bigg] G_{\phi} (\xi) \equiv D^{(4)} G_{\phi} (\xi) = 2\lambda_*^2  (N + 2)  G_{\phi}^3.
\end{split}
\end{equation}   
Again expanding this operator into powers of $\epsilon$ as $D^{(4)} = D^{(4)}_0 + \epsilon D^{(4)}_1 + \epsilon^2 D^{(4)}_2  $, we get following differential equation for the second order correction 
\begin{equation}
D^{(4)}_0 G_2(\xi) = \frac{(N + 2)}{2 (N + 8)^2} (G_{0} (\xi))^3  - D^{(4)}_1 G_1(\xi) - D^{(4)}_2 G_0 (\xi) \,.
\end{equation} 
This can be solved to give the following result
\begin{equation}
\label{eq:SolEOMFourthOrder}
G_2(\xi) = \frac{d_1}{\xi} + \frac{d_2}{1 + \xi} +    \frac{d_3 \log \xi}{1 + \xi} +   \frac{ d_4 \log (1 + \xi)}{\xi} + \frac{c_2}{2} \frac{\log (1 + \xi)}{(1 + \xi)} -\frac{N + 2}{4 (N + 8)^2} \frac{\log \xi}{\xi}  + \frac{\log^2(\xi)}{8 \xi} \,.
\end{equation}  
Fixing the normalization of $\phi$ sets $d_1 = 0$. Note that $c_2$ above is the constant that appeared in the second order correction to the Green's function. So now we have four undetermined constants, namely $c_2, d_2, d_3$ and $d_4$. Next we will relate these coefficients to the interface CFT data by expanding the two-point function in the bulk and interface channels.

Let us start with the bulk channel, so we look at the correlator in the $\xi \rightarrow 0$ limit. The first operator that appears in the bulk OPE is the $\phi^2$ operator and using \eqref{eq:Two-point-Decom} \footnote{Note that in this normalization we have $\mathcal{N}_{\phi} = 2^{2 \Delta_{\phi}}$.}, we can write its contribution as follows
\begin{equation}
\begin{split}
G_{\phi}(\xi) &= \xi^{-\Delta_{\phi}} + \lambda_{\phi; \phi; \phi^2} a_{\phi^2} \xi^{\frac{1}{2} ( \Delta_{\phi^2} - 2 \Delta_{\phi})} \  + \  \textrm{higher orders in} \  \xi \\
&= \xi^{-\Delta_{\phi}} +  \epsilon (\lambda_{\phi; \phi; \phi^2} a_{\phi^2})^{(1)} + \epsilon^2 \left( (\lambda_{\phi; \phi; \phi^2} a_{\phi^2})^{(2)} + (\lambda_{\phi; \phi; \phi^2} a_{\phi^2})^{(1)} \left( \frac{\gamma^{(1)}_{\phi^2}}{2} - \gamma^{(1)}_{\phi} \right)    \log \xi  \right) \, ,
\end{split}
\end{equation}
where the superscripts indicate the order of perturbation theory in $\epsilon$ we are working at. Comparing this expansion to what we found above using the equations of motions gives us the following relations between CFT data and the undetermined coefficients
\begin{equation}
c_2 = (\lambda_{\phi; \phi; \phi^2} a_{\phi^2})^{(1)}, \hspace{1cm} d_2 = (\lambda_{\phi; \phi; \phi^2} a_{\phi^2})^{(2)} \,.
\end{equation}
We may then use the following bulk CFT data for the Wilson-Fisher CFT, which can be found in the literature (see e.g. \cite{Kleinert:2001ax})
\begin{equation}
\gamma^{(2)}_{\phi} = \frac{N + 2}{4 (N + 8)^2}, \ \ \gamma^{(1)}_{\phi^2} =\frac{N + 2}{N + 8}, \ \ \gamma^{(1)}_{\phi} = 0 \, ,
\end{equation}
and this fixes the coefficient of $\log \xi$ in the bulk OPE limit 
\begin{equation}
d_3 = \frac{(\lambda_{\phi; \phi; \phi^2} a_{\phi^2})^{(1)} (N + 2)}{2(N + 8)} = \frac{c_2(N + 2)}{2(N + 8)} \,.
\end{equation}

Next we look at the interface channel. Recall from \eqref{eq:Two-point-Decom} that the two-point function must have following decomposition in the interface channel 
\begin{equation}
G_{\phi} (\xi) = \sum_l \mu_{\phi l}^2 f_{\mathrm{bdy}} (\hat{\Delta}_l; \xi) \,.
\end{equation}
The interface channel blocks are eigenfunctions of the second-order and fourth-order differential operators with eigenvalues
\begin{equation}
\begin{split}
D^{(2)} f_{\mathrm{bdy}} (\hat{\Delta}_l; \xi) &= \frac{(d - 2 \hat{\Delta}_l)(d - 2 - 2 \hat{\Delta}_l)}{4} f_{\mathrm{bdy}} (\hat{\Delta}_l; \xi), \\
D^{(4)} f_{\mathrm{bdy}} (\hat{\Delta}_l; \xi) &= \frac{(d - 2 \hat{\Delta}_l)^2(d - 2 - 2 \hat{\Delta}_l)^2}{16} f_{\mathrm{bdy}} (\hat{\Delta}_l; \xi) \,.
\end{split}
\end{equation} 
Plugging in this decomposition into the differential equation in \eqref{eq:SecondOrderEOM} tells us that to order $\epsilon$, we have 
\begin{equation}
\sum_l \frac{ \mu_{\phi l}^2 (d - 2 \hat{\Delta}_l)(d - 2 - 2 \hat{\Delta}_l)}{4} f_{\mathrm{bdy}} (\hat{\Delta}_l; \xi) = 0 \,.
\end{equation}
So to order $\epsilon$, we only have two operators in the interface OPE of $\phi$, namely $\hat{\phi}$ and $\widehat{\partial \phi}$ with dimension $d/2 - 1$ and $d/2$. At next order, plugging in the interface channel decomposition in \eqref{eq:FourthOrderEOM}, we get 
\begin{equation}
\sum_l \frac{ \mu_{\phi l}^2 (d - 2 \hat{\Delta}_l)^2(d - 2 - 2 \hat{\Delta}_l)^2}{16} f_{\mathrm{bdy}} (\hat{\Delta}_l; \xi) = \frac{(N + 2) \epsilon^2}{2 (N + 8)^2 \xi^3} \, .
\end{equation}
The fact that the right hand side is proportional to $1/\xi^3$ tells us that the leading two operators with dimensions $d/2 - 1$ and $d/2$ are protected to this order. This agrees with what we found in the previous subsection while analyzing the two-point function of $\phi$ on the free side: that the operators $\hat{\phi}$ and $\widehat{\partial \phi}$ are protected operators. The above equation also fixed the bulk-interface OPE coefficient of the next operator with scaling dimension $3$, $\mu_{\phi 3}^2 = \frac{(N + 2) \epsilon^2}{8 (N + 8)^2}$. 

The fact that the operator with dimension $d/2 - 1$ is protected also constrains the $\log \xi/ \xi$ terms that can appear in \eqref{eq:SolEOMFourthOrder} in the limit $\xi \rightarrow \infty$. This constrains the coefficients as follows 
\begin{equation}
d_3 + d_4 = \frac{(N + 2)}{4(N + 8)^2} \implies d_4 = \frac{(N + 2)}{4(N + 8)^2} -  \frac{c_2(N + 2)}{2(N + 8)}\,.
\end{equation}
As a result, we have fixed the two-point function of the field $\phi$ on the interacting side up to two coefficients, namely $c_2$ and $d_2$, which are both related respectively to the order $\epsilon$ and $\epsilon^2$ pieces in the bulk-one point function of $\phi^2$. In fact, we expect to have $c_2 = 0$ since, as pointed out in \cite{Gliozzi:2015qsa}, the one-point functions coefficient $a_{\phi^2}$ should start at $O(\epsilon^2)$, since we have to bring down two interaction terms to compute the leading order diagram. 
\subsubsection{Free energy}
To compute the free energy in both the $O(N)$ model and other models, we make repeated use of the following integral 
\begin{equation}\label{eq:I1EpsHd}
\begin{split}
I_1 (\Delta,d) &= \int d^d x_1 d^d x_2 \sqrt{g_{x_1}} \sqrt{g_{x_2}}  \frac{1}{(4\xi_{12})^{\Delta}} \\
&= \text{vol}(H^d) \Omega_{d-1} \int_0 ^\infty d\eta (\sinh(\eta))^{d-1} \frac{1}{(4\sinh^2 (\eta/2))^\Delta}  \\
&=   \pi ^{d-\frac{3}{2}} 2^{d-2 \Delta } \sin \left(\tfrac{\pi  d}{2}\right) \Gamma \left(\tfrac{1}{2}-\tfrac{d}{2}\right) \Gamma \left(\tfrac{d}{2}-\Delta \right) \Gamma (-d+\Delta +1) .
\end{split}
\end{equation}
where in the second line we used hyperbolic coordinates \eqref{eq:HyperbolicCoords} and fixed one of the points to be at the center 
as in \eqref{eq:CrossRatioEta}. The integral \eqref{eq:I1EpsHd} corresponds to a generalized free scalar propagator with no boundary condition imposed at the interface. For the $O(N)$ interface described by \eqref{eq:ActionON}, the free energy of the theory is given by
\begin{equation}
F_{O(N)} = N F_s + F_{\rm int}^{O(N)}
\end{equation}
where $N F_s$ is the free energy of the $N$ free scalars on $S^d$ and $F_{\rm int}^{O(N)}$ is the contribution from the interaction. We can directly use perturbation theory to compute the contribution of the interaction term to the free energy. To leading order in $\epsilon$, it is given by 
\begin{equation}
\begin{split}
 F_{\rm int}^{O(N)}  &= - \frac{\lambda^2}{32} \int d^d x d^d \sqrt{g_{x_1}} \sqrt{g_{x_2}} \langle \phi^4 (x_1) \phi^4(x_2) \rangle \\ 
&= - \frac{\lambda^2 N (N + 2) }{4 } \int d^d x_1 d^d x_2 \sqrt{g_{x_1}} \sqrt{g_{x_2}} G_{\phi}(x_1, x_2)^4 
\end{split}
\end{equation} 
where the two-point function on the hyperbolic space takes the form
\begin{equation}
G_{\phi} = \frac{\Gamma \left( \frac{d}{2} - 1 \right)}{4 \pi^{\frac{d}{2}} (4 \xi)^{d/2 - 1}}. 
\end{equation} 
Thus the free energy contribution from the interaction is given by 
\begin{equation}
\begin{split}
F_{\rm int}^{O(N)}  &= -\frac{N (N+2) \epsilon ^2}{16 \pi ^4 (N+8)^2} I_1(2d-4,4-\epsilon) \\
&=  \frac{N (N + 2)}{1152(N + 8)^2} \epsilon^2 + O(\epsilon^3)  \xrightarrow{\text{Large } N} \frac{\epsilon ^2}{1152} + O(\epsilon^3)
\label{Fint-ON}
\end{split}
\end{equation}
To compare to the large $N$ $O(N)$ model near four dimensions, we specify \eqref{eq:LargeNFreeEnergy} to $\Delta=  d - 2$ and $d = 4 - \epsilon$, 
which gives 
\begin{equation}
\begin{split}
F_{\sigma}(\Delta=d-2) &= \frac{1}{\Gamma (d+1) \sin (\pi  d)} \int_0^{ \frac{d}{2} - 2} du \ u \  \Gamma \left(\frac{d}{2} + u \right) \Gamma \left(\frac{d}{2} - u \right) \sin \pi \left(\frac{d}{2} - u \right) \\
&\stackrel{d=4-\epsilon}{=}   \frac{\epsilon^2}{1152} + \frac{13 \epsilon^3}{13824} + O(\epsilon^4),
\end{split}
\end{equation}
which is indeed consistent with the epsilon expansion result (\ref{Fint-ON}) to the leading order. 

\subsection{Gross-Neveu model}
We now consider $N_f$ Dirac fermions (denoting $N = N_f \text{tr}\mathbf{1}$), which are free on one side of the interface and have a four-fermi interaction on the other side, corresponding to a Gross-Neveu (GN) model
\begin{equation}
S = \int d^d x \left( \bar{\psi}_i \gamma^{\mu} \partial_{\mu} \psi^i + \frac{g \ \theta(z)}{2} ( \bar{\psi}_i \psi^i)^2 \right).
\end{equation}
This model has a fixed point in $d = 2 + \epsilon$ dimensions at the coupling
\begin{equation}
g_* = \frac{2 \pi }{N - 2} \epsilon .
\end{equation}
\subsubsection{Free energy}
The free energy of the theory is given by
\begin{equation}
F_{GN} = N F_f + F_{\rm int}^{GN}.
\end{equation}
where $F_f$ is the free energy of a single free fermion component on $S^d$, computed in \cite{Giombi:2014xxa}, and $F_{\rm int}^{GN}$ is the contribution from the four-fermi interaction. We compute $F_{\rm int}^{GN}$ to leading order in perturbation theory
\begin{equation}
\begin{split}
F_{\rm int}^{GN} &=  - \frac{g^2}{8} \int d^d x_1 d^d x_2 \sqrt{g_{x_1}} \sqrt{g_{x_2}} \langle ( \bar{\psi}_i \psi^i)^2  (x_1) ( \bar{\psi}_i \psi^i)^2(x_2) \rangle \\
&= - \frac{N (N - 1) \epsilon^2}{16 (N - 2)^2 \pi^2}   \int d^d x_1 d^d x_2 \sqrt{g_{x_1}} \sqrt{g_{x_2}} \frac{1}{(4 \xi)^{2 (d - 1)}} \\
&=  -\frac{(N-1) N \epsilon ^2}{16 \pi ^2 (N-2)^2} I_1 (2d-2, 2+\epsilon) \\
&=  - \frac{N (N - 1) \epsilon^2}{96 (N - 2)^2}  + O(\epsilon^3) \xrightarrow{\text{Large } N} - \frac{\epsilon^2}{96} + O(\epsilon^3) .
\end{split}
\end{equation}
In comparison, the contribution to the large $N$ free energy for the Gross-Neveu model is given by plugging in $\Delta = d - 1$ to \eqref{eq:LargeNFreeEnergy} and specifying $d = 2 + \epsilon$. In the limit of small $\epsilon$, it indeed agrees with the epsilon expansion result derived above:
\begin{equation}
\begin{split}
F_{\sigma}(\Delta = d-1) &= \frac{1}{\Gamma (d+1) \sin (\pi  d)} \int_0^{\frac{d}{2} - 1} du \ u \  \Gamma \left(\frac{d}{2} + u \right) \Gamma \left(\frac{d}{2} - u \right) \sin \pi \left(\frac{d}{2} - u \right) \\
&\overset{d\to 2+\epsilon}{=}  - \frac{\epsilon^2}{96} + \frac{\epsilon^3}{64} + O(\epsilon^4). \\
\end{split}
\end{equation}
\subsection{Gross-Neveu-Yukawa model}\label{sec:GNY}
Let us now consider an interface between $N_f$ free fermions and a Gross-Neveu-Yukawa (GNY) model, which has the following action
\begin{equation}\label{eq:GNYAction}
S = \int d^d x \left( \bar{\psi}_i \gamma^{\mu}  \partial_{\mu} \psi^i  + \ \theta(z)\left( \frac{1}{2} (\partial_{\mu} s)^2 + g_1 s \bar{\psi}_i \psi^i  + \frac{g_2}{24 } s^4 \right) \right).
\end{equation} 
Since the scalar field $s$ only lives in the half space $z>0$, we will assume Dirichlet boundary conditions for that field at $z=0$. 

The interacting GNY model on the right side of the interface has perturbative IR fixed points in $d = 4 - \epsilon$ at the following values of the couplings \cite{Zinn-Justin:1991ksq,Fei:2016sgs}
\begin{equation}
\begin{split}
(g_1^*)^2& = \frac{(4 \pi)^2}{N + 6} \epsilon \\
g_2^* &= \frac{(4\pi)^2 \left( - 2 N + 3  + \sqrt{4 N ^2 + 132 N + 9} \right)}{3 (4 N + 6)} \epsilon .
\end{split}
\end{equation}
\subsubsection{Free energy}
The free energy of the theory is given by 
\begin{equation}\label{eq:FGNY}
F_{GNY} = N F_f + F_s^D + F_{\rm int}^{GNY}
\end{equation}
where, as in the Gross-Neveu model, the first term $N F_f $ is the contribution from $N_f$ free Dirac fermions and the last term $F_{\rm int}^{GNY}$ is the contribution from the interaction terms. However, now we have an additional term $F_s^D$, which is the contribution from a single free scalar on a hemisphere with Dirichlet boundary conditions, which arises from the scalar still being present on half space when the couplings are sent to zero. We may use perturbation theory to compute the interface contribution to free energy at $d = 4-\epsilon$. Because the scalar field has Dirichlet boundary conditions, we require an additional integral beyond \eqref{eq:I1EpsHd}, which we denote $I_2$:
\begin{equation}\label{eq:I2EpsHd}
\begin{split}
I_2(\Delta_1,\Delta_2,d)  &\equiv \int d^d x_1 d^d x_2 \sqrt{g_{x_1}} \sqrt{g_{x_2}} \frac{1}{(4\xi_{12})^{\Delta_1}} \frac{1}{(4(\xi_{12}+1))^{\Delta_2}}  \\
&= \text{vol}(H^d) \Omega_{d-1} \int_0 ^\infty d\eta (\sinh(\eta))^{d-1} \frac{1}{(4\sinh^2 (\eta/2))^{\Delta_1} (4\cosh^2 (\eta/2))^{\Delta_2}}  \\
 &=  \frac{\pi ^{\frac{d-1}{2}+\frac{d}{2}} 2^{d-2 \Delta _1-2 \Delta _2} \Gamma \left(\frac{1-d}{2}\right) \Gamma \left(\frac{d}{2}-\Delta _1\right) \Gamma \left(-d+\Delta_1+\Delta _2+1\right)}{\Gamma \left(\frac{d}{2}\right) \Gamma \left(-\frac{d}{2}+\Delta_2+1\right)}.
\end{split}
\end{equation}
Thus, the free energy contribution to from the interaction is, to leading order, 
\begin{equation} 
\begin{split}
F_{\rm int}^{GNY} &= - \frac{g_1^2}{2} \int d^d x_1 d^d x_2 \sqrt{g_{x_1}} \sqrt{g_{x_2}}  \langle s \bar{\psi}_i \psi^i(x_1) s \bar{\psi}_i \psi^i(x_2) \rangle \\
&= - \frac{N \epsilon}{2 (N + 6) \pi^4} \int d^d x_1 d^d x_2 \sqrt{g_{x_1}} \sqrt{g_{x_2}} \frac{1}{(4 \xi)^{d - 1}} \left( \frac{1}{(4 \xi)^{\frac{d}{2} - 1}} - \frac{1}{(4 (1 +  \xi))^{\frac{d}{2} - 1}}\right)  \\
&= -\frac{N \epsilon }{2 \pi ^4 (N+6)} \left(I_1(\tfrac{3d}{2}-2,4-\epsilon) -I_2(d-1,\tfrac{d}{2}-1,4-\epsilon)  \right) \\
& = \frac{N \epsilon }{96 (N + 2)}  \xrightarrow{\text{Large } N} \frac{\epsilon }{96}
\end{split}
\end{equation}
Note that the contribution from the $I_2$ integral is actually zero due to the $\Gamma(0)$ in the denominator of \eqref{eq:I2EpsHd}, so computing the $F_{\rm int}^{GNY}$ with the Dirichlet propagator for $\sigma$ yields the same result as using the $\sigma$ propagator with no boundary condition.

To see how this provides a check of the large $N$ result $F_\sigma (\Delta)$, note that the IR fixed point of the GNY model is expected to be equivalent to the CFT at the UV fixed point of the GN model \cite{Zinn-Justin:1991ksq}. In other words, we expect the following to be true
\begin{equation}
F_{GNY} |_{g_1^*, g_2^*} = F_{GN} |_{g^*}  = N_f F_f + F_\sigma (\Delta = d-1)  + O(1/N)
\end{equation}
and by comparing the above equation to \eqref{eq:FGNY}, we see that $F_{\rm int}^{GNY}$ should be related to $F_\sigma$ as
\begin{equation}
F_{\rm int}^{GNY} = F_\sigma (\Delta = d-1) - F_s^D .
\label{FGNY-expected}
\end{equation}
We first obtain $F_\sigma$ by plugging in $\Delta = d - 1$ and $d = 4-\epsilon$ into \eqref{eq:LargeNFreeEnergy}:
\begin{equation}
\begin{split}
F_{\sigma}(\Delta = d-1) &= \frac{1}{\Gamma (d+1) \sin (\pi  d)} \int_0^{1 - \frac{\epsilon}{2}} du \ u \  \Gamma \left(\frac{d}{2} + u \right) \Gamma \left(\frac{d}{2} - u \right) \sin \pi \left(\frac{d}{2} - u \right) \\
&\overset{d\to 4-\epsilon}{=} \frac{1}{180 \epsilon} + \frac{240 \log (A)-480 \zeta '(-3)-\frac{180 \zeta (3)}{\pi ^2}-29-16 \gamma }{2880} + 0.00727 \epsilon + O(\epsilon^2)
\end{split}
\end{equation}
where $A$ is Glaisher's constant and $\gamma$ is the Euler-Mascheroni constant. To obtain $F_s^D$, note that the free energy of the scalar on the whole sphere, $F_s$, is given by summing over Dirichlet and Neumann modes:
\begin{equation}
F_s = F_s^D + F_s^N =  -\frac{1}{\sin(\pi d/2)\Gamma(1+d)} \int_0^1 du \ u \sin(\pi u) \Gamma\left(\frac{d}{2}+u\right)\Gamma\left(\frac{d}{2}-u\right)
\end{equation}
where $F_s$ is given in \cite{Giombi:2014xxa}. On the other hand, the change in free energy flowing from Neumann to Dirichlet boundary conditions is given by \cite{Giombi:2020rmc}
\begin{equation}
F_s^N - F_s^D = -\frac{1}{\sin(\tfrac{\pi(d-1)}{2})\Gamma(d)} \int_0^{\tfrac{1}{2}} du \ u \sin(\pi u) \Gamma\left(\frac{d-1}{2}+u\right) \Gamma\left(\frac{d-1}{2}-u\right).
\end{equation}
So the free energy contribution from the scalar on the hemisphere with Dirichlet boundary conditions is just
\begin{equation}
\begin{split}
F_s^D  &= \frac{1}{2}(F_s -(F_s^N - F_s^D )) \\
&\overset{d\to4-\epsilon}{=} \frac{1}{180 \epsilon}+\frac{240 \log (A)-480 \zeta'(-3)-29-16 \gamma }{2880}-\frac{\zeta (3)}{16 \pi ^2}- 0.003149 \epsilon +O(\epsilon^2).
\end{split}
\end{equation}
Then, according to (\ref{FGNY-expected}), we find that the contribution from the GNY interaction term at large $N$, specified to $d=4-\epsilon$ should be
\begin{equation}
F_{\rm int}^{GNY} = F_\sigma (\Delta = d-1) - F_s^D = \frac{\epsilon}{96} + O(\epsilon^2),
\end{equation}
in agreement with the epsilon expansion result. 
\subsection{Cubic $O(N)$ model}
We now consider the cubic $O(N)$ scalar theory given by
\begin{equation}
S = \int d^d x \left( \frac{1}{2}  (\partial_{\mu} \phi^i )^2 + \ \theta(z)\left( \frac{1}{2} (\partial_{\mu} s)^2 + \frac{g_1}{2} s \phi^i \phi^i  + \frac{g_2}{6 } s^3 \right) \right).
\end{equation}
For $N>N_{crit}$, the interacting model has unitary IR stable perturbative fixed points at $d = 6-\epsilon$. At leading order in large $N$, the couplings at the fixed point are given by \cite{Fei:2014yja} 
\begin{equation}
\begin{split}
g_1^* &= \sqrt{\frac{6 \epsilon (4 \pi)^3}{ N} } \\
g_2^* &= 6 g_1^* .
\end{split}
\end{equation}
\subsubsection{Free energy}
The free energy of the theory is given by 
\begin{equation}\label{eq:FCubic}
F_{\text{cubic }O(N)} = N F_s + F_s^D + F_{\rm int}^{\text{cubic }O(N)}
\end{equation}
Using ordinary perturbation theory at $d = 6-\epsilon$, we compute the contribution to the free energy from the interaction term, which gives
\begin{equation} \label{eq:CubicFreeEnergy}
\begin{split}
F_{\rm int}^{\text{cubic }O(N)}  &= - \frac{g_1^2}{8} \int d^d x_1 d^d x_2 \langle s \phi^i \phi^i (x_1) s \phi^i \phi^i (x_2) \rangle \\
&= - \frac{3 \epsilon}{2  \pi^6}\int d^d x_1 d^d x_2 \sqrt{g_{x_1}} \sqrt{g_{x_2}} \frac{1}{(4 \xi)^{d - 2}} \left( \frac{1}{(4 \xi)^{\frac{d}{2} - 1}} - \frac{1}{(4 (1 +  \xi))^{\frac{d}{2} - 1}}\right) \\
&=  -\frac{3 \epsilon }{2 \pi ^6} \left(I_1 (\tfrac{3d}{2}-3, 6-\epsilon)-I_2(d-2,\tfrac{d}{2}-1,6-\epsilon) \right)  \\
&= -\frac{\epsilon}{960} + O(\epsilon^2)  .
\end{split}
\end{equation}
Though we imposed Dirichlet boundary conditions on $\sigma$, we again find that the contribution from the $I_2$ integral is zero due to the $\Gamma(0)$ in the denominator of \eqref{eq:I2EpsHd}. 

To provide a check of the large $N$ result $F_\sigma (\Delta)$, note that the IR fixed point of the cubic $O(N)$ model is equivalent to the CFT at the UV fixed point of the $O(N)$ model \cite{Fei:2014yja}. Thus, we expect
\begin{equation}
F_{\text{cubic }O(N)} |_{g_1^*, g_2^*} = F_{O(N)} |_{\lambda^*}  = N F_s + F_\sigma (\Delta = d-2) + O(1/N)
\end{equation}
and by comparing this to \eqref{eq:FCubic}, we see $F_{\rm int}^{\text{cubic }O(N)}$ can be expressed in terms of $F_\sigma$ as
\begin{equation}
F_{\rm int}^{GNY} = F_\sigma (\Delta = d-2) - F_s^D .
\end{equation}
We first obtain $F_\sigma$ by plugging in $\Delta = d - 2$ and $d = 6-\epsilon$ to \eqref{eq:LargeNFreeEnergy}:
\begin{equation}
\begin{split}
F_\sigma (\Delta = d-2) &= \frac{1}{\Gamma (d+1) \sin (\pi  d)} \int_0^{1 - \frac{\epsilon}{2}} du \ u \  \Gamma \left(\frac{d}{2} + u \right) \Gamma \left(\frac{d}{2} - u \right) \sin \pi \left(\frac{d}{2} - u \right) \\
&\overset{d\to 6-\epsilon}{=}-\frac{1}{1512 \epsilon }+ \frac{-378 \log (A)-378 \zeta '(-5)+47+30 \gamma }{45360}+\frac{\zeta (5)}{64 \pi ^4}+\frac{\zeta (3)}{192 \pi ^2} \\
&\qquad -0.000835423 \epsilon + O(\epsilon^2).
\end{split}
\end{equation}
The free energy contribution from a free scalar on the hemisphere with Dirichlet boundary conditions was found in \eqref{sec:GNY}, and in $d = 6-\epsilon$ it is given by
\begin{equation}
\begin{split}
F_s^D  &\underset{d=6-\epsilon}{=} - \frac{1}{1512 \epsilon} + \frac{-378 \log (A)-378 \zeta '(-5)+47+30 \gamma }{45360}+\frac{\zeta (5)}{64 \pi ^4}+\frac{\zeta (3)}{192 \pi ^2} \\
&\qquad + 0.000206243817 \epsilon +O(\epsilon^2).
\end{split}
\end{equation}
Thus, we have
\begin{equation}
\delta F^{\text{cubic }O(N)}  = F_{\sigma}(\Delta = d-2) -  F_s^D =- \frac{\epsilon}{960} + O(\epsilon^2),
\end{equation}
again in agreement with the epsilon expansion result. 

\section{Holographic setup}\label{Sec:HolographicSetup}
\begin{figure}[h]
\centering
\includegraphics[width=.5\textwidth]{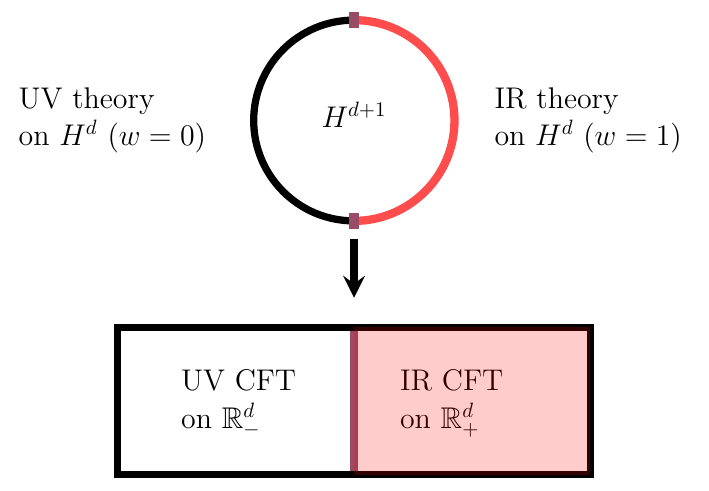}
\caption{Schematic depiction of the holographic setup describing the double-trace interfaces. The boundary of $H^{d+1}$ is divided into two $H^d$ regions, which correspond to the (Weyl-rescaled) left and right halves of the interface CFT. The bulk scalar field $\varphi$ dual to the operator $O$ has inhomogeneous boundary conditions corresponding to dual dimension $\Delta$ near the left (black) boundary, and to dual dimension $d-\Delta$ near the right (red) boundary. The two $H^d$ regions are joined through their common boundary, which corresponds to the interface on the CFT side.}
\label{holographic-sketch}
\end{figure}

In this section we review the holographic dual of the large $N$ double trace interface discussed in \cite{Melby-Thompson:2017aip}. This is essentially a generalization of the well-known dictionary \cite{Klebanov:1999tb} relating double-trace deformations to alternate boundary conditions in AdS. A schematic summary of the setup is shown in Figure \ref{holographic-sketch}. In the next section we will then apply these results to show equivalence between AdS and CFT calculations of various correlation functions. 

Following \cite{Melby-Thompson:2017aip}, it is convenient to use the following Janus coordinates for the $H^{d+1}$ bulk\footnote{An alternative, perhaps more familiar, form of the metric is obtained by setting $w=\frac{1}{2}(1+\tanh(\rho))$, which gives $ds^2_{H^{d+1}}=d\rho^2 + \cosh^2\rho\, ds^2_{H^d}$, where $-\infty < \rho < \infty$ and the two $H^d$ boundaries are at $\rho =\pm \infty$.}
\begin{equation}
ds_{H^{d+1}}^2 = \frac{dw^2}{4w^2 (1-w)^2}+\frac{ds^2_{H^d}}{4w(1-w)}, \qquad w\in (0,1)
\end{equation}
where $H^{d+1}$ is foliated by $H^d$ slices, and the two $H^d$ boundary slices at $w=0$ and $w=1$ are expected to respectively correspond, upon Weyl rescaling, to the half-spaces where the UV and IR CFTs live. The interface is at the boundary of $H^d$ (this is the common boundary of the two halves of the $H^{d+1}$ boundary). 
We denote $X = (w,x)\in H^{d+1}$ where $x\in H^d$ can be written in Poincare coordinates as in \eqref{eq:WeylRescale}. We will use the notation $\langle \dots \rangle_{H^d}$ to denote correlation functions in the Weyl-rescaled CFTs, i.e. $\langle O(x) O(x') \rangle_{H^d} = z^\Delta (z')^{\Delta} \langle O(x) O(x') \rangle$.

In the absence of the interface, the bulk scalar field $\varphi$ dual to the operator $O$ has the standard bulk propagator $G_\Delta (X, X')$ on a pure AdS background, satisfying the equation
\begin{equation}
(-\nabla^2_{H^{d+1}}+ m^2) G_\Delta (X,X') = \delta(X,X')
\end{equation}
and with the following homogeneous boundary conditions ($\sim (w')^{\Delta/2}$ or $(1-w')^{\Delta/2}$)\footnote{These boundary conditions describe the ``UV CFT" where $O$ has scaling dimension $\Delta$. The case of homogeneous boundary conditions with $\Delta\rightarrow d-\Delta$ would describe the double-trace deformation turned on over the full space.}  in terms of Janus coordinates as $w' \to 0$ and $w' \to 1$: 
\begin{equation}\label{eq:GAsymptotics}
\begin{split}
 G_\Delta (X;w',x') & = \begin{cases}
    -\frac{1}{d-2\Delta} K_\Delta (X; x')  (4w')^{\tfrac{\Delta}{2}} + O( w'^{\tfrac{\Delta+2}{2}} ) & w' \to 0  \\ 
    -\frac{1}{d-2\Delta} K_\Delta (X; x')  (4(1-w'))^{\tfrac{\Delta}{2}} + O( (1-w')^{(\tfrac{\Delta+2}{2}} ) & w' \to 1
 \end{cases} .
\end{split}
\end{equation}
Above, $K_\Delta$ is the standard scalar bulk-to-boundary propagator, given in Janus coordinates by
\begin{equation}\label{eq:BulkBoundaryStandard}
K_{\Delta} (w,x; x') = \begin{cases} \mathcal{C}_\Delta  \left(\dfrac{\sqrt{w(1-w)}}{2(w+\xi)} \right)^{\Delta} , & w' = 0 \\   \mathcal{C}_\Delta  \left(\dfrac{\sqrt{w(1-w)}}{2(1-w+\xi)} \right)^{\Delta} , & w' = 1 \end{cases} 
\end{equation}
where the normalization constant is
\begin{equation}\label{eq:A}
\mathcal{C}_\Delta = \frac{ \Gamma (\Delta )}{\pi^{d/2}\Gamma \left(\Delta -\frac{d}{2}\right) } \,.
\end{equation}
In these conventions for the bulk-to-boundary propagator, the two-point function of the dual CFT operator (in the absence of the interface) is 
given by $\langle O(x_1)O(x_2)\rangle_0 = (2\Delta-d)\mathcal{C}_{\Delta}/x_{12}^{2\Delta}$ \cite{Freedman:1998tz, Klebanov:1999tb}. This means that in these conventions, we can identify $C_O = (2\Delta-d)\mathcal{C}_{\Delta}$, where $C_O$ is the two-point normalization constant defined in (\ref{G0-norm}). 

In the presence of the double trace interface, one should instead impose inhomogeneous boundary conditions on the bulk scalar field, corresponding to dual dimension $\Delta$ or $d-\Delta$ in the left and right half of the boundary, respectively. The corresponding Green's function, denoted $\tilde{G}$ below, satisfies the usual equation
\begin{equation}
(-\nabla^2_{H^{d+1}}+ m^2) \tilde{G} (X,X') = \delta(X,X')
\end{equation}
but now with different asymptotics (either $\sim (w')^{\Delta/2}$ or $\sim (1-w')^{(d-\Delta)/2}$) as one point of the propagator is sent towards the $w'=0$ or $w'=1$ $H^d$ slice: 
\begin{equation}\label{eq:TildeGAsymptotics}
\begin{split}
&\tilde{G}(X;w',x') \\
& =  \begin{cases}   -\frac{1}{d-2\Delta} \tilde{K}_{w'=0}(X; x')  (4w')^{\tfrac{\Delta}{2}} + 0 \cdot  (w')^{\tfrac{d-\Delta}{2}} +  O( (w')^{\tfrac{\Delta+2}{2}} ) , & w' \to 0 \\  
	  0\cdot (1-w')^{\tfrac{\Delta}{2}}  + \frac{1}{d-2\Delta} \tilde{K}_{w'=1}(X;x')  (4(1-w'))^{\tfrac{d-\Delta}{2}} + O((1-w')^{\tfrac{d-\Delta+2}{2}} )  , & w' \to 1 \end{cases} .
\end{split}
\end{equation}
Above, the bulk-to-boundary propagators\footnote{In the notation of \cite{Melby-Thompson:2017aip}, $\tilde{K} _{w'=0}(w,x;x')$ corresponds to ``$K_L^{-+}(w,x; x')$'' and $\tilde{K} _{w'=1}(w,x;x')$ corresponds to ``$K_R^{-+}(w,x; x')$''.} $\tilde{K} _{w'=0}(w,x;x')$ and $\tilde{K} _{w'=1}(w,x;x')$ depend on whether the boundary insertion is at $w'=0$ or $w'=1$:
\begin{equation}\label{eq:TildeK}
\begin{split}
\tilde{K}_{w'=0} (w,x; x') &= \dfrac{\sin\pi(\Delta-\tfrac{d}{2})}{\pi} \dfrac{\Gamma(\tfrac{d}{2})}{\pi^{d/2}} \dfrac{[4w(1-w)]^{\tfrac{\Delta}{2}}}{4^{\Delta} (1-w)^{\Delta-d/2}} (1+\xi)^{-\tfrac{d}{2}} {}_2 F_1 \left( 1, \frac{d}{2}, 1+\tfrac{d}{2}- \Delta;  \dfrac{1-w}{1+\xi} \right) \\
\tilde{K}_{w'=1} (w,x; x') &= \dfrac{\sin\pi(\tfrac{d}{2}-\Delta)}{\pi} \dfrac{\Gamma(\tfrac{d}{2})}{\pi^{d/2}} \dfrac{[4w(1-w)]^{\tfrac{d-\Delta}{2}}}{4^{d-\Delta} w^{d/2-\Delta}} (1+\xi)^{-\tfrac{d}{2}} {}_2 F_1 \left( 1, \tfrac{d}{2},  \Delta-\tfrac{d}{2}+1;  \dfrac{w}{1+\xi} \right) .
\end{split}
\end{equation}
Note that we can go from $\tilde{K}_{w'=0} (w,x; x') $ to $\tilde{K}_{w'=1} (w,x; x') $ by taking $w \to 1- w$ and $\Delta \to d-\Delta$. The bulk propagator $\tilde{G} (X; X') $ solved for in \cite{Melby-Thompson:2017aip} takes a complicated form that we include in Appendix \ref{App:HolographicPropagator}. The asymptotics of $\tilde{K}_{w'=0}$ and  $\tilde{K}_{w'=1}$ as we take $w \to 0, 1$ are related to the various two-point functions involving the operators $O$ (of dimension $\Delta$) and $\sigma$ (of dimension $d-\Delta$) in the dual CFT. For example, the $w \to 0$ limit of $\tilde{K}_{w'=0} $ is given by
\begin{equation}
\begin{split}
\lim_{w \to 0} &\tilde{K}_{w_L=0}(x_L; w,x) =- (4w)^{\Delta /2} \frac{\Gamma (\tfrac{d}{2}) \sin (\tfrac{\pi  (d-2 \Delta )}{2} ) }{ \pi^{\frac{d}{2}+1}(\xi +1)^{d/2} }  \,_2F_1\left(1,\frac{d}{2};\frac{d}{2}-\Delta +1;\frac{1}{\xi +1}\right)+O(w^{\Delta/2+1}) \\
&= - (4w)^{\Delta /2} \frac{\Gamma (\tfrac{d}{2}) \sin (\tfrac{\pi  (d-2 \Delta )}{2} ) }{ \pi^{\frac{d}{2}+1}(\xi)^{d/2} } \,_2F_1\left(\frac{d}{2},\frac{d}{2}-\Delta ;\frac{d}{2}-\Delta +1;-\frac{1}{\xi }\right)+O(w^{\Delta/2+1}) \\
&=  (4 w)^{\Delta/2}\frac{ z_1^\Delta z_2^\Delta  \langle O(x_1) O(x_2) \rangle}{2\Delta -d}+O(w^{\Delta/2+1})
\end{split}
\end{equation}
where we applied a hypergeometric identity\footnote{$\,_2 F_1 (a,b,c,z) = (1-z)^{-b} \,_2 F_1 (c-a,b,c, \frac{z}{z-1})$} in the second line. To get the last line, we recall that the normalization conventions we are using on the AdS side correspond to the choice $C_O = (2\Delta-d)\mathcal{C}_{\Delta}$, as explained below (\ref{eq:A}). The rest of the asymptotics are obtained in a similar way:
\begin{equation}\label{eq:TildeKAsymptotics}
\begin{split}
&\tilde{K}_{w_L =0} (x_L; w,x) \\
=&\begin{cases}
\frac{1}{2\Delta-d}\langle O(x_L) O(x) \rangle_{H^d}  (4w)^{\tfrac{\Delta}{2}} + \delta(x-x') (4w)^{\tfrac{d-\Delta}{2}} +O(w^{\tfrac{\Delta+2}{2}} ), &  w\to 0 \\
0 \cdot (1-w)^{\tfrac{\Delta}{2}} -  \langle \sigma(x) O(x_L) \rangle_{H^d}^\text{folded} (4(1-w))^{\tfrac{d-\Delta}{2}} +O((1-w)^{\tfrac{d-\Delta+2}{2}} ), & w\to 1 
\end{cases}   \\
&\tilde{K}_{w_R =1} (w,x; x_R) \\
=&\begin{cases}
 \langle O(x)\sigma(x_R) \rangle_{H^d}^\text{folded}  (4w)^{\tfrac{\Delta}{2}} +0 \cdot  w^{\tfrac{d-\Delta}{2}} +O( w^{\tfrac{\Delta+2}{2}} ), & w\to 0 \\
\delta(x-x')  (4(1-w))^{\tfrac{\Delta}{2}} 
-(2\Delta-d)\langle \sigma(x) \sigma(x_R) \rangle_{H^d} (4(1-w))^{\tfrac{d-\Delta}{2}} +O( (1-w)^{\tfrac{d-\Delta+2}{2}} ), & w\to 1
\end{cases} .
\end{split}
\end{equation}
One can see that the two-point functions defined by the above limits of the bulk-boundary propagator (\ref{eq:TildeK}) indeed agree with those computed in Section \ref{Sec:LargeN} using the large $N$ expansion on the CFT side. 
In the expressions above, we introduced the ``folded'' superscript to denote correlators that differ as a function of $\xi$ when going from the folded to the unfolded theory, so it is only necessary to include when the corresponding flat space correlation function contains operators on opposite sides of the interface. For example, in the UV theory we have
\begin{equation}
\begin{split}
\langle O(\mathbf{x}_1, z_1) O(\mathbf{x}_2, z_2) \rangle_{0, H^d}  &= C_O (4\xi)^{-\Delta}, \qquad \text{sign}(z_1 z_2) = 1  \\
\langle O(\mathbf{x}_1, z_1) O(\mathbf{x}_2, z_2) \rangle_{0, H^d}^\text{folded}  &=  C_O (-4\xi)^{-\Delta} \bigg |_{z_1 \to -z_1} = C_O  (4(1+\xi))^{-\Delta}, \qquad \text{sign}(z_1 z_2) = -1. 
\end{split}
\end{equation}  
For comparison with $\tilde{K}$, we also write the asymptotics of the standard bulk-to-boundary propagator $K_\Delta$:
\begin{equation} \label{eq:KAsymptotics}
\begin{split}
& K_\Delta (w,x; w', x') \\
&= \begin{cases} 
\frac{1}{2\Delta-d}\langle O(x') O(x) \rangle_{0, H^d}  (4w)^{\tfrac{\Delta}{2}} + \delta(x-x') (4w)^{\tfrac{d-\Delta}{2}} +O(w^{\tfrac{\Delta+2}{2}} ), & w'=0, w\to 0 \\
\frac{1}{2\Delta-d}\langle O(x') O(x) \rangle_{0, H^d}^\text{folded} (4(1-w))^{\tfrac{\Delta}{2}} + 0 \cdot  (1-w)^{\tfrac{d-\Delta}{2}} +O((1-w)^{\tfrac{\Delta+2}{2}} ), & w'=0, w\to 1 . \\
\end{cases} 
\end{split}
\end{equation}
With an explicit expression for the bulk propagator at coincident points $\tilde{G}(X;X)$, one can compute the one-loop correction to the free energy on the sphere. This is calculated in \cite{Melby-Thompson:2017aip} as
\begin{equation}
\begin{split}
&\frac{(F - F_{0,\text{UV}})+(F - F_{0,\text{IR}})}{2} = -\log \left(\frac{\text{det}(-\nabla^2_{H^{d+1}}+ m^2) \left[\text{det}(-\nabla^2_{H^{d+1}}+ m^2)\right]_{\substack{w \to 1-w \\ \Delta \to d- \Delta }} }{\text{det}\left(-\nabla^2_{H^{d+1}}+ m^2 \right)_{0, \text{UV}} \text{det}\left(-\nabla^2_{H^{d+1}}+ m^2 \right)_{0,\text{IR}}} \right)^{-\frac{1}{2}} \\
&\frac{\partial}{\partial m^2} \left[ \frac{(F - F_{0,\text{UV}})+(F - F_{0,\text{IR}})}{2} \right] = \frac{1}{2} \int d^{d+1} X \sqrt{g_{H^{d+1}}} \biggr[ \tilde{G}_\Delta (X;X) + \tilde{G}_{d-\Delta} (X;X) \\
&\qquad\qquad\qquad\qquad\qquad\qquad\qquad\qquad\qquad\qquad\qquad\qquad\qquad - G_\Delta (X;X) - G_{d-\Delta} (X;X) \biggr]
\end{split}
\end{equation}
where, as in Section \ref{Sec:LargeNFreeEnergy}, we explicitly write the ``UV'' subscript to distinguish the UV and IR theories on the full sphere. As discussed in Section \ref{Sec:LargeNFreeEnergy}, the above AdS calculation matches our large $N$ CFT result in Eq. \eqref{F-final}. 

\section{Equivalence of CFT and AdS calculations}
\label{sec:equivalence}
In this section, we will show there is a manifest equivalence between calculations of correlation functions in the $H^{d+1}$ bulk and those in the interface CFT. As in the simpler case of a double-trace deformation over the whole space \cite{Hartman:2006dy, Giombi:2011ya, Giombi:2018vtc}, the equivalence follows from certain identities involving the bulk two-point functions and bulk-to-boundary propagators discussed in the previous section. We begin by describing the identities, then proceed to apply them to the explicit examples of one-point and three-point functions to show the equivalence of AdS and CFT calculations. The extension to general correlation functions should follow along similar lines. 

\subsection{Identity for \texorpdfstring{$\tilde{G} -G_\Delta $}{TEXT}}
\begin{figure}[h!]
\centering
\includegraphics[scale=1.5]{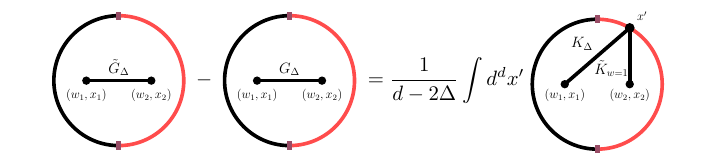}
\caption{Diagrammatic representation of \eqref{eq:GIdentity}}
\label{fig:GTildeIdentity}
\end{figure}

As pointed out in \cite{Melby-Thompson:2017aip}, the difference $\tilde{G} - G_\Delta$ can be expressed as a convolution on the boundary between $K_\Delta$ and a function proportional to $\tilde{K}_{w=1}$:
\begin{equation}\label{eq:GIdentity}
\tilde{G}(w_1, x_1; w_2, x_2) - G_\Delta (w_1, x_1; w_2, x_2) = \frac{1}{d-2\Delta} \int_{H^d (w'=1)} d^d x'  K_\Delta (w_1, x_1; x') \tilde{K}_{w'=1}(w_2, x_2; x'). 
\end{equation}
This identity is depicted diagrammatically in Figure \ref{fig:GTildeIdentity}. 
To prove the identity, one can notice that each side of \eqref{eq:GIdentity} satisfies the scalar equation of motion in the $H^{d+1}$ bulk (this is immediate to see), and has the same boundary conditions. To see the matching boundary conditions, we can take $w_2 \to 0$ using the asymptotic expressions given in \eqref{eq:TildeGAsymptotics}, \eqref{eq:GAsymptotics}, \eqref{eq:TildeKAsymptotics}, and \eqref{eq:KAsymptotics}, which gives
\begin{align}
&\text{\underline{$w_2 \to 0$:}}   \nonumber \\
&LHS(\ref{eq:GIdentity})  \to   w_2^{\Delta/2}  \frac{1}{d-2\Delta} \left( -\tilde{K}_{w_2=0} (w_1, x_1; x_2) + K_\Delta (w_1, x_1;x_2) \right) \\
&\qquad\qquad\qquad + w_2^{(d-\Delta)/2} \cdot 0  +\dots  \nonumber \\
&RHS(\ref{eq:GIdentity}) \to   w_2^{\Delta/2}\frac{1}{d-2\Delta} \int_{H^d (w'=1)} d^d x' \sqrt{g} K_\Delta(w_1,x_1; x') \langle \sigma(x') O(x_2)  \rangle_{H^d} \\
&\qquad\qquad\qquad + w_2^{(d-\Delta)/2} \cdot 0  + \dots   \nonumber  
\end{align}
where one only needs the vanishing $w_2^{(d-\Delta)/2}$ asymptotics to specify boundary conditions. Similarly, taking $w_2 \to 1$ gives\footnote{Again, 
when we write bulk results in terms of CFT correlators, we specialize to the 
normalization choice $C_O=(2\Delta-d)\mathcal{C}_{\Delta}$, which follows from our bulk-to-boundary propagator conventions.}
\begin{align}
&\text{\underline{$w_2 \to 1$:}}   \nonumber \\
&LHS(\ref{eq:GIdentity})  \to   (1-w_2)^{\Delta/2} \frac{1}{d-2\Delta} K_\Delta (w_1, x_1; x_2)  \\
&\qquad\qquad +  (1-w_2)^{(d-\Delta)/2}  \frac{1}{d-2\Delta} \tilde{K}_{w_2=1} (w_1, x_1; x_2)   +  \dots  \nonumber \\
&RHS(\ref{eq:GIdentity})  \to (1-w_2)^{\Delta/2} \frac{1}{d-2\Delta} K_\Delta (w_1, x_1; x_2 ) \\
&\qquad
(1-w_2)^{(d-\Delta)/2} 
\int_{H^d (w'=1)} d^d x' \sqrt{g} K_\Delta(w_1,x_1; x') \langle \sigma(x_2) \sigma(x') \rangle_{H^d} + \dots  \nonumber
\end{align}
where one only needs the $(1-w_2)^{\Delta/2} $ asymptotics to specify boundary conditions. Clearly there are matching boundary conditions for each side of \eqref{eq:GIdentity}, verifying the validity of the expression. In addition, by matching the coefficients of $w_2^{\Delta/2}$ as $w_2 \to 0$ and $(1-w_2)^{(d-\Delta)/2}$ as $w_2 \to 1$, we arrive at two identities relating $\tilde{K}$ to a convolution on the boundary between $K_\Delta$ and a CFT correlator:
\begin{align}
\tilde{K}_{w_2=0} (w_1, x_1;  x_2) &= K^{w_2=0}_\Delta (w_1, x_1;  x_2)- \int_{H^d (w'=1)} d^d x' \sqrt{g} K^{w'=1}_\Delta(w_1,x_1; x') \langle \sigma(x')  O(x_2) \rangle_{H^d} \label{eq:TildeK0Identity} \\
\tilde{K}_{w_2=1}  (w_1, x_1; x_2)  &= 
\ (d-2\Delta) 
\int_{H^d (w'=1)} d^d x' \sqrt{g} K^{w'=1}_\Delta(w_1,x_1; x') \langle  \sigma(x')  \sigma(x_2)\rangle_{H^d} \label{eq:TildeK1Identity} . 
\end{align}
These identities are depicted diagrammatically in Figure \ref{fig:KTildew0} and \ref{fig:KTildew1}. 
In Appendix \eqref{App:KTildeIdentities} we perform explicit calculations to verify these identities. 
\begin{figure}[h!]
\centering
\includegraphics[scale=1.5]{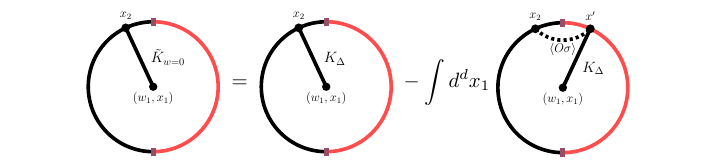}
\caption{Diagrammatic representation of \eqref{eq:TildeK0Identity}}
\label{fig:KTildew0}
\end{figure}

\begin{figure}[h!]
\centering
\includegraphics[scale=1.5]{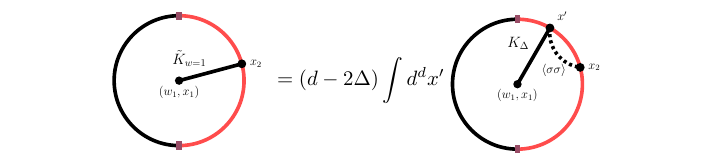}
\caption{Diagrammatic representation of \eqref{eq:TildeK1Identity}}
\label{fig:KTildew1}
\end{figure}

The identities \eqref{eq:GIdentity}, \eqref{eq:TildeK0Identity}, and \eqref{eq:TildeK1Identity} are generalizations of those that appear in \cite{Giombi:2018vtc}, which considers a CFT on the whole space perturbed by a double-trace operator. 

\subsection{One-point functions}
\subsubsection{$\langle \Phi \rangle$}
We now discuss the holographic computation of one-point functions for various operators. 
In Section \ref{Sec:InterfaceData} we considered the one-point function coefficients of $\Phi$, an operator other than $O$ or $\sigma$, when inserted on either the UV or IR side of the interface. In the $H^{d+1}$ bulk, the (Weyl-rescaled) one-point function for $\Phi$ inserted on the IR (``right'') side of the interface is given to leading order by a one-loop tadpole Witten diagram
\begin{equation}\label{eq:PhiOnePointWitten}
\begin{split}
\langle \Phi (x_R) \rangle_{H^d} &=  \vcenter{\hbox{\includegraphics[scale=.85]{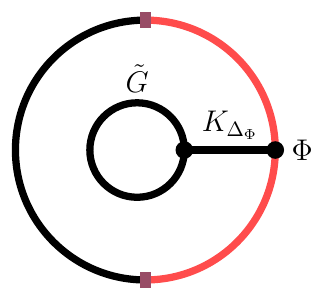}}} +\sum_{h_i \neq \Delta}\vcenter{\hbox{\includegraphics[scale=.85]{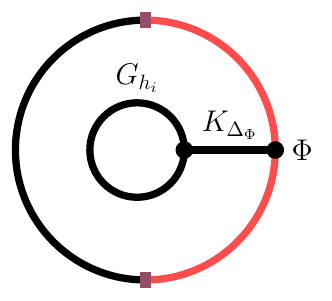}}}  \\
&= -\int_{H^{d+1}}d^{d+1}X K^{w_R=1}_{\Delta_\Phi} (X; x_R) \left( \lambda_{\Phi OO} \tilde{G}(X; X) + \sum_{h_i \neq \Delta}  \lambda_{\Phi h_i h_i} G_{h_i} (X; X) \right) \\
&=  -\lambda_{\Phi OO}  \int_{H^{d+1}}d^{d+1}X K^{w_R=1}_{\Delta_\Phi} (X; x_R) \left(\tilde{G}(X; X)  - G_\Delta (X;X)  \right)  \\
& - \int_{H^{d+1}}d^{d+1}X K^{w_R=1}_{\Delta_\Phi} (X; x_R)  \left( \lambda_{\Phi OO}  G_\Delta (X; X) + \sum_{h_i \neq \Delta}  \lambda_{\Phi h_i h_i}  G_{h_i} (X; X)  \right)
\end{split}
\end{equation}
where $h_i$ labels all fields in the bulk with homogeneous boundary conditions. Note that the term in the last line is just $\langle \Phi \rangle_{0, H^d}$, the one-point function of a scalar in the UV theory on whole space, which is zero by conformal invariance. Then we may apply the identities \eqref{eq:GIdentity} and \eqref{eq:TildeK1Identity} to the other term, which puts the bulk computation in a form closer to the CFT calculation: 
\begin{equation}
\begin{split}
&\langle \Phi (x_R) \rangle_{H^d} = 
-\int_{H^d (w'=1)} d^d x_1 d^d x_2  \langle \sigma(x_1) \sigma(x_2) \rangle_{H^d} \\
&\qquad\qquad\qquad \times  \lambda_{\Phi OO} \int_{H^{d+1}}d^{d+1}X K^{w_R=1}_{\Delta_\Phi} (X; x_R) K^{w_1=1}_\Delta(X; x_1) K^{w_2=1}_\Delta (X; x_2) . \\
\end{split}
\end{equation}
The integral in the last line is just the Witten diagram for propagators with homogeneous bulk boundary conditions, so it corresponds to a (Weyl-rescaled) three-point function in the UV theory on full space: 
\begin{equation}\label{eq:PhiOOWittenHomogeneous}
\begin{split}
&-2 \lambda_{\Phi OO}^\text{UV} \int_{H^{d+1}}d^{d+1}X K^{w_R=1}_{\Delta_\Phi} (X; x_R) K^{w_1=1}_\Delta(X; x_1) K^{w_2=1}_\Delta (X; x_2) \\
&=  \langle \Phi (x_R) O (x_1) O(x_2) \rangle_{0, H^d}  
\end{split}
\end{equation}
The equivalence with the CFT calculation is clear (see eq. (\ref{eq:PhiOnePoint}), and recall we need to perform a Weyl rescaling since we are working with $H^d$ at the boundary) :
\begin{equation}\label{eq:EquivCFTForm}
\begin{split}
\langle \Phi (x_R) \rangle_{H^d} &= \frac{1}{2} \int \frac{d^d x_1 d^d x_2}{z_1^d z_2^d} (z_1 z_2)^{d-\Delta} \langle \sigma(x_1) \sigma(x_2) \rangle z_R^{\Delta_\Phi}  (z_1 z_2)^\Delta \langle \Phi (x_R) O (x_1) O(x_2) \rangle_{0}   \\
&= z_R^{\Delta_\Phi} \frac{1}{2} \int_{\mathbb{R}_+^d} d^d x_1 d^d x_2  \langle \sigma(x_1) \sigma(x_2) \rangle \langle \Phi (x_R) O (x_1) O(x_2) \rangle_{0} . \\
\end{split}
\end{equation}
When $\Phi$ is placed on the unperturbed (``left'') side of the interface (at a point $x_L = (\mathbf{x}_L, -z_L)$, $z_L > 0$), we have
\begin{equation}\label{eq:PhiOnePointLeftH}
\langle \Phi (x_L) \rangle =  \frac{1}{2} \int_{\mathbb{R}_+^d} d^d x_1 d^d x_2 \langle \Phi (x_L) O (x_1) O(x_2) \rangle_{0} \langle \sigma(x_1) \sigma(x_2) \rangle.
\end{equation}
In this case, the corresponding tadpole Witten diagram is the same as \eqref{eq:PhiOnePointWitten} after replacing $K^{w=1}_{\Delta_\Phi} $ with $K^{w=0}_{\Delta_\Phi} $. We can show the equivalence with the CFT calculation by applying the same identities that we used for the $\langle \Phi (x_R) \rangle$ case.
\subsubsection{$\langle O \rangle$ and $\langle \sigma \rangle$}
In the bulk, the tadpole Witten diagram that corresponds to the $O$ one-point function is
\begin{equation}
\begin{split}
&\langle O (x_L) \rangle_{H^d} =  \vcenter{\hbox{\includegraphics[scale=.85]{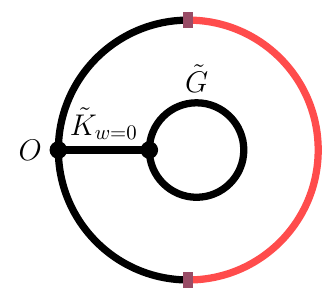}}} +\sum_{h_i \neq \Delta}\vcenter{\hbox{\includegraphics[scale=.85]{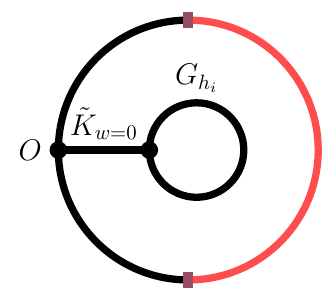}}} \\
&= -\int_{H^{d+1}}d^{d+1}X  \tilde{K}_{w=0} (X; x_L) \left(3 \lambda_{OOO} \tilde{G}(X; X) + \sum_{h_i \neq \Delta} \lambda_{O h_i h_i }  G_{h_i} (X; X) \right) \\
& = -3\lambda_{OOO}  \int_{H^{d+1}}d^{d+1}X  \tilde{K}_{w=0} (X; x_L) \left( \tilde{G}(X; X) -G_\Delta(X; X) \right) \\
& -  \int_{H^{d+1}}d^{d+1}X K_\Delta (X; x_L) \left(3 \lambda_{OOO}  G_\Delta(X; X)+ \sum_{h_i \neq \Delta} \lambda_{O h_i h_i }  G_{h_i} (X; X) \right) \\
& +  \int_{H^d}  d^d x_1 \langle \sigma(x_1) O(x_L) \rangle  \int_{H^{d+1}}d^{d+1}X K_\Delta (X; x_1) \left(3 \lambda_{OOO}   G_\Delta(X; X)+ \sum_{h_i \neq \Delta} \lambda_{O h_i h_i }  G_{h_i} (X; X) \right) \\
\end{split}
\end{equation}
where we applied the identity \eqref{eq:TildeK0Identity}. Note that the last two lines produce $\langle O \rangle_{0}$ upon integrating over the bulk, which is zero by conformal invariance. Then after applying the identities  \eqref{eq:GIdentity} and \eqref{eq:TildeK1Identity}, we have 
\begin{equation}
\begin{split}
&\langle O (x_L) \rangle_{H^d} \\
&= -3 \lambda_{OOO}\biggr[ \int_{H^d}  d^d x_1 d^d x_2 \langle \sigma(x_1) \sigma(x_2) \rangle_{H^d} \int_{H^{d+1}} d^{d+1} X K_\Delta (X; x_L) K_\Delta (X; x_1) K_\Delta (X; x_2)  \\
& - \int_{H^d}  d^d x_1 d^d x_2 d^d x_3 \langle \sigma(x_1) O(x_L) \rangle_{H^d} \langle \sigma(x_2) \sigma (x_3) \rangle_{H^d}  \int_{H^{d+1}} d^{d+1} X K_\Delta (X; x_1) K_\Delta (X; x_2) K_\Delta (X; x_3) \biggr]
\end{split}
\end{equation}
The integral in the last line is just the Witten diagram for propagators with homogeneous boundary conditions and relates to the corresponding CFT three-point function by
\begin{equation}\label{eq:OOOWittenHomogeneous}
\begin{split}
&-6 \lambda_{OOO}^\text{UV} \int_{H^{d+1}}d^{d+1}X K^{w_1=1}_\Delta(X; x_1) K^{w_2=1}_\Delta (X; x_2) K^{w_2=1}_\Delta (X; x_3)   \\
&=  \langle O (x_1) O (x_2) O(x_3) \rangle_{0, H^d}  
\end{split}
\end{equation}
This puts the bulk computation in an equivalent form as the CFT calculation, see eq. (\ref{O1pt-CFT}) (recall again that we are working in conventions where $C_O\equiv (2\Delta -d) \mathcal{C}_\Delta$,  as explained above). 

The tadpole Witten diagram corresponding to the $\sigma$ one-point function to leading order at large $N$ is\footnote{There is a factor of $d-2\Delta$ on the LHS because of the differing normalizations between the field $\sigma$ defined on the CFT side (whose two-point function in the absence of the interface has a normalization given by $C_{\sigma}$ in (\ref{eq:Csig-norm})), and the operator $O_{d-\Delta}$ dual to the bulk scalar field with $d-\Delta$ boundary conditions, whose two-point function obtained from the bulk has the normalization (in the absence of the interface) $\langle O_{d-\Delta}(x_1)O_{d-\Delta}(x_2)\rangle = (d-2\Delta)\mathcal{C}_{d-\Delta}/x_{12}^{2(d-\Delta)}$, see \cite{Klebanov:1999tb,Melby-Thompson:2017aip}. The two operators are related by $O_{d-\Delta}=(d-2\Delta)\sigma$.}
\begin{equation}
\begin{split}
&(d-2\Delta) \langle \sigma (x_R) \rangle_{H^d} =  \vcenter{\hbox{\includegraphics[scale=.85]{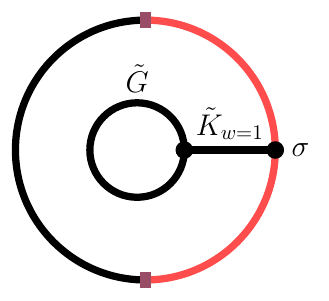}}} +\sum_{h_i \neq \Delta}\vcenter{\hbox{\includegraphics[scale=.85]{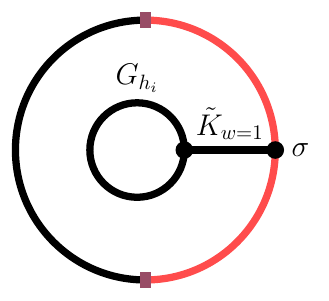}}} \\\\
&= -  \int_{H^{d+1}}d^{d+1}X  \tilde{K}_{w=1} (X; x_R) \left(3 \lambda_{OOO} \tilde{G}(X; X) + \sum_{h_i \neq \Delta} \lambda_{\sigma h_i h_i} G_{h_i} (X; X) \right) \\
&=(2\Delta-d)\int_{H^{d+1}}d^{d+1}X  \int  d^d x_1 K_\Delta(X; x_1) \langle  \sigma(x_1)  \sigma(x_R)\rangle_{H^d}  \\
&\qquad\times \left[ 3\lambda_{OOO} G_\Delta (X; X) +  \sum_{h_i \neq \Delta} \lambda_{\sigma h_i h_i} G_{h_i} (X; X) + \frac{3 \lambda_{OOO}}{d-2\Delta} \int d^d x_2 \tilde{K}_{w=1}(X; x_2) K_\Delta(X; x_2) \right] \\
&=(2\Delta-d)
3\lambda_{OOO} \int d^d x_1 d^d x_2 d^d x_3 \langle \sigma(x_R) \sigma (x_1) \rangle  \langle \sigma(x_2) \sigma (x_3) \rangle \\
&\qquad\qquad\qquad\times  \int_{H^{d+1}}d^{d+1}X K_\Delta (X; x_1) K_\Delta (X; x_2) K_\Delta (X; x_3) 
\end{split}
\end{equation}
where, as before, we applied the identities \eqref{eq:GIdentity} and \eqref{eq:TildeK1Identity} and used the fact that $\langle O \rangle_{0}$ vanishes. After replacing the 3-point Witten diagram of homogeneous bulk-to-boundary propagators with its CFT equivalent \eqref{eq:OOOWittenHomogeneous}, this again matches the CFT calculation in eq. (\ref{sigma1pt-CFT}). 
\subsection{Three-point functions}
We also discuss equivalence of the three point functions $\langle \Phi \Phi O \rangle$, $\langle \Phi O O \rangle$, and $\langle O O O \rangle$ in AdS and CFT. From the CFT perspective, the three-point function $\langle \Phi \Phi O \rangle$ takes the form
\begin{equation}
\begin{split}
\langle \Phi (x_1) \Phi (x_2) O(x_L) \rangle &= \langle \Phi (x_1) \Phi (x_2) O(x_L) \rangle_0 \\
&+ \int_{\mathbb{R}_+^d } d^dx_1' d^dx_2' \langle \Phi(x_1)\Phi (x_2) O(x_1')\rangle_0 \langle O(x_L) O(x_2') \rangle_0 \langle \sigma(x_1') \sigma (x_2') \rangle \\
&= \langle \Phi (x_1) \Phi (x_2) O(x_L) \rangle_0 - \int_{\mathbb{R}_+^d }  d^d x_1'  \langle \Phi(x_1)\Phi (x_2) O(x_1')\rangle_0 \langle \sigma(x_1') O(x_L) \rangle .
\end{split}
\end{equation}
In AdS, the three-point function is given by the Witten diagram
\begin{equation}
\begin{split}
&\langle \Phi (x_1) \Phi (x_2) O(x_L) \rangle_{H^d} =  \vcenter{\hbox{\includegraphics[scale=.85]{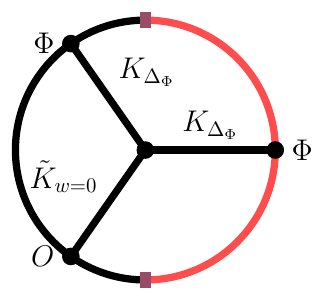}}} \\
&=  -2\lambda_{\Phi \Phi O} \int_{H^{d+1}}d^{d+1}X K_{\Delta_\Phi} (X; x_1) K_{\Delta_\Phi} (X; x_2) \tilde{K}_{w_L=0} (X; x_L) \\
&=  -2\lambda_{\Phi \Phi O} \biggr[ \int_{H^{d+1}}d^{d+1}X K_{\Delta_\Phi} (X; x_1) K_{\Delta_\Phi} (X; x_2) K_\Delta (X; x_L)  \\
& - \int_{H^d} \langle O(x_L) \sigma(x') \rangle_{H^d} \int_{H^{d+1}}d^{d+1}X K_{\Delta_\Phi} (X; x_1) K_{\Delta_\Phi} (X; x_2) K_\Delta^{w_L=0} (X; x_L) \biggr]
\end{split}
\end{equation}
where we used identity \eqref{eq:TildeK0Identity} in the last equality. The remaining integrals over the AdS bulk are standard Witten diagrams corresponding to (Weyl-rescaled) three-point functions in the UV theory in the expected way
\begin{equation}\label{eq:PhiPhiOWittenHomogeneous}
\begin{split}
&-2 \lambda_{\Phi \Phi O}^\text{UV} \int_{H^{d+1}}d^{d+1}X K_{\Delta_\Phi} (X; x_1) K_{\Delta_\Phi} (X; x_2) K_\Delta (X; x_R) \\
&=\langle \Phi (x_1) \Phi (x_2) O(x_R) \rangle_{0, H^d} 
\end{split}
\end{equation}

The CFT calculation for $\langle \Phi \Phi \sigma \rangle $ is given by
\begin{equation}
\begin{split}
\langle \Phi(x_1) \Phi(x_2) \sigma(x_R) \rangle &= -\int_{\mathbb{R}_+^d} d^dx' \langle \Phi (x_1) \Phi(x_2) O(x') \rangle_0 \langle \sigma (x') \sigma(x_R) \rangle
\end{split}
\end{equation}
The Witten diagram is given by
\begin{equation}
\begin{split}
&(d-2\Delta)\langle \Phi(x_1) \Phi(x_2) \sigma(x_R) \rangle_{H^d}  =  \vcenter{\hbox{\includegraphics[scale=.85]{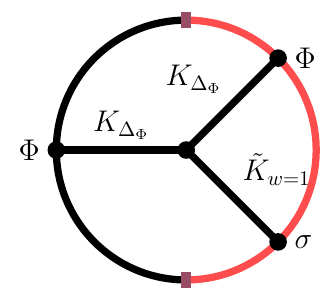}}} \\
&= -2 \lambda_{\Phi \Phi O} \int_{H^{d+1}} K_{\Delta_\Phi} (X; x_1) K_{\Delta_\Phi} (X; x_2) \tilde{K}_{w_R=1} (X; x_R) \\
&=(2\Delta-d)\int_{H^d (w'=1)} d^dx' \langle \Phi (x_1) \Phi(x_2) O(x') \rangle_{0, H^d} \langle \sigma(x') \sigma(x_R) \rangle_{H^d}  
\end{split}
\end{equation}
where we used the relation between the UV CFT three point function and Witten diagram in the homogeneous theory given in \eqref{eq:PhiPhiOWittenHomogeneous}.
The CFT calculation for $\langle \Phi O O \rangle$ is given by
\begin{equation}
\begin{split}
&\langle \Phi(x_1) O(x_{L2}) O(x_{L3}) \rangle = \langle \Phi(x_1) O(x_{L2}) O(x_{L3}) \rangle_0 \\
&+ \frac{1}{2} \int_{\mathbb{R}_+^d } d^dx_1' d^dx_2'  \langle \Phi(x_1) O(x_{L2}) O(x_{L3}) O(x_1') O(x_2') \rangle_0 \langle \sigma(x_1') \sigma (x_2') \rangle \\
&+ \frac{1}{4!} \int_{\mathbb{R}_+^d } d^dx_1' \dots d^dx_4'  \langle \Phi(x_1) O(x_{L2}) O(x_{L3}) O(x_1') \dots O(x_4') \rangle_0 \langle \sigma(x_1') \sigma (x_2')  \sigma(x_3') \sigma (x_4') \rangle .
\end{split}
\end{equation}
After Wick contractions and using the definition of $\langle O \sigma \rangle$, we have
\begin{equation}
\begin{split}
\langle \Phi(x_1) O(x_{L2}) O(x_{L3}) \rangle &= \langle \Phi(x_1) O(x_{L2}) O(x_{L3}) \rangle_0 \\
&- \int_{\mathbb{R}_+^d } d^dx_1' \left[  \langle \Phi(x_1) O(x_{L2}) O(x_1')  \rangle_0 \langle \sigma(x_1') O (x_{L3}) \rangle + (L2 \leftrightarrow L3) \right] \\
&+ \int_{\mathbb{R}_+^d } d^dx_1' d^dx_2'  \langle \Phi(x_1) O(x_1') O(x_2') \rangle_0 \langle \sigma(x_1') O (x_{L2})  \rangle \langle \sigma(x_2') O (x_{L3}) \rangle .
\end{split}
\end{equation}
The corresponding Witten diagram is 
\begin{equation}
\begin{split}
&\langle \Phi(x_1) O(x_{L2}) O(x_{L3}) \rangle_{H^d} =  \vcenter{\hbox{\includegraphics[scale=.85]{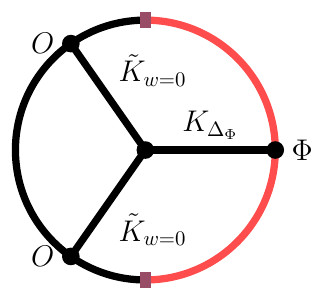}}} \\
&=  -2\lambda_{\Phi OO} \int_{H^{d+1}}d^{d+1}X K^{w_1 =0}_{\Delta_\Phi} (X; x_1) \tilde{K}_{w_{L2}=0} (X; x_{L2}) \tilde{K}_{w_{L3}=0} (X; x_{L3}) \\
&=\biggr[ \langle \Phi(x_1) O(x_{L2}) O(x_{L3}) \rangle_{0,H^d} \\
&- \int_{\mathbb{R}_+^d } d^dx_1' \left[  \langle \Phi(x_1) O(x_{L2}) O(x_1')  \rangle_{0,H^d} \langle \sigma(x_1') O (x_{L3}) \rangle_{H^d} + (L2 \leftrightarrow L3) \right] \\
&+ \int_{\mathbb{R}_+^d } d^dx_1' d^dx_2'  \langle \Phi(x_1) O(x_1') O(x_2') \rangle_{0,H^d} \langle \sigma(x_1') O (x_{L2})  \rangle_{H^d} \langle \sigma(x_2') O (x_{L3}) \rangle_{H^d} \biggr]  . 
\end{split}
\end{equation}
The CFT calculation for $\langle \Phi \sigma \sigma \rangle$ is given by 
\begin{equation}
\begin{split}
\langle \Phi (x_1) \sigma(x_{R2}) \sigma(x_{R3}) \rangle &= \int_{\mathbb{R}_+^d} d^dx_1' d^d x_2' \langle \Phi (x_1) O(x_1') O(x_2') \rangle_0 \langle \sigma(x_{R2}) \sigma(x_1') \rangle \langle \sigma( x_{R3}) \sigma (x_2' ) \rangle
\end{split}
\end{equation}
while the corresponding Witten diagram is given by
\begin{equation}
\begin{split}
&(2\Delta-d)^2\langle \Phi (x_1) \sigma(x_{R2}) \sigma(x_{R3}) \rangle_{H^d} =  \vcenter{\hbox{\includegraphics[scale=.85]{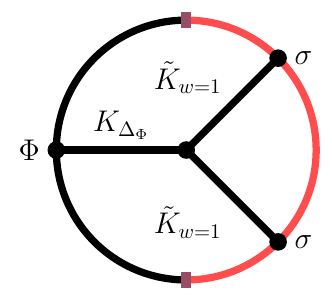}}} \\
&= -2\lambda_{\Phi OO}  \int_{H^{d+1}}d^{d+1}X K_{\Delta_\Phi} (X; x_1) \tilde{K}_{w_{L2}=0} (X; x_{L2}) \tilde{K}_{w_{L3}=0} (X; x_{L3}) \\
&=(2\Delta-d)^2
 \int_{H^d} d^dx_1' d^d x_2' \langle \Phi (x_1) O(x_1') O(x_2') \rangle_{0,H^d} \langle \sigma(x_{R2}) \sigma(x_1') \rangle_{H^d} \langle \sigma( x_{R3}) \sigma (x_2' ) \rangle_{H^d}.
\end{split}
\end{equation}
The CFT calculation for $\langle OOO \rangle$ involves several Wick contractions, and the final result at leading order in large $N$ is given by
\begin{equation}
\begin{split}
& \langle O (x_{L1}) O(x_{L2}) O(x_{L3} \rangle = \langle O(x_{L1}) O(x_{L2}) O(x_{L3}) \rangle_0 \\
&- \left[ \int_{\mathbb{R}_+^d} d^dx_1'  \langle O(x_{L1}) O(x_{L2}) O(x_1') \rangle_0 \langle O(x_{L3}) \sigma(x_1') \rangle + (L3 \leftrightarrow L1) +(L3\leftrightarrow L2) \right] \\
&+ \biggr[\int_{\mathbb{R}_+^d} d^dx_1' d^d x_2' \langle O(x_{L3}) O(x_1') O(x_2')  \rangle_0 \langle O(x_{L1}) \sigma(x_1') \rangle \langle O(x_{L2}) \sigma(x_1') \rangle \\
& + (L3 \leftrightarrow L1) +(L3\leftrightarrow L2)  \biggr] \\
&- \int_{\mathbb{R}_+^d} d^dx_1' d^d x_2' d^d x_3' \langle O(x_1') O(x_2') O(x_3')  \rangle_0 \langle O(x_{L1}) \sigma(x_1') \rangle \langle O(x_{L2}) \sigma(x_2') \rangle \langle O(x_{L3}) \sigma(x_3') \rangle .
\end{split}
\end{equation}
The corresponding Witten diagram is 
\begin{equation}
\begin{split}
&\langle O(x_{L1}) O(x_{L2}) O(x_{L3}) \rangle_{H^d} =  \vcenter{\hbox{\includegraphics[scale=.85]{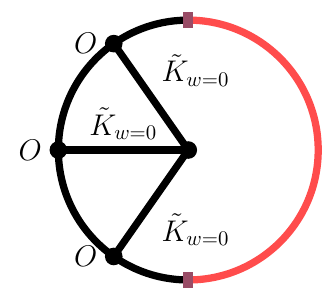}}} \\
&= -6 \lambda_{OOO} \int_{H^{d+1}}d^{d+1}X \tilde{K}_{w_{L1} =0} (X; x_{L1}) \tilde{K}_{w_{L2}=0} (X; x_{L2}) \tilde{K}_{w_{L3}=0} (X; x_{L3}) 
\end{split}
\end{equation}
which, through straightforwardly applying identity \eqref{eq:TildeK0Identity}, can be seen to agree with the CFT calculation.
Finally, the CFT computation of $\langle \sigma \sigma \sigma \rangle$ is given by
\begin{equation}
\begin{split}
&\langle \sigma(x_{R1})\sigma(x_{R2})\sigma(x_{R3}) \rangle =  \\
&= - \int_{\mathbb{R}_+^d} d^dx_1' d^dx_2' d^d x_3' \langle O(x_1') O(x_2') O(x_3') \rangle_0 \langle \sigma(x_{R1}) \sigma(x_1') \rangle \langle \sigma(x_{R2}) \sigma(x_2') \rangle \sigma(x_{R3}) \sigma(x_3') \rangle
\end{split}
\end{equation}
and the corresponding Witten diagram is
\begin{equation}
\begin{split}
&(d-2\Delta)^3\langle \sigma(x_{R1})\sigma(x_{R2})\sigma(x_{R3}) \rangle_{H^d} 
=\vcenter{\hbox{\includegraphics[scale=.85]{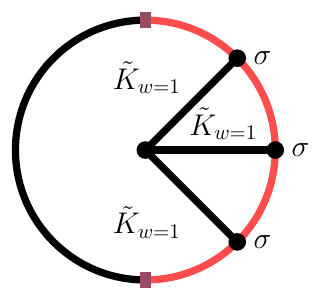}}} \\
&= - 6\lambda_{OOO} \int_{H^{d+1}}d^{d+1}X \tilde{K}_{w_{R1} =1} (X; x_{R1}) \tilde{K}_{w_{R2}=1} (X; x_{R2}) \tilde{K}_{w_{R3}=1} (X; x_{R3}) \\
&=(2\Delta-d)^3
\int_{H^d} d^dx_1' d^d x_2' d^d x_3' \langle O (x_1') O(x_2') O(x_3') \rangle_{0,H^d} \langle \sigma(x_{R1}) \sigma(x_1') \rangle_{H^d} \\
&\qquad\times \langle \sigma( x_{R2}) \sigma (x_2' ) \rangle_{H^d} \langle \sigma( x_{R3}) \sigma (x_3' ) \rangle_{H^d}
\end{split}
\end{equation}
which agrees with the CFT calculation. 
\section*{Acknowledgments}
The work of SG and EH is supported in part by the US NSF under Grant No.~PHY-2209997. The work of HK is supported in part by the US NSF under Grant No.~PHY-2210420. We are grateful to Marco Meineri for useful correspondence. 

\appendix

\section{Two Point functions}\label{App:CorrelationFunctions}
\subsection{$\langle \sigma \sigma\rangle$ }\label{App:SigmaPropagator}
In this section we compute the sigma two-point function by inverting on half space the two-point function of $O$ from the UV theory on whole space. We have
\begin{equation}\label{eq:InversionProblem}
\int_0^{\infty} dz \int d^{d - 1} \mathbf{x}  \langle O (x_1) O (x) \rangle_0 \langle \sigma (x) \sigma (x_2) \rangle  = -\delta^d (x_1 - x_2)
\end{equation}
where the correlation functions are constrained by conformal symmetry to take the form
\begin{equation}\label{eq:ConflSym}
\langle O (x) O (x') \rangle_0 = \frac{C_O}{(x-x')^{2 \Delta}}\equiv \frac{g (\xi)}{(4 z z' )^{\Delta}}, \hspace{1cm} \langle \sigma (x) \sigma (x') \rangle = \frac{h(\xi)}{(4 z z' )^{d - \Delta}}.
\end{equation}
We perform the inversion using the boundary CFT methods in \cite{McAvity:1995zd, Giombi:2020rmc}, which we review here.  We would like to restate the inversion problem \ref{eq:InversionProblem} in terms of a single variable, so we first introduce another integral over coordinates parallel to the interface $\int d^{d-1} \mathbf{x}$, giving
\begin{equation}
\begin{split}
 \int d^{d-1} \mathbf{x}_1  \int_0^\infty dz  \int d^{d - 1} \mathbf{x} \frac{g (\xi_1)}{(4 z z_1 )^{\Delta}} \frac{h(\xi_2)}{(4 z z_2 )^{d - \Delta}} &= -\delta (z_1 - z_2)  
\end{split}
\end{equation}
where we defined 
\begin{equation}
\xi_i = \frac{(\mathbf{x}-\mathbf{x}_i)^2+(z-z_i)^2}{4 z z_i},\qquad \rho_i = \frac{(z-z_i)^2}{4z z_i}.
\end{equation}
The act of integrating out  $\int d^{d-1} \mathbf{x}$ corresponds to an integral transform, which has an inverse. We call this transform the ``hat transform'' and define it as follows
\begin{equation}\label{eq:HatTransform}
\begin{split}
\hat{f}(\rho_1) &\equiv \frac{1}{(4z z_1)^{(d-1)/2}} \int d^{d-1} \mathbf{x} f(\xi_1 ) \\
&= \frac{\pi^{(d-1)/2}}{\Gamma(\tfrac{d-1}{2})} \int du \ u^{(d-3)/2} f(u+\rho_1) 
\end{split}
\end{equation}
The inverse of the hat transform is given by 
\begin{equation}\label{eq:InverseHatTransform}
f(\xi_1 ) = \frac{1}{\pi^{(d-1)/2}\Gamma(\tfrac{1-d}{2})} \int_0 ^\infty d\rho_1 \ \rho_1 ^{-(d+1)/2} \hat{f}(\rho_1+\xi_1) 
\end{equation} 
and this will be important for obtaining $\langle \sigma \sigma \rangle$. The integrals over $\mathbf{x}_1$ and $\mathbf{x}$ produce the hat transforms of $g(\xi_1)$ and $h(\xi)$, so we have
\begin{equation}\label{eq:InversionProblem2}
\int_0^\infty \frac{dz}{2z} \hat{g} (\rho_1) \hat{h} (\rho_2) = -2 z_1 \delta(z_1-z_2) .
\end{equation}
We can then use a change of variables to rewrite the inversion problem in Eq. \ref{eq:InversionProblem2} over positive $z$ as a convolution over the real line, which has a simple Fourier transform. Using the variables
\begin{equation}
z_1 = e^{2 \theta_1}, \hspace{0.5cm} z_2 = e^{2 \theta_2}, \hspace{0.5cm} z = e^{2 \theta}
\end{equation}
Eq. \ref{eq:InversionProblem2} becomes
\begin{equation}
\int_{-\infty}^\infty d\theta \hat{g}(\sinh^2(\theta_1-\theta)) \hat{h}(\sinh^2 (\theta-\theta_2)) = -\delta(\theta_1 - \theta_2). 
\end{equation}
Then Fourier transforming, we find that the Fourier transforms of $\hat{h}$ and $\hat{g}$ are related by
\begin{equation}
\begin{split}
\int d\theta_{12} d\theta e^{i k \theta_{12}} \hat{g}(\sinh^2(\theta_{12}-\theta)) \hat{h}(\sinh^2 (\theta)) &= - 1 \\
\tilde{\hat{h}}(k) = - \frac{1}{\tilde{\hat{g}}(k)}. 
\end{split}
\end{equation}
We can straightforwardly compute $\tilde{\hat{g}}(k)$ by applying the two integral transforms to $g(\xi)$:
\begin{equation}
\begin{split}
&g(\xi) = \frac{C_O}{\xi^{\Delta}} \\
&\hat{g} (\rho) = \frac{ C_O \pi^{\frac{d-1}{2}} \rho ^{\frac{1}{2} (d-2 \Delta -1)} \Gamma \left(\frac{1 - d}{2}+\Delta \right)}{\Gamma (\Delta )} \\
&\tilde{\hat{g}} (k) = \frac{C_O \pi^{\frac{d-1}{2}} \Gamma \left( d- 2 \Delta \right) \Gamma \left(\frac{ 1 - d}{2}+\Delta \right)}{2^{d - 2 \Delta}\Gamma (\Delta )} \left( \frac{\Gamma \left( \frac{1 - d + 2 \Delta}{2} - i \frac{k}{2} \right)}{\Gamma \left(\frac{1 + d - 2 \Delta}{2} - i \frac{k}{2} \right)} + c. c. \right) .
\end{split}
\end{equation}
Then $\tilde{\hat{h}}(k)$ is
\begin{equation}\label{eq:hhat}
 \tilde{\hat{h}} (k) = \frac{- 2^{d - 2 \Delta - 1}\Gamma (\Delta ) }{ C_O \pi^{\frac{d - 3}{2}} \Gamma \left( d- 2 \Delta \right) \Gamma \left(\frac{ 1 - d}{2}+\Delta \right) \Gamma \left(\frac{1 - d + 2 \Delta - i k}{2} \right) \Gamma \left(\frac{1 - d + 2 \Delta + i k}{2} \right) \sin \pi \left(\frac{1-d + 2 \Delta}{2} \right) \cosh \left(\frac{\pi k }{2} \right)} .
\end{equation}
Now we can work backwards to obtain the expression for $h(\xi)$ through an inverse Fourier transform followed by an inverse hat transform. The inverse Fourier transform can be computed by summing over residues of $k$. We have
\begin{equation}\label{eq:sumoverres}
\begin{split}
&-\frac{\pi ^{\frac{3}{2}-\frac{d}{2}} 2^{d-2 \Delta -1} \Gamma (\Delta ) \sec \left(\frac{1}{2}\pi  (d-2 \Delta )\right)}{C_O \Gamma (d-2 \Delta ) \Gamma \left(-\frac{d}{2}+\Delta
   +\frac{1}{2}\right)} \int_{-\infty}^\infty \frac{dk}{2\pi} e^{- i k \theta} \frac{\text{sech}\left(\frac{\pi  k}{2}\right)}{\Gamma \left(\tfrac{-d-i k+1}{2} +\Delta \right) \Gamma \left(\tfrac{-d+i k+1}{2} +\Delta \right)} \\
&= -\frac{2 \pi ^{-\frac{d}{2}-1} e^{-\theta } \Gamma (\Delta ) \sin \left(\frac{1}{2} \pi  (d-2\Delta )\right) \, _2F_1\left(1,\frac{1}{2} (d-2 \Delta +2);-\frac{d}{2}+\Delta +1;e^{-2\theta }\right)}{C_O \Gamma \left(\Delta -\frac{d}{2}\right)}
\end{split}
\end{equation}
After applying the hypergeometric identity\footnote{We are temporarily assuming $\theta >0$, so $e^{-2\theta} < 1$ and the identity is valid. We actually made this assumption when closing the contour in the lower half $k$-plane in Eq. \ref{eq:sumoverres}, which converges for $\theta > 0$. If we instead consider $\theta <0$ and close the contour in the upper half plane of $k$, the result contains a hypergeometric function with the same arguments, except for the last argument which is $e^{2\theta}$ instead of $e^{-2\theta}$. Thus, the identity \ref{eq:DLMF15815} still applies, and in fact the final answer for $\hat{h}(\theta)$ is the same function for $\theta >0$ and $\theta <0$.\label{commentOnResidues}}
\begin{equation}\label{eq:DLMF15815}
\, _2F_1(a,b;a-b+1;z) = (z+1)^{-a} \, _2F_1\left(\frac{a}{2},\frac{a+1}{2};a-b+1;\frac{4 z}{(z+1)^2}\right), \qquad |z| < 1
\end{equation}
we have
\begin{equation}
\begin{split}
\hat{h}(\theta) &= -\frac{2 \pi ^{-\frac{d}{2}-1} e^{\theta } \Gamma (\Delta ) \sin \left(\frac{1}{2} \pi  (d-2\Delta )\right) \, _2F_1\left(\frac{1}{2},1;-\frac{d}{2}+\Delta +1;\text{sech}^2(\theta)\right)}{C_O \left(e^{2 \theta }+1\right) \Gamma \left(\Delta -\frac{d}{2}\right)} \\
\hat{h}(\rho) &= \frac{ \Gamma (\Delta) \sin \pi\left( \Delta - \frac{d}{2} \right)}{C_O \pi^{\frac{d}{2} + 1} \Gamma \left( \Delta - \frac{d}{2} \right) \sqrt{1 + \rho}} {}_2 F_1 \left( \frac{1}{2}, 1, 1 - \frac{d}{2} + \Delta, \frac{1}{1 + \rho} \right) \\
\end{split}
\end{equation}
We now rewrite the hypergeometric function as a sum, apply the inverse hat transform, and perform the sum
\begin{equation}
\begin{split}
h(\xi) &= \frac{ (d-2 \Delta ) \Gamma (\Delta ) \sin \pi\left( \tfrac{d-2 \Delta}{2} \right)}{2 \pi^{d+1} C_O \Gamma \left(\frac{1}{2}-\frac{d}{2}\right)} \sum_{n=0}^\infty \frac{\Gamma \left(n+\frac{1}{2}\right)}{\Gamma \left(-\frac{d}{2}+n+\Delta +1\right)} \int_0^\infty d\rho \rho^{-\frac{d+1}{2}} (1+ \rho +\xi)^{-n-\frac{1}{2}} \\
 &= \frac{\Gamma (\Delta) \sin \pi\left( \Delta - \frac{d}{2} \right) \Gamma \left( \frac{d}{2} \right)  }{C_O \pi^{d + 1} \Gamma \left( \Delta - \frac{d}{2} \right) \xi^{\frac{d}{2}}} {}_2 F_1 \left( \frac{d}{2},  - \frac{d}{2} + \Delta, 1 - \frac{d}{2} + \Delta, -\frac{1}{\xi} \right).
\end{split}
\end{equation}
In the last line we used we used a hypergeometric identity\footnote{$\, _2F_1(a,b;c;z) = (1-z)^{-b} \, _2F_1\left(b,c-a;c;\frac{z}{z-1}\right)$}. The full result for the $\sigma$ propagator at leading order in large $N$ is thus
\begin{equation}\label{eq:SigmaPropagatorApp}
\begin{split}
\langle \sigma (x) \sigma( x') \rangle &= \frac{h(\xi)}{(4 z z' )^{d - \Delta}}\\
&= \frac{1}{(4 z z' )^{d - \Delta}} \frac{\Gamma (\Delta) \sin \pi\left( \Delta - \frac{d}{2} \right) \Gamma \left( \frac{d}{2} \right)  }{C_O \pi^{d + 1} \Gamma \left( \Delta - \frac{d}{2} \right) \xi^{\frac{d}{2}}} {}_2 F_1 \left( \frac{d}{2},  - \frac{d}{2} + \Delta, 1 - \frac{d}{2} + \Delta, -\frac{1}{\xi} \right).
\end{split}
\end{equation}
\subsection{$\langle O \sigma \rangle$} \label{App:OSigmaPropagator}
In this section we derive the correlator $\langle O (x_1) \sigma (x_2) \rangle$, where $x_1 \in R_-^d$ and $x_2 \in R_+^d$ are on opposite sides of the interface. We will work in the folded theory for convenience, taking $z_1 \to -z_1$ and $\xi_1 \to -(1+ \xi_1)$. We can perform similar manipulations as in Appendix \eqref{App:SigmaPropagator}. The definition of $\langle O (x_1) \sigma (x_2) \rangle_\text{folded}$ is
\begin{equation}\label{eq:defOsigma}
\langle O(x_1) \sigma (x_2) \rangle_\text{folded} = -  \int_0^\infty dz \int d^{d-1} \mathbf{x}  \langle O(x_1) O(x) \rangle_{0, \text{folded}} \langle \sigma (x) \sigma( x_2) \rangle
\end{equation}
The correlators $\langle O\sigma \rangle_\text{folded}$ and $\langle OO \rangle_\text{folded}$ are constrained by conformal symmetry to take the form
\begin{equation}\label{eq:ConflSym2}
\langle O(x) \sigma (x') \rangle_\text{folded} \equiv \frac{k(\xi)}{(2 z)^{\Delta}(2 z')^{d - \Delta}}, \hspace{1cm} \langle O (x) O (x') \rangle_{0, \text{folded}} \equiv \frac{g_f (\xi)}{(4 z z' )^{\Delta}} =  \frac{C_O}{(4 z z' )^{\Delta}}\frac{1}{(1+\xi)^\Delta}
\end{equation}
while $\langle \sigma \sigma \rangle$ was just derived in \ref{eq:SigmaPropagatorApp}. Plugging this into \ref{eq:defOsigma} and integrating both sides by $\int d^{d-1} \mathbf{x}_1$ gives
\begin{equation}
\begin{split}
\int d^{d-1} \mathbf{x}_1 \frac{k(\xi_{12})}{(2 z_1)^{\Delta}(2 z_2)^{d - \Delta}}  &= - \int d^{d-1} \mathbf{x}_1 \int_0^\infty dz \int d^{d-1} \mathbf{x} \frac{g_f(\xi_1)}{(4z_1 z)^\Delta}\frac{h(\xi_2)}{(4 z z_2 )^{d - \Delta}} \\
\hat{k}(\rho_{12}) &= -\int_0^{\infty} \frac{d z}{2 z} \hat{g}_f (\rho_1) \hat{h}(\rho_2)  .
\end{split}
\end{equation}
where 
\begin{equation}
\rho_{12} = \frac{(z_1-z_2)^2}{4 z_1 z_2}, \hspace{0.5cm} \rho_1 = \frac{(z_1 - z)^2}{4 z_1 z}, \hspace{0.5cm} \rho_2 = \frac{(z_2 - z)^2}{4 z_2 z}.
\end{equation}
Using the same change of variables as before
\begin{equation}
z_1 = e^{2 \theta_1} \hspace{0.5cm} z_2 = e^{2 \theta_2},
\end{equation}
we have
\begin{equation}
\hat{k}(\sinh^2 \theta_{12}) = -\int_{-\infty}^{\infty} d\theta \ \hat{g}_f (\sinh^2(\theta_{12}-\theta)) \hat{h}(\sinh^2(\theta)).
\end{equation}
The Fourier transform over $\theta_{12}$ of this convolution is just
\begin{equation}\label{eq:relationBetweenFourier}
\begin{split}
\tilde{\hat{k}} (k) &= - \int d\theta_{12} e^{i k \theta_{12}} \int d\theta \ \hat{g}_f (\sinh^2(\theta_{12}-\theta)) \hat{h}(\sinh^2(\theta)) \\
&= - \tilde{\hat{g}}_f (k) \tilde{\hat{h}}(k) .
\end{split}
\end{equation}
While $\tilde{\hat{h}}(k)$ is given in \ref{eq:hhat}, we still need to compute $\tilde{\hat{g}}_f (k)$ by applying a hat transform and Fourier transform. We have
\begin{equation}\label{eq:gfResult}
\begin{split}
g_f (\xi) &= \frac{1}{(1+\xi)^\Delta} \\
\hat{g}_f (\rho) &= \frac{C_O \pi ^{\frac{d-1}{2}} \Gamma \left(-\frac{d}{2}+\Delta+\frac{1}{2}\right)}{\Gamma (\Delta )} (\rho +1)^{\frac{1}{2} (d-2 \Delta -1)} \\
\tilde{\hat{g}}_f (k) &= \frac{\text{Co} \pi ^{d/2} \Gamma \left(\frac{1}{2} (-d-i k+1)+\Delta \right) \Gamma \left(\frac{1}{2} (-d+i k+1)+\Delta \right)}{\Gamma (\Delta ) \Gamma \left(-\frac{d}{2}+\Delta +1\right)}
\end{split}
\end{equation}
In the last line we used the following result
\begin{equation}\label{eq:Fouriergf}
\begin{split}
\int_{-\infty}^\infty d\theta \ e^{i k \theta} (\cosh^2(\theta))^{\frac{d-2\Delta-1}{2}} &= \frac{2 \, _2F_1\left(1,\frac{1}{2} (d+i k-2 \Delta +1);\frac{1}{2} (-d+i k+3)+\Delta ;-1\right)}{-d+2 \Delta +i k+1} \\
&\qquad -\frac{2 \, _2F_1\left(1,\frac{1}{2} (d-i k-2 \Delta +1);\frac{1}{2} (-d-i k+3)+\Delta ;-1\right)}{d-2 \Delta +i k-1} \\
&= \frac{2^{2 \Delta -d} \Gamma \left(-\frac{d}{2}-\frac{i k}{2}+\Delta +\frac{1}{2}\right) \Gamma \left(-\frac{d}{2}+\frac{i k}{2}+\Delta +\frac{1}{2}\right)}{\Gamma (-d+2 \Delta +1)}
\end{split}
\end{equation}
where we applied the hypergeometric identity 
\begin{equation}\label{eq:DLMF1582}
\begin{split}
\, _2F_1(a,b;c;z) &= \frac{(-z)^{-a} \Gamma (c) \Gamma (b-a) \, _2F_1\left(a,a-c+1;a-b+1;\frac{1}{z}\right)}{\Gamma(b) \Gamma (c-a)} \\
&\qquad +\frac{(-z)^{-b} \Gamma (c) \Gamma (a-b) \,_2F_1\left(b,b-c+1;-a+b+1;\frac{1}{z}\right)}{\Gamma (a) \Gamma (c-b)}
\end{split}
\end{equation}
to the second hypergeometric function in the first line of \ref{eq:Fouriergf}. Plugging the results for $\tilde{\hat{g}}_f$ and $\tilde{\hat{h}}$ into \ref{eq:relationBetweenFourier}, we have simply
\begin{equation}
\begin{split}
\tilde{\hat{k}} (k) &=  \sin \pi (\tfrac{d}{2}- \Delta ) \text{sech}\left(\tfrac{\pi  k}{2}\right).
\end{split}
\end{equation}
It is then straightforward to compute the inverse Fourier transform and inverse hat transform:
\begin{equation}\label{eq:kResult}
\begin{split}
\hat{k}(\rho) &= \frac{\sin \left(\frac{1}{2} \pi  (d-2 \Delta )\right)}{\pi  \sqrt{\rho +1}}  \\
k(\xi) &= \frac{ \Gamma \left( \frac{d}{2} \right) \sin \pi \left( \frac{d}{2} - \Delta \right) }{ \pi^{d/2 + 1}  (1+\xi)^{\frac{d}{2}} }.
\end{split}
\end{equation}
So the final result for the $\langle O\sigma \rangle$ correlator is given by 
\begin{equation}
\begin{split}
\langle O(x_1) \sigma (x_2) \rangle_\text{folded} &= \frac{\Gamma \left( \frac{d}{2} \right) \sin \pi \left( \frac{d}{2} - \Delta \right)  }{ \pi^{d/2 + 1} (2 z_1)^{\Delta}(2 z_2)^{d - \Delta} (1+ \xi_{12})^{\frac{d}{2}} } \\
\langle O(x_1) \sigma (x_2) \rangle &=  \frac{\Gamma \left( \frac{d}{2} \right) \sin \pi \left( \frac{d}{2} - \Delta \right)  }{ \pi^{d/2 + 1} (-2 z_1)^{\Delta}(2 z_2)^{d - \Delta} (-\xi_{12})^{\frac{d}{2}} }
\end{split}
\end{equation}
\subsection{$\langle OO \rangle$} \label{App:OPropagator}
Starting from the $O \sigma$ two point function we now look at deriving the $\< O(x_1) O(x_2) \>$ correlator on the UV side of the interface. Again, we use the folded theory for convenience and take $z_1 \to - z_1$, $z_2 \to - z_2$:
\begin{equation}\label{eq:OOfolded}
\langle O(x_1) O (x_2) \rangle_\text{folded} =  \frac{C_O}{(x_{12})^{2 \Delta}} - \int d^d x  \langle O(x_1) \sigma (x) \rangle_\text{folded} \langle O(x) O(x_2) \rangle_\text{folded}
\end{equation}
Since $\langle OO \rangle$ is constrained by conformal symmetry, the second term of \ref{eq:OOfolded} must take the form
\begin{equation}
\frac{ \ell (\xi)}{(2 z_1)^{\Delta}(2 z_2)^{\Delta}} =  - \int d^d x  \frac{k(\xi_1)}{(2 z_1)^{\Delta}(2 z)^{d - \Delta}} \frac{g_f (\xi_2)}{(4 z z_2 )^{\Delta}} 
\end{equation}
where $k(\xi)$ and $g_f (\xi)$ are defined in \ref{eq:ConflSym2}. We find $\ell(\xi)$ the exact same way we found $k(\xi)$ in \ref{App:OSigmaPropagator}, so we need not repeat all the steps. We have
\begin{equation}
\begin{split}
\tilde{\hat{\ell}} (k) &= - \tilde{\hat{g}} (k) \tilde{\hat{k}} (k) \\
&= -\frac{C_O \pi ^{d/2} \text{sech}\left(\frac{\pi  k}{2}\right) \sin \left(\frac{1}{2} \pi (d-2 \Delta )\right) \Gamma \left(\frac{1}{2} (-d-i k+1)+\Delta \right) \Gamma \left(\frac{1}{2} (-d+i k+1)+\Delta \right)}{\Gamma (\Delta ) \Gamma \left(-\frac{d}{2}+\Delta +1\right)}
\end{split}
\end{equation}
where we plugged in $\tilde{\hat{g}} (k) $ and $ \tilde{\hat{k}} (k)$ from \ref{eq:gfResult} and \ref{eq:kResult}. We can perform the inverse Fourier transform by summing over residues\footnote{Note that the comment in footnote \footref{commentOnResidues} also applies here.} 
\begin{equation}
\begin{split}
\hat{\ell}(\rho) &= - \frac{ C_O \pi^{\frac{d -1}{2}} \Gamma \left( \Delta + \frac{1 - d}{2} \right) }{\Gamma \left( \Delta \right) \rho^{\Delta + \frac{1 - d}{2}}}  + \frac{C_O \pi^{d/2}}{\Gamma(\Delta) \Gamma(1+\tfrac{d}{2}-\Delta)\rho^{1/2}} \, _2F_1\left(\frac{1}{2},\frac{d}{2}-\Delta ;\frac{d}{2}-\Delta +1;-\frac{1}{\rho }\right) \\
\end{split}
\end{equation}
where we applied hypergeometric identities \ref{eq:DLMF15815} and
\begin{equation}\label{eq:DLMF1581}
\, _2F_1(a,b;c;z) = (1-z)^{-a} \, _2F_1\left(a,c-b;c;\frac{z}{z-1}\right)
\end{equation}
to simplify the expression. Then applying the inverse hat transform gives
\begin{equation}
\ell (\xi) = - \frac{C_O}{ \xi^{\Delta}} + \frac{C_O  \Gamma \left(\frac{d}{2}\right) \,_2F_1\left(\frac{d}{2},\frac{1}{2} (d-2 \Delta );\frac{1}{2} (d-2 \Delta +2);-\frac{1}{\xi}\right)}{\Gamma (\Delta ) \Gamma \left(\frac{d}{2}-\Delta +1\right) \xi ^{d/2}}
\end{equation}
where, to evaluate the second term, we rewrote the hypergeometric as a sum, performed the integral, and then performed the sum. 
Thus, our final expression for $\langle OO \rangle$ is 
\begin{equation}
\begin{split}
\langle O(x_1) O (x_2) \rangle &= \frac{1}{(4z_1 z_2)^\Delta}\frac{C_O  \Gamma \left( \frac{d}{2} \right)  }{\Gamma(\Delta) \Gamma \left( \frac{d}{2} + 1 - \Delta \right) \xi_{12}^{ \frac{d}{2}}} {}_2 F_1 \left( \frac{d}{2},  \frac{d}{2} - \Delta, 1 + \frac{d}{2} - \Delta, -\frac{1}{\xi} \right).
\end{split}
\end{equation}
So it is related to the $\sigma$ two point function up to a change of $\Delta \rightarrow d - \Delta$ and change of the coefficient $C_{O} \rightarrow C_{\sigma}$ along with the reflection $z_{1,2} \rightarrow - z_{1,2}$.
\section{Details of $\langle \Phi \rangle$ calculation}\label{App:PhiOnePoint}
In this appendix, we extract the coefficient of the one-point function of $\Phi$ on the free (left) side of the interface, $a_{\Phi, L}$ by multiplying the first and second line of \eqref{eq:PhiOnePointLeft} by $z_L^{\Delta_{\Phi}-d}$, integrating out $(\mathbf{x}_L, z_L)$, and dividing by the regularized volume of hyperbolic space at the end. We have
 \begin{equation}\label{eq:OnePointLeftFirstStep}
\begin{split}
&\frac{\sqrt{\mathcal{N}_\Phi} a_{\Phi, L}}{2^{\Delta_\Phi}} \text{vol}(H^d)=  \frac{C_{\Phi OO} \tilde{C}_\sigma}{2}\int_0^\infty \frac{dz_L}{z_L^{d-\Delta_{\Phi}}} \int d^{d-1}\mathbf{x}_L (\dots) \\
&= \frac{C_{\Phi OO} \tilde{C}_\sigma}{2} \frac{\pi ^{\frac{d-1}{2}} 2^{\Delta _{\Phi }-2 d} \Gamma \left(-\frac{d}{2}+\Delta _{\Phi}+\frac{1}{2}\right)}{\Gamma \left(\frac{\Delta _{\Phi }}{2}\right){}^2} \int_0^\infty \frac{dz_1 dz_2 }{z_1^d z_2^d} \int d^{d-1} \mathbf{x}_1 d^{d-1} \mathbf{x}_2   \\
&\times \, _2F_1\left(\tfrac{d}{2},\tfrac{2\Delta -d}{2};\tfrac{2\Delta + 2 -d}{2};-\frac{1}{\xi _{12}}\right)  \int_0^1 d\alpha \left(\alpha -\alpha ^2\right)^{\frac{\Delta _{\Phi }}{2}-1} \\
&\times \int_{\alpha z_1 + (1-\alpha) z_2}^\infty dz_L \frac{(z_1 z_2)^{\Delta_{\Phi}/2}}{\left(z_L-(\alpha z_1 + (1-\alpha) z_2)\right)^{d-\Delta_{\Phi}} (z_L^2 + \alpha (1-\alpha)x_{12}^2)^{\Delta_{\Phi}+(1-d)/2}} .
\end{split}
\end{equation}
The integral over $z_L$ produces a hypergeometric function which may be written as a Mellin-Barnes integral, after which we may integrate out the Feynman parameter, which gives
\begin{equation}
\begin{split}
&\frac{\sqrt{\mathcal{N}_\Phi} a_{\Phi, L}}{2^{\Delta_\Phi}} \text{vol}(H^d)=  \frac{C_{\Phi OO} \tilde{C}_\sigma}{2} \frac{\pi ^{d/2} 2^{-d-\Delta _{\Phi }} \Gamma \left(-d+\Delta _{\Phi }+1\right)}{\Gamma \left(\frac{\Delta _{\Phi }}{2}\right){}^2}
\int_{-\infty}^\infty \frac{ds}{2\pi i}  \frac{\Gamma (-s) \Gamma \left(s+\frac{\Delta _{\Phi }}{2}\right){}^2}{\Gamma \left(-\frac{d}{2}+s+\Delta _{\Phi }+1\right)} \\
&\times \int_0 ^\infty \frac{dz_1 dz_2}{z_1^d z_2^d} \int d^{d-1} \mathbf{x}_1 d^{d-1} \mathbf{x}_2 \, _2F_1\left(\frac{d}{2},\Delta -\frac{d}{2};-\frac{d}{2}+\Delta +1;-\frac{1}{\xi _{12}}\right) \xi _{12}^{\frac{1}{2} \left(-d+\Delta _{\Phi}-2 \Delta +2 s\right)} . 
\end{split}
\end{equation}
This leaves us with a double integral over hyperbolic space with an integrand that depends only on the hyperbolic distance $\xi_{12}$. We can integrate over this distance using hyperbolic ball coordinates by setting one point at the center of the ball ($\mathbf{x} = 0, z = 1$), where the hyperbolic ball coordinates are defined as
\begin{equation}
ds^2 = \frac{4}{(1-u^2)^2} (du^2 + u^2 d \Omega_{d-1}^2 ).
\end{equation}
Here the chordal distance is given by $\xi_{12} = \frac{u^2}{1-u^2}$. The expression then becomes
\begin{equation}
\begin{split}
&\frac{\sqrt{\mathcal{N}_\Phi} a_{\Phi, L}}{2^{\Delta_\Phi}} \text{vol}(H^d) = \frac{C_{\Phi OO} \tilde{C}_\sigma}{2} \frac{\pi ^{d/2} \left(\Delta -\frac{d}{2}\right) 2^{-\Delta _{\Phi }} \Gamma \left(-d+\Delta _{\Phi }+1\right)}{\Gamma \left(\frac{d}{2}\right)
   \Gamma \left(\frac{\Delta _{\Phi }}{2}\right){}^2} \text{vol}(H^d) \\
&\qquad\times \int_{-\infty}^\infty \frac{ds dt}{(2\pi i)^2}  \frac{\Gamma (-s) \Gamma \left(s+\frac{\Delta _{\Phi }}{2}\right){}^2}{\Gamma \left(-\frac{d}{2}+s+\Delta _{\Phi }+1\right)} \frac{\Gamma (-t) \Gamma \left(\frac{d}{2}+t\right) \Gamma \left(-\frac{d}{2}+t+\Delta \right)}{\Gamma \left(-\frac{d}{2}+t+\Delta +1\right)}\\
&\qquad\times \int d^{d-1} \Omega \int_0 ^1 du \  u^{\Delta _{\Phi }-2 \Delta +2 s-2 t-1} \left(1-u^2\right)^{-\frac{d}{2}-\frac{\Delta _{\Phi }}{2}+\Delta -s+t} \\
&\frac{\sqrt{\mathcal{N}_\Phi} a_{\Phi, L}}{2^{\Delta_\Phi}} =  \frac{C_{\Phi OO} \tilde{C}_\sigma}{2} \frac{\pi ^d \left(\Delta -\frac{d}{2}\right) 2^{-\Delta _{\Phi }} \Gamma \left(-d+\Delta _{\Phi }+1\right)}{\Gamma \left(1-\frac{d}{2}\right)
   \Gamma \left(\frac{d}{2}\right)^2 \Gamma \left(\frac{\Delta _{\Phi }}{2}\right){}^2} \int_{-\infty}^\infty \frac{ds }{(2\pi i)}  \frac{\Gamma (-s) \Gamma \left(s+\frac{\Delta _{\Phi }}{2}\right){}^2}{\Gamma \left(-\frac{d}{2}+s+\Delta _{\Phi }+1\right)} \\
&\times \int_{-\infty}^\infty \frac{ dt}{(2\pi i)}  \frac{\Gamma (-t) \Gamma \left(\frac{d}{2}+t\right) \Gamma \left(\tfrac{2\Delta -d}{2}+t \right) \Gamma \left(s-t +\tfrac{\Delta _{\Phi}-2\Delta}{2}\right) \Gamma \left(\frac{2+2\Delta-\Delta_\Phi-d}{2}-s+t \right)}{\Gamma \left(-\frac{d}{2}+t+\Delta +1\right)} .
\end{split}
\end{equation}
The $t$-integral is in the appropriate form to apply Barnes' second lemma, and after applying this we have
\begin{equation}
\begin{split}
&\frac{\sqrt{\mathcal{N}_\Phi} a_{\Phi, L}}{2^{\Delta_\Phi}} =  \frac{C_{\Phi OO} \tilde{C}_\sigma}{2} \frac{\pi ^d 2^{-\Delta _{\Phi }} \Gamma \left(-\frac{d}{2}+\Delta +1\right) \Gamma \left(-d+\Delta _{\Phi }+1\right)}{\Gamma
   \left(\frac{d}{2}\right) \Gamma (-d+\Delta +1) \Gamma \left(\frac{\Delta _{\Phi }}{2}\right){}^2}  \int_{-\infty}^\infty \frac{ds }{(2\pi i)}  \\
&\times \frac{\Gamma (-s) \Gamma \left(s+\frac{\Delta _{\Phi }}{2}\right) \Gamma \left(-\frac{d}{2}-s+\Delta -\frac{\Delta _{\Phi }}{2}+1\right) \Gamma
   \left(-\frac{d}{2}+s+\frac{\Delta _{\Phi }}{2}\right) \Gamma \left(\frac{d}{2}+s-\Delta +\frac{\Delta _{\Phi }}{2}\right)}{\Gamma \left(-\frac{d}{2}+s+\Delta _{\Phi }+1\right)} .
\end{split}
\end{equation}
The $s$-integral can also be performed using Barnes' second lemma. This gives the final result
\begin{equation}
\begin{split}
\sqrt{\mathcal{N}_\Phi} a_{\Phi, L}  &= \frac{\tilde{C}_{\sigma} C_{\Phi OO} \pi^d \Delta _{\Phi } \Gamma \left(-\frac{d}{2}+\Delta +1\right)^2 \Gamma
   \left(\frac{d}{2}-\Delta +\frac{\Delta _{\Phi }}{2}\right) \Gamma \left(\Delta _{\Phi }-d\right)}{2 \Gamma \left(\frac{d}{2}\right) \Gamma
   \left(\frac{\Delta _{\Phi }}{2}+1\right){}^2 \Gamma \left(-d+\Delta +\frac{\Delta _{\Phi }}{2}+1\right)} \\
   &=-\frac{C_\sigma C_{\Phi OO}  \pi ^d 2^{-\Delta _{\Phi }-1} \Delta _{\Phi } \Gamma
   \left(\frac{d}{2}-\Delta \right) \Gamma \left(-\frac{d}{2}+\Delta +1\right)^2 \Gamma \left(\Delta
   _{\Phi }-d\right) \Gamma \left(\frac{1}{2} \left(d-2 \Delta +\Delta _{\Phi
   }\right)\right)}{\Gamma \left(\frac{d}{2}-\Delta +1\right) \Gamma (d-\Delta ) \Gamma \left(\Delta
   -\frac{d}{2}\right) \Gamma \left(\frac{\Delta _{\Phi }}{2}+1\right){}^2 \Gamma \left(-d+\Delta
   +\frac{\Delta _{\Phi }}{2}+1\right)}
\end{split}
\end{equation}
where in the last line we expressed the result in terms of the normalization $C_\sigma$ of $\langle \sigma \sigma \rangle$ on whole space. 
\section{Green's functions in AdS with inhomogeneous boundary conditions}\label{App:HolographicPropagator}
The scalar propagator given inhomogeneous boundary conditions on $H^{d+1}$ was computed in \cite{Melby-Thompson:2017aip}. For $w_1 < w_2$, it is given by\footnote{This corresponds to ``$G^{- +}(X_1, X_2)$'' in the notation of \cite{Melby-Thompson:2017aip}.} 
\begin{equation}\label{eq:AdSPropagator}
\begin{split}
\tilde{G}(X_1, X_2) &= \frac{2 \pi ^{-\frac{d}{2}-1} \csc \left(\frac{1}{2} \pi  (d-2 \Delta )\right)}{(d-2 \Delta ) \Gamma\left(\frac{d}{2}\right)} \left(1-w_2\right){}^{\Delta /2} w_2^{\Delta /2} \left(1-w_1\right){}^{d-\frac{3 \Delta }{2}}
   w_1^{\frac{d-\Delta }{2}}  \\
   & \times \int_0^\infty d \nu \ \nu  \sinh (2 \pi  \nu ) \Gamma \left(\frac{1}{2} (d-2 i \nu -1)\right) \Gamma \left(\frac{1}{2} (d+2 i \nu -1)\right) \\
   & \times \, _2F_1\left(\frac{1}{2}-i \nu ,i \nu +\frac{1}{2};-\frac{d}{2}+\Delta
   +1;w_1\right) \\
   &\times  \, _2F_1\left(\frac{1}{2} (d-2 \Delta -2 i \nu +1),\frac{1}{2} (d-2 \Delta +2 i \nu
   +1);\frac{1}{2} (d-2 \Delta +2);1-w_2\right)   \\
   &\times \, _2F_1\left(\frac{1}{2} (d-2 i \nu -1),\frac{1}{2} (d+2 i \nu -1);\frac{d}{2};-\xi
   _{12}\right)
\end{split}
\end{equation}
where $X_1, X_2 \in H^{d+1}$, $\xi$ is defined in the same way as before $\xi = \tfrac{(x-x')^2}{4 z z'}$, and $H^d$ is described by the following Poincare metric 
\begin{equation}
ds_{H^{d}}^2 = \frac{dz^2 + d\mathbf{x}^2}{z^2} .
\end{equation}
Note that Eq. \eqref{eq:AdSPropagator} is not symmetric under $w_1 \leftrightarrow w_2$, but under both $w_1 \leftrightarrow w_2$ and $\Delta \to d-\Delta$. The standard holographic propagator with homogeneous $\Delta$ boundary conditions is given by 
\begin{equation}
\begin{split}
G_\Delta (X_1, X_2) &= \frac{2 \pi ^{-d/2} \csc \left(\frac{1}{2} \pi  (d-2 \Delta )\right) \Gamma \left(-\frac{d}{2}+\Delta+1\right)}{(d-\Delta ) \Gamma \left(\frac{d}{2}\right) \Gamma \left(\frac{d}{2}-\Delta +1\right)} \left(1-w_1\right){}^{d-\frac{3 \Delta }{2}} w_1^{\frac{d-\Delta }{2}}
   \left(1-w_2\right){}^{\frac{d-\Delta }{2}} w_2^{\frac{d-\Delta }{2}} \\
   &\times \int_0^\infty d\nu \frac{\nu  \sinh (\pi  \nu ) \Gamma \left(\frac{1}{2} (d-2 i \nu -1)\right) \Gamma
   \left(\frac{1}{2} (d+2 i \nu -1)\right)}{\Gamma \left(-\frac{d}{2}+\Delta -i \nu
   +\frac{1}{2}\right) \Gamma \left(-\frac{d}{2}+\Delta +i \nu +\frac{1}{2}\right)} \\
   &\times \, _2F_1\left(\frac{1}{2}-i \nu ,i \nu +\frac{1}{2};-\frac{d}{2}+\Delta +1;w_1\right) \\
   &\times \, _2F_1\left(-\frac{d}{2}+\Delta -i \nu +\frac{1}{2},-\frac{d}{2}+\Delta +i \nu
   +\frac{1}{2};-\frac{d}{2}+\Delta +1;1-w_2\right) \\
   &\times \, _2F_1\left(\frac{1}{2} (d-2 i \nu -1),\frac{1}{2} (d+2 i \nu -1);\frac{d}{2};-\xi
   _{12}\right)
\end{split}
\end{equation}

\section{Explicit calculation for \texorpdfstring{$\tilde{K}_{w=1}$}{TEXT} and \texorpdfstring{$\tilde{K}_{w=0}$}{TEXT} identities} \label{App:KTildeIdentities}
\subsection{\texorpdfstring{$\tilde{K}_{w=0}$}{TEXT} identity}
To explicitly verify the identity relating $\tilde{K}_{w=0}$ to the standard bulk-to-boundary propagator $K_\Delta$, we start with the following ansatz and solve for $A_1$ and $A_2$:
\begin{equation}
\begin{split}
 \tilde{K} (w_1,x_1; w_2=0, x_2) &= A_1 K_{\Delta}(w_1,x_1; w_2=0, x_2) \\
 & + A_2 \int_{H^d (w=1)} d^d x \sqrt{g} K_{\Delta}(w_1,x_1; w=1, x)   \langle \sigma(x) O (x_2) \rangle_{H^d (w=w_2=1)} . \\ 
 \end{split}
\end{equation}
We will simplify this expression in the same way we manipulated the $\tilde{K}_{w=1}$ identity, starting by integrating both sides by $\int d^{d-1}\mathbf{x}_1$. Since the integral is over the $H^d$ slice at $w=1$, we should fold the CFT correlator to $\mathbb{R}_+^d$ before Weyl rescaling:
\begin{equation}
\begin{split}
 \langle \sigma(x) O (x_2) \rangle_{\text{folded}} &=  \langle \sigma(x) O (x_2) \rangle \underset{z_2 \to -z_2}{\bigg |}  =  - \frac{2^{-d} \pi ^{-\frac{d}{2}-1} \Gamma \left(\tfrac{d}{2}\right) \sin \left(\tfrac{\pi  (d-2 \Delta )}{2} \right)}{(2z)^{d-\Delta} (2z_2)^{\Delta}} \left(\xi _2+1\right)^{-d/2} \\
 \langle \sigma(x) O (x_2) \rangle_{H^d (w=w_2=1)}  &= z^{d-\Delta}z_2^{\Delta} \langle \sigma(x) O (x_2) \rangle_{\text{folded}}\\
 & = 2^{-d} \pi ^{-\frac{d}{2}-1}  \Gamma \left(\tfrac{d}{2}\right) \sin \left(\tfrac{1}{2} \pi  (d-2 \Delta )\right) \left(\xi _2+1\right){}^{-d/2}. 
\end{split}
\end{equation}
The proposed identity is then
\begin{equation}
\begin{split}
&(1-w_1)^{\tfrac{d-\Delta}{2}} w_1^{\tfrac{\Delta}{2}} \left(1+\xi _{12}\right)^{-d/2}  \, _2F_1\left(1,\frac{d}{2};\frac{d}{2}-\Delta +1;\frac{1-w_1}{1+\xi_{12}}\right) \\
&= -A_1 \frac{\pi  \Gamma (\Delta ) \csc \left(\frac{1}{2} \pi  (d-2 \Delta )\right)}{\Gamma \left(\frac{d}{2}\right) \Gamma \left(\Delta-\frac{d}{2}\right)} (1-w_1)^{\Delta /2} w_1^{\Delta /2} \left(\xi _{12}+w_1\right){}^{-\Delta } \\
& + A_2 \frac{2^{-d} \pi ^{-d/2} \Gamma (\Delta )}{\Gamma \left(\Delta -\frac{d}{2}\right)} (1-w_1)^{\tfrac{\Delta}{2}}w_1^{\tfrac{\Delta}{2}} \int_{H^d (w=1)} d^{d-1}\mathbf{x} dz \ z^{-d}  \left(\xi _1 +1- w_1 \right){}^{-\Delta } \left(\xi _2+1\right){}^{-d/2}  .
\end{split}
\end{equation}
Integrating over $\int  d^{d-1} \mathbf{x}_1$ gives
\begin{equation}
\begin{split}
&\frac{(1-w_1)^{\frac{1}{2} (d-2 \Delta )} \, _2F_1\left(\frac{1}{2},1;\frac{d}{2}-\Delta +1;\frac{1-w_1}{\rho _{12}+1}\right)}{\sqrt{\rho_{12}+1}} \\
&= -A_1 \frac{\sqrt{\pi } \csc \left(\frac{1}{2} \pi  (d-2 \Delta )\right) \Gamma \left(-\frac{d}{2}+\Delta +\frac{1}{2}\right)}{\Gamma \left(\Delta
   -\frac{d}{2}\right)} \left(\rho _{12}+w_1\right){}^{\frac{1}{2} (d-2 \Delta -1)} \\
& + A_2 \frac{\Gamma \left(-\frac{d}{2}+\Delta +\frac{1}{2}\right)}{2 \sqrt{\pi } \Gamma \left(\Delta -\frac{d}{2}\right)} \int_0^\infty \frac{dz}{z} \frac{\left(\rho _1+1- w_1\right){}^{\frac{1}{2} (d-2 \Delta -1)}}{\sqrt{\rho _2+1} } . 
\end{split}
\end{equation}
Again using the change of variables $z = e^{2\theta}, z_1 = e^{2\theta_1}, z_2 = e^{2\theta_2}$, we have
\begin{equation}
\begin{split}
&(1-w_1)^{\frac{1}{2} (d-2 \Delta )} \text{sech}\left(\theta _{12}\right)  \, _2F_1\left(\frac{1}{2},1;\frac{1}{2} (d-2 \Delta +2);(1-w_1)
   \text{sech}^2\left(\theta _{12}\right)\right) \\
&= -A_1\frac{\sqrt{\pi } \csc \left(\frac{1}{2} \pi  (d-2 \Delta )\right) \Gamma \left(-\frac{d}{2}+\Delta +\frac{1}{2}\right)}{\Gamma \left(\Delta
   -\frac{d}{2}\right)} \left(\cosh ^2\left(\theta _{12}\right)+w_1\right){}^{\frac{1}{2} (d-2 \Delta -1)} \\
&  +  A_2 \frac{\Gamma \left(-\frac{d}{2}+\Delta +\frac{1}{2}\right)}{2 \sqrt{\pi } \Gamma \left(\Delta -\frac{d}{2}\right)} \int_{-\infty}^{\infty} (2d\theta) \frac{\left(\sinh ^2\left(\theta -\theta _{12}\right)+1-w_1\right){}^{\frac{1}{2} (d-2 \Delta -1)}}{\sqrt{\sinh ^2(\theta )+1}} .
\end{split}
\end{equation}
We expand the LHS and the first term on the RHS about $w_1 = 1$, then Fourier transform and sum over infinite powers of $w_1$. For the convolution on the RHS, we expand one factor about $w_1 = 0$, Fourier transform, and sum at the end. The Fourier transforms of the functions in the convolution are
\begin{equation}
\begin{split}
&\int_{-\infty}^{\infty} d\theta \ e^{i k \theta} \left(\sinh ^2\left(\theta \right)+1-w_1\right)^{\frac{1}{2} (d-2 \Delta -1)} \\
&\qquad = \frac{2^{2 \Delta -d} \Gamma \left(\tfrac{-d-ik +1+2\Delta}{2} \right) \Gamma \left(\tfrac{-d+ik +1+2\Delta}{2}\right) }{\Gamma(-d+2 \Delta +1)} \, _2F_1\left(\begin{array}{cc} \tfrac{-d-i k+1 + 2\Delta}{2} & \tfrac{-d+i k+1 + 2\Delta}{2} \\ -\tfrac{d}{2}+\Delta +1 & \end{array} \bigg | w_1\right) \\
&\qquad =  \frac{2^{2 \Delta -d} \cosh (\tfrac{\pi  k}{2})  \Gamma (\tfrac{-d-ik +1 + 2\Delta}{2}) \Gamma (\tfrac{-d+ik +1+ 2\Delta}{2}) }{\sin (\tfrac{\pi  (d-2 \Delta )}{2} ) \Gamma (-d+2 \Delta +1)} \, _2F_1\left(\begin{array}{cc}\tfrac{-d-i k+1 + 2\Delta}{2} & \tfrac{-d+i k+1}{2} \\ -\tfrac{d}{2}+\Delta +1 & \end{array} \biggr| 1-w_1\right)\\
&\qquad\qquad -\frac{\pi ^{3/2} \csc \left(\tfrac{\pi  (d-2 \Delta )}{2} \right) \left(1-w_1\right){}^{\tfrac{d}{2}-\Delta }}{\Gamma \left(\tfrac{d}{2}-\Delta +1\right) \Gamma \left(-\tfrac{d}{2}+\Delta +\tfrac{1}{2}\right)}  \, _2F_1\left(\begin{array}{cc} \tfrac{1}{2} (i k+1) & -\tfrac{1}{2} i (k+i) \\ \tfrac{d}{2} - \Delta +1 & \end{array} \biggr|1-w_1\right) \\
\end{split}
\end{equation}
and 
\begin{equation}
\int_{-\infty}^{\infty} d\theta \ e^{i k \theta}  \frac{1}{\sqrt{\sinh ^2(\theta )+1}} =  2 \pi  \text{sech}\left(\frac{\pi  k}{2}\right).
\end{equation}
Note that we simplified the Fourier transform of $ \left(\sinh ^2\left(\theta \right)+1-w_1\right)^{\frac{1}{2} (d-2 \Delta -1)}$ using the hypergeometric identity\footnote{valid for $c-a-b \not\in \mathbb{Z}$}
\begin{equation}
\begin{split}
\, _2F_1(a,b;c;z) &= \frac{\Gamma (c) (1-z)^{-a-b+c} \Gamma (a+b-c) \, _2F_1(c-a,c-b;-a-b+c+1;1-z)}{\Gamma (a) \Gamma (b)} \\
&\qquad +\frac{\Gamma (c) \Gamma (-a-b+c) \, _2F_1(a,b;a+b-c+1;1-z)}{\Gamma (c-a) \Gamma (c-b)}  .
\end{split}
\end{equation}
Plugging these results into the original expression, we have 
\begin{equation}
\begin{split}
& \Gamma \left(\tfrac{1+ik}{2} \right) \Gamma \left(\tfrac{1-ik}{2} \right) (1-w_1)^{\frac{d}{2}-\Delta } \, _2F_1\left(\begin{array}{cc} \frac{1}{2}-\frac{i k}{2} & \frac{i k}{2}+\frac{1}{2} \\ \frac{d}{2}-\Delta +1 & \end{array} \bigg | 1-w_1\right) \\
& = A_1 \frac{\Gamma \left(\tfrac{d}{2}-\Delta +1\right) \Gamma \left(\tfrac{-d- i k +1 + 2\Delta}{2} \right) \Gamma\left(\tfrac{-d+ ik +1 + 2\Delta}{2}\right) }{\Gamma \left(-\frac{d}{2}+\Delta +1\right)} \, _2F_1\left(\begin{array}{cc} \tfrac{-d-i k+1+2\Delta}{2}  & \tfrac{-d+i k+1+2\Delta}{2}  \\  -\frac{d}{2}+\Delta +1 & \end{array} \bigg | 1-w_1\right) \\
& + A_2 \biggr[- \Gamma \left(\tfrac{1+i k}{2} \right) \Gamma \left(\tfrac{1-ik}{2} \right) (1-w_1)^{\frac{d}{2}-\Delta } \, _2F_1\left(\begin{array}{cc} \frac{1}{2}-\frac{i k}{2} & \frac{i k}{2}+\frac{1}{2} \\ \frac{d}{2}-\Delta +1 & \end{array} \bigg | 1-w_1\right) \\
& + \frac{\Gamma \left(\tfrac{d}{2}-\Delta +1\right) \Gamma \left(\tfrac{-d- i k +1 + 2\Delta}{2} \right) \Gamma\left(\tfrac{-d+ ik +1 + 2\Delta}{2}\right) }{\Gamma \left(-\frac{d}{2}+\Delta +1\right)} \, _2F_1\left(\begin{array}{cc} \tfrac{-d-i k+1+2\Delta}{2}  & \tfrac{-d+i k+1+2\Delta}{2}  \\  -\frac{d}{2}+\Delta +1 & \end{array} \bigg | 1-w_1\right) \biggr ]
\end{split}
\end{equation}
which is true for $A_1 = 1, A_2 = -1$. Thus, we have shown the identity holds and takes the form
\begin{equation}
\begin{split}
 \tilde{K} (w_1,x_1; w_2=0, x_2) &= K_{\Delta}(w_1,x_1; w_2=0, x_2) \\
 & - \int_{H^d (w=1)} d^d x \sqrt{g} K_{\Delta}(w_1,x_1; w=1, x)   \langle \sigma(x) O (x_2) \rangle_{H^d (w=w_2=1)} . \\ 
 \end{split}
\end{equation}
\subsection{\texorpdfstring{$\tilde{K}_{w=1}$}{TEXT} identity}
To explicitly verify the identity relating $\tilde{K}_{w=1}$ to the standard bulk-to-boundary propagator $K_\Delta$, we start with the following ansatz and solve for $A$:
\begin{equation}\label{eq:BulkBoundaryIdentity}
\begin{split}
& \tilde{K} (w_1,x_1; w_2=1, x_2) = A \int_{H^d (w=1)} d^d x \sqrt{g} K_{\Delta}(w_1,x_1; w=1, x)  (z z_2)^{d- \Delta} \langle \sigma(x) \sigma (x_2) \rangle \\
&  w_1^{\Delta/2}(1-w_1)^{(d-\Delta)/2}(1+\xi_{12})^{-d/2} {}_2 F_1 \left( 1, \tfrac{d}{2},  \Delta-\tfrac{d}{2}+1;  \frac{w_1}{1+\xi_{12}} \right) = - A \frac{(2 \pi )^{-d} \Gamma (\Delta )^2}{C_O \Gamma \left(\Delta -\frac{d}{2}\right)^2}\\
&\times ((1-w_1) w_1)^{\Delta /2} \int d^{d-1} \mathbf{x} dz \ z^{-d} \left(\xi _1-w_1+1\right){}^{-\Delta }  \left(\xi _2+1\right){}^{-d/2} \, _2F_1\left(1,\tfrac{d}{2};\Delta -\tfrac{d}{2}+1;\tfrac{1}{\xi _2+1}\right).
 \end{split}
\end{equation}
where $A$ is a constant to be solved for. We can simplify the expression using the same technique from \cite{McAvity:1995zd} that we used to derive the two-point functions: first we integrate both sides by $\int d^{d-1} \mathbf{x}_1$, which allows us to integrate $\int d^{d-1} \mathbf{x}$ on the RHS:
\begin{equation}
\begin{split}
&\frac{(1-w_1)^{\frac{d}{2}-\Delta} \, _2F_1\left(\frac{1}{2},1;-\frac{d}{2}+\Delta +1;\frac{w_1}{\rho _{12}+1}\right)}{\sqrt{\rho_{12}+1}}  =  -A\frac{\pi ^{-\frac{d}{2}-\frac{1}{2}} \Gamma (\Delta ) \Gamma \left(-\frac{d}{2}+\Delta +\frac{1}{2}\right)}{2 C_O \Gamma \left(\Delta -\frac{d}{2}\right)^2} \\
&\qquad \times \int_0^\infty \frac{dz}{z} \frac{ \left(\rho _1-w_1+1\right)^{\frac{1}{2} (d-2 \Delta-1)} \, _2F_1\left(\frac{1}{2},1;-\frac{d}{2}+\Delta +1;\frac{1}{\rho _2+1}\right)}{\sqrt{\rho _2+1}}
\end{split}
\end{equation}
where $\rho_i = \frac{(z-z_i)^2}{4 z z_i}$. Then we perform a change of variables $z = e^{2\theta}, z_1 = e^{2\theta_1}, z_2 = e^{2\theta_2}$, which turns the RHS into a convolution of two functions of $\theta$ over the real line:
\begin{equation}
\begin{split}
& \frac{(1-w_1)^{\frac{d}{2}-\Delta}}{\sqrt{\cosh^2(\theta_{12})}} \, _2F_1\left(\frac{1}{2},1;-\frac{d}{2}+\Delta +1;\frac{w_1}{\cosh^2(\theta_{12})} \right) = -A\frac{\pi ^{-\frac{d}{2}-\frac{1}{2}} \Gamma (\Delta ) \Gamma \left(-\frac{d}{2}+\Delta +\frac{1}{2}\right)}{2 C_O \Gamma \left(\Delta-\frac{d}{2}\right)^2}  \\
&\times \int_{-\infty}^{\infty} (2d\theta)\left(\sinh ^2\left(\theta -\theta_{12}\right)-w_1+1\right){}^{\frac{1}{2} (d-2 \Delta -1)} \frac{1}{\sqrt{\cosh^2(\theta)}} \, _2F_1\left(\frac{1}{2},1;-\frac{d}{2}+\Delta +1; \frac{1}{\cosh^2(\theta)} \right) .
\end{split}
\end{equation}
This motivates us to Fourier transform both sides. The Fourier transform of the hypergeometric factor on the RHS integrand was already computed during the derivation of $\langle \sigma \sigma \rangle$ so it is just \eqref{eq:hhat} with a different prefactor:
\begin{equation}
\begin{split}
&\int_{-\infty}^{\infty} d\theta e^{i k \theta}  \frac{1}{\sqrt{\cosh^2(\theta)}} \, _2F_1\left(\frac{1}{2},1;-\frac{d}{2}+\Delta +1; \frac{1}{\cosh^2(\theta)} \right) \\
& = \frac{\pi ^{5/2} 2^{d-2 \Delta +1} \text{sech}\left(\frac{\pi  k}{2}\right) \csc (\pi  (d-2 \Delta )) \Gamma \left(\Delta
   -\frac{d}{2}\right)}{\Gamma (d-2 \Delta ) \Gamma \left(-\frac{d}{2}+\Delta +\frac{1}{2}\right) \Gamma \left(\frac{1}{2} (-d-i k+1)+\Delta \right) \Gamma \left(\frac{1}{2} (-d+i k+1)+\Delta \right)}.
\end{split}
\end{equation}
The Fourier transform of the other factor in the RHS integrand can be computed by expanding about $w_1=0$, integrating term by term, and summing over powers of $w_1$ at the end:
\begin{equation}
\begin{split}
&\int_{-\infty}^{\infty} d\theta e^{i k \theta} \left(\sinh ^2\left(\theta \right)-w_1+1\right){}^{\frac{1}{2} (d-2 \Delta -1)} \\
& = \frac{2^{2 \Delta -d} \Gamma \left(\tfrac{-d-i k+2 \Delta +1}{2} \right) \Gamma \left(\tfrac{-d+i k+2 \Delta +1}{2} \right) }{\Gamma (-d+2 \Delta +1)} \, _2F_1\left(\begin{array}{cc}-\frac{d}{2}-\frac{i k}{2}+\Delta +\frac{1}{2}, & -\frac{d}{2}+\frac{i k}{2}+\Delta +\frac{1}{2} \\ -\frac{d}{2}+\Delta+1 & \end{array} \biggr | w_1\right).
\end{split}
\end{equation}
The Fourier transform of the LHS can be found in a similar way, which leads to the expression
\begin{equation}
\begin{split}
&\pi   \text{sech} \left(\frac{\pi  k}{2}\right) \, _2F_1\left(\frac{-d-i k+1}{2} +\Delta ,\frac{-d+i k+1}{2} +\Delta ;-\frac{d}{2}+\Delta
   +1;w_1\right)\\
   & =- A\frac{\pi ^{-d/2} \Gamma (\Delta )}{C_O \Gamma \left(\Delta -\frac{d}{2}\right)}  \text{sech}\left(\frac{\pi  k}{2}\right) \, _2F_1\left(\frac{-d-i k+1}{2} +\Delta ,\frac{-d+i k+1}{2} +\Delta ;-\frac{d}{2}+\Delta
   +1;w_1\right) . 
\end{split}
\end{equation}
This is true for 
\begin{equation}
A =-\frac{\pi ^{d/2} C_O \Gamma \left(\Delta -\frac{d}{2}\right)}{\Gamma (\Delta )} \equiv -\frac{C_O}{\mathcal{C}_\Delta} .
\end{equation}
Thus, we have shown the identity is correct and takes the form
\begin{equation}
\begin{split}
& \tilde{K} (w_1,x_1; w_2=1, x_2) \\
 & = -\frac{C_O}{\mathcal{C}_\Delta} \int_{H^d (w=1)} d^d x \sqrt{g} K_{\Delta}(w_1,x_1; w=1, x)  (z z_2)^{d- \Delta} \langle \sigma(x) \sigma (x_2) \rangle .
 \end{split}
\end{equation}

\bibliographystyle{ssg}

\bibliography{RGInterfaceDraft-bib}

\end{document}